\documentclass[11pt]{article}
\pdfoutput=1

\usepackage{verbatim}
\usepackage{amsmath,amscd}
\usepackage{amsfonts}
\usepackage{amssymb}
\usepackage{bbm}
\usepackage{latexsym}
\usepackage{graphicx}
\usepackage{lscape} 
\usepackage{epsfig}
\usepackage{color}
\usepackage{tikz} 
\usepackage{multirow}
\usepackage{afterpage}
\usepackage{amscd}
\usepackage{cite}
\usepackage{tabu} 

\usepackage{hyperref}
\usepackage{longtable}            
\usepackage[font=normalsize]{caption}

\newcommand{\Z}[1]{\ensuremath{\mathbbm{Z}_{#1}}} 

\newcommand{\Id}{\mathbbm{1}}
\newcommand{\ra}{\rightarrow}
\newcommand{\sm}{{\,\mbox{-}}}
\newcommand{\ProjP}[2]{{{#1}^{\textsc{\tiny$#2$}}_{\textsc{\tiny $\parallel$}}}}
\newcommand{\ProjO}[2]{{{#1}^{\textsc{\tiny$#2$}}_{\textsc{\tiny $\perp$}}}}

\newcommand{\E}[1]{\mathrm{E_{#1}}}
\newcommand{\U}[1]{\mathrm{U(#1)}}
\newcommand{\SU}[1]{\mathrm{SU(#1)}}
\newcommand{\SO}[1]{\mathrm{SO(#1)}}

\newcommand{\der}{\partial}
\newcommand{\bder}{\bar\partial}
\newcommand{\brkt}[2]{\bigl[ ^{#1}_{#2} \bigr]}

\newcommand{\equ}[1]{\begin{gather} #1 \end{gather}}
\newcommand{\equa}[1]{\begin{align} #1 \end{align}}
\newcommand{\pmtrx}[1]{\begin{pmatrix} #1 \end{pmatrix}}
\newcommand{\sfrac}[2]{\mbox{$\frac{#1}{#2}$}}
\newcommand{\tabul}[2]{\begin{tabu}{#1} #2 \end{tabu}}
\newcommand{\arry}[2]{\begin{array}{#1} #2 \end{array}}
\newcommand{\items}[1]{\begin{itemize} #1 \end{itemize}}

\newcommand{\Intr}{\mathbbm{Z}}
\newcommand{\Cplx}{\mathbbm{C}}
\newcommand{\Real}{\mathbbm{R}}
\newcommand{\Ratl}{\mathbbm{Q}}
\newcommand{\Narain}{{\mathrm{I}\hspace{-.4ex}\gG}}

\newcommand{\gm}{\mu}

\newcommand{\ga}{\alpha}
\newcommand{\gb}{\beta}
\newcommand{\gd}{\delta}
\renewcommand{\ge}{\epsilon}
\newcommand{\get}{\eta}
\newcommand{\gf}{\phi}
\newcommand{\gps}{\psi}
\newcommand{\gl}{\lambda}
\newcommand{\gn}{\nu}
\newcommand{\gr}{\rho}
\newcommand{\gs}{\sigma}
\newcommand{\gt}{\tau}
\newcommand{\gp}{\pi}
\newcommand{\gth}{\theta}
\newcommand{\gch}{\chi}

\newcommand{\gTh}{\Theta}
\newcommand{\gD}{\Delta}
\newcommand{\gG}{\Gamma}

\newcommand{\bga}{{\bar \alpha}}
\newcommand{\bgth}{{\bar\theta}}
\newcommand{\bgs}{{\bar\sigma}}
\newcommand{\bgt}{{\bar\tau}}
\newcommand{\bget}{{\bar\eta}}

\newcommand{\bq}{{\bar q}}

\newcommand{\cH}{{\cal H}}
\newcommand{\cM}{{\cal M}}
\newcommand{\cN}{{\cal N}}
\newcommand{\cP}{{\cal P}}
\newcommand{\cZ}{{\cal Z}}

\numberwithin{equation}{section}

\begin{document}

\thispagestyle{empty}

\begin{flushright}
LMU-ASC 13/17 \\
TUM-HEP 1080/17 
\\
\end{flushright}
\vskip .2 cm
\begin{center}
{\Large {\bf T-duality orbifolds of heterotic Narain compactifications} 
}
\\[0pt]

\bigskip
\bigskip {\large
{\bf Stefan Groot Nibbelink$^{a,}$}\footnote{
E-mail: groos@hr.nl},
{\bf Patrick K.S. Vaudrevange$^{b,c,}$}\footnote{
E-mail: patrick.vaudrevange@tum.de},
\bigskip }\\[0pt]
\vspace{0.23cm}
${^a}$ {\it School of Engineering and Applied Sciences, Rotterdam University of Applied Sciences, \\ 
G.J.\ de Jonghweg 4 - 6, 3015 GG Rotterdam, Netherlands}\\[1ex]
${^b}$ {\it Arnold Sommerfeld Center for Theoretical Physics,\\
~~Ludwig-Maximilians-Universit\"at M\"unchen, Theresienstra\ss e 37, 80333 M\"unchen, Germany}\\[1ex]
$^c$ {\it Physik Department T30, Technische Universit\"at M\"unchen, \\
James--Franck--Stra\ss e, 85748 Garching, Germany\\}

\bigskip
\end{center}

\subsection*{\centering Abstract}

To obtain a unified framework for symmetric and asymmetric heterotic orbifold constructions we 
provide a systematic study of Narain compactifications orbifolded by finite order $T$-duality 
subgroups. We review the generalized vielbein that parametrizes the Narain moduli space (i.e.\ 
the metric, the $B$-field and the Wilson lines) and introduce a convenient basis of generators of the 
heterotic $T$-duality group. Using this we generalize the space group description of orbifolds to 
Narain orbifolds. This yields a unified, crystallographic description of the orbifold twists, 
shifts as well as Narain moduli. In particular, we derive a character formula that counts the 
number of unfixed Narain moduli after orbifolding. Moreover, we develop new machinery that 
may ultimately open up the possibility for a full classification of Narain orbifolds. This is done 
by generalizing the geometrical concepts of $\Ratl$-, $\Intr$- and affine classes from the 
theory of crystallography to the Narain case. Finally, we give a variety of examples illustrating 
various aspects of Narain orbifolds, including novel $T$-folds.

\newpage 
\setcounter{page}{1}
\setcounter{footnote}{0}

\tableofcontents

\section{Introduction and conclusions}
\label{SecIntro} 


Since the early days of superstring theory, the heterotic string~\cite{Gross:1984dd,Gross:1985fr,Gross:1985rr} 
has served as a promising candidate theory for a unified quantum description of particle physics 
as well as gravity, see e.g.~\cite{Ibanez:2012zz} for a textbook introduction to string 
phenomenology. One of the main obstacles lies in the fact that the heterotic string is 
conventionally defined in a ten-dimensional space-time. Hence, six spatial dimensions have to be 
compactified in order to make contact to the observable four-dimensional world. 


One possibility is to compactify on a six-dimensional (symmetric) toroidal 
orbifold~\cite{Dixon:1985jw, Dixon:1986jc} which is the quotient of a six-torus $T^6$ by some of 
its discrete isometries, see~\cite{Fischer:2012qj} for a full classification with $\mathcal{N}\geq 1$ 
supersymmetry in four dimensions. For example, one can use an Abelian rotational symmetry $\Z{K}$ 
and define the orbifold geometrically as the quotient space $T^6/\Z{K}$. Especially, in the presence 
of discrete Wilson lines~\cite{Ibanez:1986tp} orbifold compactifications have been used to construct 
(minimal) supersymmetric extensions of the Standard Model (MSSM) from the heterotic 
string~\cite{Ibanez:1987sn,Casas:1988hb,Casas:1987us,Font:1989aj,Bailin:1999nk,Forste:2004ie,Kobayashi:2004ya,Buchmuller:2005jr,Buchmuller:2006ik,Kim:2006hw,Lebedev:2006kn,Kim:2007mt,Lebedev:2007hv,Lebedev:2008un,Blaszczyk:2009in,Pena:2012ki,Nibbelink:2013lua,Carballo-Perez:2016ooy}\footnote{For related MSSM model-building using compactifications of the heterotic 
string on Calabi-Yaus see e.g.~\cite{Braun:2005ux,Braun:2005bw,Blumenhagen:2005ga,Anderson:2012yf,Anderson:2013xka,Nibbelink:2015vha}.}.
These constructions can be considered to be promising directions to connect string theory to particle 
physics: Beside reproducing MSSM-like models, they offer an appealing geometrical interpretation, 
in which many properties of the elementary particles depend on their localization in extra dimensions~\cite{Forste:2004ie,Buchmuller:2005sh,Kobayashi:2006wq,Nilles:2014owa}. 
Unfortunately, these constructions generically leave a number of moduli, like the compactification 
radius $R$, unfixed. 


A possibility to stabilize moduli is to generalize the construction of symmetric orbifolds to 
asymmetric ones: In this case one quotients the compactification space not only geometrically, 
but also by a genuine stringy symmetry~\cite{Mueller:1986yr}. The most famous example of such a 
symmetry of string theory is $T$-duality: In its simplest form, $T$-duality is a $\Z{2}$ 
transformation that identifies a string compactification on a circle with small radius $R$ with 
another compactification on a circle with large radius $1/R$. This is a full quantum duality on the 
string worldsheet as this can be described as field redefinitions in a path integral 
approach~\cite{Buscher:1987sk,Tseytlin:1990nb,Siegel:1993th}. Now, in order to be able to perform 
the quotient by this $T$-duality transformation the radius $R$ can no longer be a free parameter, 
but it has to be fixed at the so-called self-dual value $R=1$ (in string units). This promotes the 
$T$-duality transformation $R \mapsto 1/R$ to a symmetry of the theory. On the left- and 
right-moving coordinate fields $X_\text{l}$ and $X_\text{r}$ this $T$-duality transformation 
is realized by $X_\text{l} \mapsto +X_\text{l}$ and $X_\text{r} \mapsto -X_\text{r}$. 
Hence, in general, such $T$-dualities act differently on the left- and right-moving degrees of 
freedom of the string and the resulting quotient spaces are often called asymmetric 
orbifolds~\cite{Narain:1986qm}. Asymmetric orbifolds provide specific examples of non-geometric 
string backgrounds~\cite{Hellerman:2002ax,Dabholkar:2002sy,Shelton:2005cf} or so-called 
$T$-folds~\cite{Hull:2004in,Hull:2006va}. More recently double field 
theory~\cite{Hull:2009mi,Hohm:2010jy,Aldazabal:2013sca} was introduced as an attempt to obtain a 
setting with doubled geometry to describe such $T$-folds using geometrical tools inspired by 
a string field-theoretical description of the left- and right-moving string coordinates. Hence, 
asymmetric string constructions are of increasing interest in the connection to non-geometric 
flux backgrounds~\cite{Lust:2010iy,Condeescu:2012sp}. Various aspects of asymmetric orbifolds have 
been studied in the past~\cite{Harvey:1987da,Ibanez:1987pj,Narain:1990mw,Imamura:1992np,Imamura:1992bz,Sasada:1994yf,Sasada:1994iv,Erler:1996zs,Aoki:2004sm} 
and with recent renewed interest~\cite{Tan:2015nja,Satoh:2016izo} and in particular also in the context on 
non-supersymmetric constructions~\cite{Taylor:1987uv,Satoh:2015nlc,Sugawara:2016lpa}.


In contrast to symmetric orbifolds the phenomenological prospects of heterotic asymmetric orbifolds 
are far less studied. The main asymmetric activities in this direction concentrated up to now on the 
free fermionic construction of the heterotic string~\cite{Kawai:1986ah,Antoniadis:1986rn}. 
These free fermionic models naturally incorporate both, asymmetric as well as symmetric $\Z{2}$ 
twists~\cite{Athanasopoulos:2016aws} and successful MSSM model-building has been carried 
out~\cite{Faraggi:1989ka,Cleaver:1998saa,Faraggi:1991jr,Faraggi:1992fa}. Furthermore, there has 
been some recent activities on model-building using asymmetric $\Z{3}$ 
orbifolds~\cite{Beye:2013moa,Beye:2013ola,Beye:2016snr}. 
Finally, asymmetric string constructions can be further generalized in the covariant lattice 
approach~\cite{Lerche:1986cx} which generalizes the Narain lattice~\cite{Narain:1985jj}, in 
phenomenologically promising Gepner 
models~\cite{Gepner:1987qi,Gepner:1987vz,GatoRivera:2010gv,GatoRivera:2010xn,Schellekens:2016hhf} 
and further with asymmetric CFTs~\cite{Israel:2013wwa,Israel:2015efa,Blumenhagen:2016rof}.

\subsubsection*{Main results}

In this work we develop a generalized space group description of Narain orbifolds and utilize this 
formalism throughout this work to study various aspects of symmetric and asymmetric orbifolds in a 
unified fashion. To define the generalized space group, we first perform a concise investigation of 
the heterotic $T$-duality group: We decompose its generators into geometrical and non-geometric 
ones and use them to parametrize the maximal compact subgroup of the $T$-duality group. This is 
important, as the maximal compact subgroup contains the finite subgroups that can be used to build 
(a-)symmetric orbifolds. Hence, the generalized space group provides a unified framework to study 
symmetric and asymmetric orbifolds in a systematic manner. 

We apply our understanding of the $T$-duality group to derive conditions for the stabilization of 
Narain moduli by orbifolding. This leads us to a closed character formula to count the number of 
unstabilized Narain moduli. In particular, this formula shows that all Narain moduli are 
fixed, if the left- and right-moving twists do not have any irreducible representations of the point 
group in common. We use our findings on moduli stabilization to formulate sufficient conditions for 
a Narain orbifold to exist crystallographically by reducing this question to the question 
whether certain Riccati equations admit solutions. Hence, using our generalized space group 
description one can check that a Narain orbifold exists at least crystallographically and one can 
identify the associated Narain torus that is compatible with the orbifold action. 

Moreover, in this paper we lay the foundation for a classification of Narain orbifolds. 
Even though asymmetric orbifolds have been studied essentially since the birth of superstring 
theory, they have been analyzed so far essentially on a case-by-case basis. Based on our 
definition of the generalized space group we identify equivalence relations for Narain orbifolds. 
These equivalences extend the notations of $\Ratl$-, $\Intr$- and affine-equivalences from theory 
of crystallography to the Narain case leading to the notions of Narain $\Ratl$-, $\Intr$- and 
Poincar\'e-classes. This can be seen as a first step towards a classification of symmetric as well 
as asymmetric Narain orbifolds, which includes -- besides the information on the 
six-dimensional compactification space -- also the anti-symmetric Kalb-Ramond $B$-field, the 
(discrete) Wilson lines and the orbifold shift-vectors in a unified fashion.

Finally, we construct a non-trivial set of (two-dimensional and more general) Narain orbifolds by 
specifying their generalized space groups. We use these examples to illustrate many aspects of our 
study, like the stabilization of Narain moduli and the equivalence classes for Narain orbifolds.

\subsubsection*{Outlook}

In this work we investigated necessary conditions for a Narain orbifold to exist. However, we 
ignored possible extra conditions coming from modular invariance, as they have been studied in the 
past, see e.g.\cite{Narain:1990mw}. However, it would be advantageous to check for full modular 
invariance on the level of the generalized space group and, ultimately, to incorporate modular 
invariance in the definition of generalized space groups such that generalized space groups yield 
modular invariant Narain orbifolds by construction.

Moreover, we can imagine various applications of our work: The space group formulation of Narain 
orbifolds allows for a systematic construction of large sets of examples in various dimensions and 
in both, the $(D,D)$ case as well as the heterotic $(D,D+16)$ case. In addition, using our 
definitions of Narain $\Ratl$-, $\Intr$- and Poincar\'e classes one can unambiguously decide 
whether two Narain orbifold models are physically identical or not. This might proof to be very 
useful for systematic investigations and classifications for various reasons: 

First of all, in the traditional approach two (symmetric) orbifold models are often said to be 
equivalent if their massless matter spectra agree. However, this is neither necessary nor 
sufficient: For example, two different string constructions might possess identical massless 
spectra but different couplings, or the massless spectrum of a given toroidal orbifold 
compactification can be enhanced at specific points in its moduli space. Precisely here the 
Narain Poincar\'e classes would come to the rescue and decide for (in-)equivalence.
However, our new definition of equivalence might be computationally very intensive and, hence, 
further studies might be necessary in order to apply it practically for large computer scans.

Second, having an unambiguous criterion for two Narain orbifolds to be inequivalent, our work can 
be used to classify Narain orbifolds, both symmetric and asymmetric ones. Such a classification 
would automatically include the orbifold twists and shifts as well as the background fields, i.e.\ 
the torus metric, the $B$-field and (discrete) Wilson lines.

Finally, one can use our definitions of Narain $\Ratl$-, $\Intr$- and Poincar\'e classes to decide 
whether a Narain orbifold is genuine asymmetric or only seemingly. Hence, our approach 
might be also very helpful in the study of non-geometrical backgrounds for string theory in 
general, since it has been proven to be quite difficult to obtain concrete, yet true, examples of 
such backgrounds.

\subsubsection*{Paper outline} 

In Section~\ref{SecNarain} we recall the basics of the Narain description of heterotic torus 
compactifications with continuous Wilson lines $A$, the anti-symmetric Kalb-Ramond $B$-field and 
the metric $G$. In this section we exploit the fact that the moduli space of Narain compactifications 
is concisely described as the coset of the continuous $T$-duality group over its maximal compact 
subgroup and the discrete $T$-duality group $\mathrm{O}_{\widehat{\get}}(D,D+16;\Intr)$. 

Given this prominent roles of continuous and discrete $T$-duality groups, we reserve 
Section~\ref{SecTDualityGroup} to study their properties. In particular, we list a complete set of 
generators of $\mathrm{O}_{\widehat{\get}}(D,D+16;\Real)$, which are chosen such that 
they parametrize the discrete $T$-duality group if their parameters are restricted to specific, 
quantized values. In addition, we give the non-linear transformations of the moduli $G,B,A$ under 
arbitrary $T$-duality group elements. 

After these preparations, Section~\ref{SecOrbifolds} sets up a generalized space group description 
of Narain orbifolds involving combined shift- and twist-elements. In this section various properties 
of Narain orbifolds are uncovered. In particular, we show that the shifts of the generalized space 
group are quantized in the directions in which the twists act trivially. Moreover, we emphasize that 
the amount of preserved target-space supersymmetry is solely decided by the twists $\gth_{\ga\,r}$ 
that acts on the right-moving sector. 

Section~\ref{sec:ModuliStabilization} investigates two related questions: i) under what conditions 
does a Narain orbifold exist and ii) how many Narain moduli, $G,B,A$, are fixed. To facilitate this 
discussion the lattice basis is introduced in which the twists are represented by integral 
matrices $\widehat\gr_\ga \in \mathrm{O}_{\widehat{\get}}(D,D+16;\Intr)$. Some properties of these 
twists in the lattice basis can concisely be characterized using the generalized metric $\cH$ and 
the associated $\Intr_2$-grading $\cZ$. By exploiting the coset structure of the Narain moduli 
space, we show that a Narain orbifold exists provided that certain Ricatti equations, i.e.\ 
coupled matrix equations, have a solution. Deformations of such a solution correspond to the 
unconstrained moduli of a Narain orbifold. Using some results collected in 
Appendix~\ref{sec:ModuliDeformations} we derive a character formula to count their number. 

All these results are used in Section~\ref{SecClassification} to lay the foundations for a 
classification of Narain orbifolds. Given that the concepts of $\Ratl$-, $\Intr$- and 
affine-classes proved to be very useful for the classification of symmetric orbifolds, we extend 
these concepts to Narain orbifolds. 

To illustrate the power of the generalized space group description of Narain orbifolds we study 
symmetric orbifolds in Section~\ref{sec:SymmetricZKExample} in this language. Even though the main 
interest of Narain orbifolds lies in the construction of asymmetric orbifolds (or $T$-folds), we 
show in this section that the language of Narain orbifolds gives a convenient, unified description 
of the geometry and the (discrete) Wilson lines.

Finally, in Section~\ref{sec:spacegroupexamples} we employ the Narain $\Ratl$- and $\Intr$-classes 
to study two-dimensional Abelian $\Intr_K$ Narain orbifolds. We provide a large table with many 
examples of previously unknown two-dimensional Narain orbifolds. By an explicit construction we 
show that it is possible to have a \Z{12} two-dimensional Narain orbifold, while it is well-known 
that the largest order of Euclidean $\Z{K}$ twists is $K=6$ in two dimensions. Moreover, $\Ratl$- 
and $\Intr$-classes are particularly useful to distinguish seemingly asymmetric from truly asymmetric 
orbifolds as we illustrate by various examples.

\subsubsection*{Acknowledgments}

This work was partially supported by the DFG cluster of excellence ``Origin and Structure of the 
Universe'' (www.universe-cluster.de) and by the Deutsche Forschungsgemeinschaft (SFB1258).

\newpage


\section[Heterotic Narain torus compactifications]{Heterotic Narain torus compactifications}
\label{SecNarain}

This section reviews the Narain formulation of heterotic torus compactifications~\cite{Narain:1985jj} 
and sets the notation used throughout this work. The moduli space can be described using the generalized 
vielbein $E$, which is parametrized by continuous Wilson lines $A$, the anti-symmetric Kalb-Ramond 
$B$-field and the metric $G$. This vielbein characterizes coordinate field boundary conditions as 
well as the momenta that appear in the representation of the Narain torus partition function as a 
lattice sum.

\subsection[Worldsheet field content of the heterotic string]{Worldsheet field content of the heterotic string}

We parametrize the two-dimensional string worldsheet by (real) coordinates $\gs$ and $\bgs$, defined by 
\equ{
\gs ~=~ \gs_1+ \gs_0~, 
\qquad  
\bgs ~=~ \gs_1 - \gs_0~,
}
where $\gs_0$ and $\gs_1$ denote the worldsheet time and space coordinate, respectively. Worldsheet 
fields that solely depend on $\gs$ or $\bgs$ are called left-moving or right-moving fields, 
respectively. They are correspondingly labelled by a subscript l or r (or in capital letters L/R). 
The heterotic string is closed because of the identification $(\gs_0, \gs_1) \sim (\gs_0, \gs_1 +1)$. 
Hence, $(\gs_0, \gs_1)$ are coordinates on a worldsheet cylinder for the freely propagating string.

The heterotic string~\cite{Gross:1984dd,Gross:1985fr,Gross:1985rr} is described by a conformal 
field theory on the worldsheet with 26 left-moving real bosonic fields and ten right-moving real 
bosonic and fermionic fields. 

The easiest approach to connect this theory to particle physics in $d$ dimensions (for example 
$d=4$) is to perform a stepwise compactification: In the first step one compactifies the 16 surplus 
left-moving bosonic fields on a 16-dimensional torus in order to match the number of left- and 
right-moving bosonic fields to ten. The resulting theory corresponds to a ten-dimensional theory 
with a gauge group dictated by modular invariance of the string partition function. For example, in 
the case of ten-dimensional $\mathcal{N}=1$ supersymmetry the gauge group is fixed to either 
$\E{8}\times\E{8}$ or $\SO{32}$. Then, in a second step one compactifies on a $D$-dimensional space, 
for example on a Calabi-Yau or an orbifold. As a result one obtains a $d$-dimensional theory, where 
$d+D = 10$, e.g.\ $4+6=10$. An alternative approach, which we use in this paper, is the 
so-called Narain construction, where the two-step compactification described above is performed in 
a single step compactification of the heterotic string directly to $d$ dimensions, see 
Section~\ref{NarainConstruction}.

In light-cone gauge two left- and right-moving uncompactified dimensions are gauge-fixed and, hence, 
eliminated. Thus, the heterotic string in light-cone gauge can be described by the following worldsheet fields: 
\begin{itemize}
\item As left-moving fields, there are 8+16=24 real bosonic fields. They are denoted by 
$x_\text{l}^\gm(\gs)$ with $\gm = 2,\ldots, d-1$ ($\gm = 0,1$ are chosen to be fixed in light-cone 
gauge) for the uncompactified and $Y_\text{L}(\gs)$ for the compactified dimensions, respectively. 
Furthermore, we set
\equ{ \label{VectorNotationL} 
Y_\text{L}(\gs) ~=~ \pmtrx{ y_\text{l}(\gs) \\[1ex] y_\text{L}(\gs) }~, 
}
where $y_\text{l}(\gs)=\big(y_\text{l}^i(\gs)\big)$ for $i=1,\ldots,D$ live on the $D$-dimensional 
compactification space. In addition, $y_\text{L}(\gs) = \big(y_\text{L}^I(\gs)\big)$ for 
$I=1,\ldots, 16$ are often referred to as the gauge degrees of freedom. 
\item As right-moving fields, there are eight real bosonic fields plus their real fermionic 
superpartners. They are denoted by $(x_\text{r}^\gm(\bgs), y_\text{r}^i(\bgs))$ and 
$\gps_\text{R}(\bgs) = (\gps_\text{R}^\gm(\bgs), \gps_\text{R}^i(\bgs))$, respectively, with 
$\gm = 2,\ldots, d-1$ and $i=1,\ldots,D$.
\end{itemize}

Left- and right-moving bosonic fields can be combined to coordinate fields $x^\gm(\gs, \bgs)$ and 
$X^i(\gs, \bgs)$ which parametrize the $d$ uncompactified and $D$ compactified dimensions, 
respectively, i.e.
\equa{ \label{LRmoversX}
x^\gm(\gs, \bgs) ~=~ \sfrac 1{\sqrt 2}\left(x_\text{r}^\gm(\bgs) +  x_\text{l}^\gm(\gs)\right)~ \quad\text{and}\quad X^i(\gs, \bgs) ~=~ \sfrac 1{\sqrt 2}\left(y_\text{r}^i(\bgs)  +  y_\text{l}^i(\gs)\right)~.
}
Their classical equations of motion read 
\equ{
\der_\gs \der_\bgs x^\gm(\gs,\bgs) ~=~ 0 \quad\text{and}\quad \der_\gs \der_\bgs X^i(\gs,\bgs) ~=~ 0~, 
}
which is solved by the general ansatz~\eqref{LRmoversX}.

Hence, collectively, we have $2D+16$ compactified bosonic worldsheet fields $Y$ nested in the 
following fashions: 
\equ{ \label{VectorNotation} 
Y(\gs,\bgs) ~=~ \pmtrx{ y_\text{r}(\bgs)  \\[1ex] y_\text{l}(\gs)  \\[1ex] y_\text{L}(\gs)  }~, 
\qquad 
y(\gs,\bgs)  ~=~ \pmtrx{ y_\text{r}(\bgs)  \\[1ex] y_\text{l}(\gs)  }~, 
\qquad 
Y_\text{L}(\gs)  ~=~ \pmtrx{ y_\text{l}(\gs)  \\[1ex] y_\text{L}(\gs)  }~.
}
We define the following dimensions: $D_\text{r} = D_\text{l} = D$ and $D_\text{L}=D_\text{l}+16=D+16$. 
We will use the same notation as in eqn.~\eqref{VectorNotation} for other types of vectors.


The separation~\eqref{LRmoversX} of the coordinate fields $X^i(\gs,\bgs)$ into 
left- and right-moving coordinates $y_\text{l}^i(\gs)$ and $y_\text{r}^i(\bgs)$ is unique up to a 
constant shift of the zero modes $\xi^i$, i.e. 
\equ{ \label{LRgaugeTrans}
Y(\gs,\bgs)  ~\sim~ Y(\gs,\bgs)  + \Xi~, 
\quad 
\Xi ~=~ (\xi, -\xi, 0)~: 
\quad 
y_\text{r}^i(\bgs) ~\sim~  y_\text{r}^i(\bgs) + \xi^i~, 
\quad 
y_\text{l}^i(\gs) ~\sim~  y_\text{l}^i(\gs) - \xi^i~,   
}
with $\xi \in \Real^D$. This has important consequences for the number of worldsheet degrees 
of freedom: If one counts left- and right-movers $y(\gs, \bgs) \in \Real^{2D}$ independently there 
seems to be a doubling of degrees of freedom on the worldsheet compared to the coordinate fields 
$X(\gs, \bgs) \in \Real^D$, see eqn.~\eqref{LRmoversX}. However, due to eqn.~\eqref{LRgaugeTrans} 
there are only $D$ independent zero-modes of $y(\gs, \bgs)$ that specify the position of the string 
and the numbers of worldsheet degrees of freedom are equal for $X(\gs, \bgs)$ and $y(\gs, \bgs)$.


\subsection[Torus partition functions as Narain lattice sums]{Torus partition functions as Narain lattice sums}\label{NarainConstruction}

We consider torus compactifications $T^{2D+16}_\Narain=\Real^{2D+16}/\Narain$ of the $2D+16$ bosonic 
worldsheet fields $Y$. $\Narain$ is a so-called $2D+16$-dimensional Narain lattice, which we will 
analyze in this section in detail. This will be of use when we discuss the more general case of 
Narain orbifolds later in Section~\ref{SecOrbifolds}. 

In the case of a Narain torus, the closed string boundary conditions of the worldsheet fields are 
given by
\equ{ \label{NarainBoundaryConditions}
x(\gs+1,\bgs+1) ~=~ x(\gs,\bgs)~, 
\quad 
\gps_\text{R}(\bgs + 1) ~=~ (-)^s\, \gps_\text{R}(\bgs)~,
\quad 
Y(\gs+1,\bgs+1) ~=~ Y(\gs,\bgs) + L~, 
}
where $s=0,1$ parametrizes the different spin structures of the right-moving fermions $\gps_\text{R}$, 
i.e.\ $s=0$ yields the so-called Ramond sector and $s=1$ the Neveu-Schwarz sector. Furthermore, 
$L \in \Narain$ denotes a lattice vector of $\Narain$.

At one-loop the partition function $Z_\text{full}(\gt, \bgt)$ is given by the string 
vacuum-to-vacuum amplitude which corresponds to a worldsheet torus. This torus is defined by two 
periodicities of worldsheet fields: $(\gs_0, \gs_1) \sim (\gs_0, \gs_1 +1)$ and 
$(\gs_0, \gs_1) \sim (\gs_0 + \tau_2, \gs_1 + \tau_1)$ for the string to close in the 
worldsheet-spatial and worldsheet-time directions, respectively. Here, $\gt=\gt_1+i\, \gt_2$ is the 
so-called modular parameter of the torus. Then, the full partition function 
$Z_\text{full}(\gt, \bgt)$ of the one-loop worldsheet torus can be factorized as follows
\equ{
\label{FullNarainPartition}
Z_\text{full}(\gt, \bgt) ~=~ Z_x(\gt, \bgt)\, Z_{\gps}(\bgt)\, Z_Y(\gt, \bgt)~.
}
The individual partition functions are given by 
\begin{subequations} 
\label{NarainPartition}
\equa{
Z_x(\gt, \bgt) & ~=~ \frac 1{\gt_2^{d/2-1}}\, \left|\frac 1{\get^{(d-2)}(\gt)}\right|^2~, \label{PFnonCompact} 
\\[2ex] 
Z_{\gps}(\bgt) & ~=~ \frac 12\, \frac 1{\bget^4(\bgt)}\, \sum_{s,s'=0}^{1} e^{\gp i( s+s' + s's)}\bgth\brkt{\frac {s}2 e_4}{\frac{s'}2 e_4}~, \label{PFfermions} 
\\[2ex] 
Z_Y(\gt, \bgt) & ~=~ \frac 1{ \bget^{D_\text{r}}(\bgt) \get^{D_\text{L}}(\gt)}\, \sum_{P\in \Narain^*}\, q^{\frac 12 P_\text{L}^2} \, \bq^{\frac 12 p_\text{r}^2}~, \label{GaugeNarainPartition} 
}
\end{subequations} 
where $q = e^{2\pi i\, \gt}$, $\bq = e^{-2\pi i\, \bgt}$ and $e_d = (1,\ldots, 1)$ denotes the 
$d$-dimensional vector with all entries equal to one. Here and in the following we often omit the 
dependencies on $\gt$ and $\bgt$ for notational ease. In addition, $\get(\gt)$ denotes the Dedekind 
function and $\gth$ the theta-function.
The vectors $P$ are from the dual lattice $\Narain^*$ which is defined as $P \in \Narain^*$ if 
\equ{\label{eqn:ConjugateMomentum}
P^T \eta\, L ~\in~ \Intr~,
}
for any $L\in \Narain$. Here, we have introduced the Lorentzian inner product of lattice vectors as
\equ{ 
P^T\, \get\, P' ~=~ - p_\text{r}^T\, p_\text{r}' + P_\text{L}^T\, P_\text{L}'~,
\quad\text{using}\quad 
P ~=~ \pmtrx{p_\text{r} \\  P_\text{L}}\quad\text{and}
\quad 
 \get ~=~ \pmtrx{
-\Id_D &  0 \\[2ex]   
0 & \Id_{16+D}
}~. 
}
The metric $\get$ should not be confused with the Dedekind function $\get(\gt)$ that appears in 
partition functions; we assume that the reader understands from the context which is meant.


The partition function $Z_\gps$ for the right-moving fermions can also be presented as a lattice 
sum, i.e. from~\eqref{PFfermions} we get
\equ{\label{FermiLatticeSum}
Z_\gps(\bgt) ~=~ \frac 1{\bget^4(\bgt)}\,  \sum_{p_\text{R}\in \gG_\gps} \bq^{\frac 12 p_\text{R}^2} (-1)^{F}~,
} 
where the lattice $\gG_{\gps}=\gG_\text{vec} \oplus \gG_\text{spin}$ consists of the vectorial and 
spinorial weight lattices, given by 
$\gG_\text{vec} = \{p_\text{R}\in \Intr^4 ~|~ p_\text{R}^Te_4 = \text{odd}\}$ and 
$\gG_\text{spin} = \{ p_\text{R} + \sfrac 12\, e_4 ~|~ p_\text{R}\in \Intr^4 \text{ and } p_\text{R}^Te_4 = \text{even}\}$. 
Furthermore, $F$ is the target-space fermion number, i.e. $F=0$ for $p_\text{R}\in\gG_\text{vec}$ 
and $F=1$ for $p_\text{R}\in\gG_\text{spin}$.

Eqn.~\eqref{FermiLatticeSum} can also be obtained as follows: the eight real worldsheet fermions 
$\gps_\text{R} =(\gps_\text{R}^\gm,\gps_\text{R}^i)$ can be grouped in four complex fermions 
$\gps_\text{R} = (\gps_\text{R}^m,\gps_\text{R}^a)$, where $m=1,\ldots, d/2-1$ and 
$a=1,\ldots,D/2$ correspond to the uncompactified and compactified dimensions, respectively. Then, 
one can bosonize the complex fermions. Consequently, the bosonized fermions carry momentum 
$p_\text{R} = (p_\text{R}^m,p_\text{R}^a)$ and the associated partition function coincides with 
eqn.~\eqref{FermiLatticeSum}. The momentum $p_R^m$ has an important target-space interpretation: A 
string state with $p_\text{R}^m$ being integer or half-integer signals a target-space boson or 
fermion in $d$ dimensions, respectively.

\subsubsection*{Modular invariance}

The full partition function is required to be modular invariant: At one-loop the worldsheet has the 
topology of a torus with modular parameter $\gt$. Not all $\gt \in \Cplx$ with $\text{Im}(\gt) > 0$ 
parametrize inequivalent worldsheet tori. Because of conformal symmetry tori related by the modular 
transformations 
\equ{ 
T~:~ \gt ~\ra~ \gt + 1~, 
\qquad 
S~:~ \gt ~\ra~ -1/\gt~, 
}
give the same physics. $T$ and $S$ generate the modular group $\text{PSL}(2,\Z{})$. Invariance of 
the partition function~\eqref{FullNarainPartition} under $T$ and $S$ transformations requires that 
\equ{ \label{NarainModInv}
\forall~  P ~\in~ \Narain~: 
\quad 
\sfrac 12\, P^T\, \get\, P ~\equiv ~0
\qquad\text{and}\qquad   
\Narain^* ~=~ \Narain~, 
}
where $a\equiv b$ means that $a$ and $b$ are equal up to 
some integer. These conditions tell us that $\Narain$ is an even self-dual lattice with 
signature $(D,D+16)$; the so-called Narain lattice. Note that vectors $P \in \Narain$ can be 
redefined as
\equ{ \label{Utrans} 
P ~\ra~ U\, P 
}
for $U\in \text{O}(D;\Real)\times\text{O}(D+16;\Real)$ without changing the partition 
function~\eqref{FullNarainPartition}. 

\subsection[Narain lattices]{Narain lattices}
\label{sec:NarainLattice}


We analyse the conditions~\eqref{NarainModInv} in more detail. To do so, we may parametrize a 
general lattice vector $P \in \Narain$ as 
\equ{ \label{NarainElement} 
P ~=~  E\, N~, 
\qquad 
N ~=~ \left(\begin{array}{c}m\\n\\q\end{array}\right) ~\in~ \Intr^{2D + 16}~,
}
in terms of an invertible matrix $E$. This matrix $E$ is called the generalized vielbein of the 
Narain lattice $\Narain$ as its columns correspond to $2D + 16$ basis vectors of the lattice 
$\Narain$. The components of the vector $N$ can be interpreted as winding numbers $m\in \Intr^D$, 
Kaluza-Klein numbers $n\in \Intr^D$ and gauge lattice numbers $q\in \Intr^{16}$. 
From the vielbein $E$ we can define the Narain metric $\widehat{\get}$ as
\equ{ \label{Minkowski}
\widehat{\get} ~=~ E^T\, \get\, E~.
}
Then, the scalar product of two vectors $P_i = E\, N_i \in \Narain$ for $i=1,2$ is given by
\equ{
P_1^T\, \get\, P_2 ~=~ N_1^T\, \left(E^T\, \get\, E\right)\, N_2 ~=~ N_1^T\, \widehat{\get}\, N_2~.
}
Hence, the lattice $\Narain$ is even if
\equ{ \label{EvenSelfDual}
P^T\, \get\, P ~=~ N^T\, \widehat{\get}\, N ~\in~ 2\Intr~.
}
Note that an even lattice is automatically integral, i.e.\ 
$P^T \get\, P' = N^T \widehat{\get} N'\, \in \Intr$. Therefore, the Narain metric $\widehat{\get}$ 
is a symmetric, integer matrix with even entries on the diagonal and signature $(D,D+16)$. 
The dual lattice $\Narain^*$ is spanned by the dual vielbein $E^*$ which is defined as
\equ{ \label{DefDualBasis} 
(E^*)^T \get \, E = \Id_{2D+16}~,
}
so that for a given $P = E^* N\in \Narain^*$ we have $P^T \get P' \equiv 0$ for all 
$P' = E\, N'\in \Narain$. By comparing this equation with~\eqref{Minkowski} one infers that the 
dual basis is given by 
\equ{ \label{DualBasis} 
E^* = E \, \widehat{\get}^{-1}~. 
}
Two lattices are identical if their vielbeins are related by a $\text{GL}(2D+16; \Intr)$ 
transformation. Hence, $\Narain$ is self-dual, $\Narain^* = \Narain$, if the Narain metric 
in eqn.~\eqref{DualBasis} satisfies
\equ{
\widehat{\get} ~\in~ \text{GL}(2D+16; \Intr)~.
}
Consequently, $\det\widehat{\get}=\pm 1$ and we see from eqn.~\eqref{Minkowski} that the volume 
of the unit cell spanned by the vielbein $E$ is given by $\text{vol}(\Narain)=\pm \det E = 1$. 

It is often convenient to choose a special representation of the Narain metric. If not stated 
otherwise we will use
\equ{ \label{eqn:etahat}
\widehat{\get} ~=~ 
\pmtrx{
0 & \Id_D & 0 \\[1ex] 
\Id_D & 0 & 0 \\[1ex]
0 & 0 & g 
}~, 
}
where $g$ is the metric of an even, self-dual 16-dimensional lattice. (Throughout this paper 
we use a hatted notation to refer to objects that are naturally defined in the lattice basis.) We 
choose it to be the Cartan matrix of $\E{8}\times\E{8}$ and write $g=\ga_\text{g}^T \ga_\text{g}$ 
where the columns of $\ga_\text{g}$ are the 16 simple root vectors of $\E{8}\times\E{8}$. 
The explicit expression for $\ga_\text{g}$ is given by
\equ{ \label{E82roots} 
\ga_\text{g} = \pmtrx{ \ga(\E{8}) & 0 \\[1ex] 0 & \ga(\E{8}) }~. 
}
The columns of $\ga(\E{8})$ represent the eight simple roots $\ga^I(\E{8})$, $I=1,\ldots,8$, of the 
exceptional Lie algebra $\E{8}$. They can be chosen as follows
\equ{ \label{E8roots} 
\ga(\E{8}) = 
\pmtrx{ 
~1 & 0 & 0 & 0   & 0 & 0 & 0 & \sm \sfrac 12 
\\ 
\sm 1 & ~1 & 0 & 0   & 0 & 0 & 0 & \sm \sfrac 12 
\\
0 &  \sm 1 & ~1 & 0  & 0 & 0 & 0 & \sm \sfrac 12 
\\ 
0 & 0 & \sm 1 & ~1   & 0 & 0 & 0 & \sm \sfrac 12 
\\ 
0 & 0 & 0 & \sm 1  & ~1 & 0 & 0 & \sm \sfrac 12 
\\ 
0 & 0 & 0 & 0 & \sm 1 & ~1 & 1 & \sm \sfrac 12 
\\ 
0 & 0 & 0 & 0 & 0 & \sm 1 & 1 & \sm \sfrac 12 
\\
0 & 0 & 0 & 0 & 0 & 0 & 0 & \sm \sfrac 12 
}~. 
}

\subsection[The Narain moduli space]{The Narain moduli space}
\label{SecNarainVielbein} 


Given the choice of a Narain metric $\widehat{\get}$ in eqn.~\eqref{eqn:etahat} it is natural to 
look for a corresponding generalized vielbein $E$, which yields this Narain metric 
$E^T \get\, E = \widehat{\get}$. We see that a particular solution $R$ to equation~\eqref{Minkowski} 
is given by
\equ{ \label{eqn:vielbeinR}
R ~=~ \pmtrx{ 
\sfrac 1{\sqrt 2} \Id_D & \sfrac {-1}{\sqrt 2} \Id_D & 0 \\[1ex]
\sfrac 1{\sqrt 2} \Id_D & \sfrac 1{\sqrt 2} \Id_D & 0 \\[1ex]  
0 & 0 & \ga_\text{g}
}
\qquad\text{with}\qquad
R^T\, \get\, R ~=~ \widehat{\get}~.
}
The general solution to~\eqref{Minkowski} can be written in terms of this particular solution as 
\equ{ \label{ExpansionE} 
E ~=~ U\, R\, \widehat{E}~,  
}
so that consequently, 
\equ{ \label{ConditionEhat} 
E^T\, \get\, E ~=~ \widehat{E}^T\, \widehat{\get}\, \widehat{E} ~=~ \widehat{\get}~, 
}
if $U \in \text{O}_{\get}(D,D+16;\Real)$ and $\widehat{E} \in \text{O}_{\widehat{\get}}(D,D+16;\Real)$, 
i.e.\ if $U^T \get\,U=\get$ and $\widehat{E}^T \widehat{\get}\,\widehat{E}=\widehat{\get}$. 

In the following we want to identify which transformations $U$ and $\widehat{E}$ in 
eqn.~\eqref{ExpansionE} map between physically inequivalent theories and which do not. Therefore, 
we will identify the moduli space of heterotic Narain constructions. To do so, we 
define\footnote{\label{ConguationImplicit}In the remainder of this paper we will use this 
conjugation with $R$ to switch between $\text{O}_\get$ and $\text{O}_{\widehat\get}$ group 
elements.} 
\equ{ \label{UhatUequivalence} 
\widehat{U} = R^{-1}\, U\, R
}
and note that $\widehat U \in \text{O}_{\widehat{\get}}(D,D+16;\Real)$ if 
$U \in \text{O}_{\get}(D,D+16;\Real)$. Now, take a general vielbein $E = U\, R\, \widehat{E}$. 
Then, one can absorb $U$ into a redefinition of $\widehat{E}$ by defining $\widehat{E}'$ as
\equ{
\widehat{E}' ~=~ \widehat{U}\, \widehat{E} ~=~ \left(R^{-1}\, U\, R\right)\,\widehat{E} ~\in~ \text{O}_{\widehat{\get}}(D,D+16;\Real) \quad\text{hence}\quad E ~=~ R\, \widehat{E}'~.
}
However, it is not useful to absorb all $U$ transformations in eqn.~\eqref{ExpansionE} into a 
redefinition of $\widehat{E}$: Consider 
$U\in \text{O}(D;\Real)\times\text{O}(D+16;\Real) \subset \text{O}_{\get}(D,D+16)$. As the 
partition function~\eqref{FullNarainPartition} depends only on $P_\text{L}^2$ and $p_\text{r}^2$ 
such transformations leave the partition function invariant. Thus, 
$U\in \text{O}(D;\Real)\times\text{O}(D+16;\Real)$ in eqn.~\eqref{ExpansionE} maps physically 
equivalent theories to each other. On the other hand, $\widehat{E}$ transformations in 
eqn.~\eqref{ExpansionE} change the partition function~\eqref{FullNarainPartition} in general. 
Therefore, $\widehat{E}$ contains the parameters (i.e.\ the moduli) that continuously deform the 
Narain lattice with vielbein $R$ to Narain lattices with vielbeins $R\,\widehat{E}$, which are in 
general physically inequivalent but share the same Narain metric $\widehat{\get}$. However, not all 
vielbeins $\widehat{E}$ are physically inequivalent: 
Consider two vielbeins $E, E'$ for two Narain lattices $\Narain,\Narain'$ satisfying \eqref{Minkowski}. 
Under what condition(s) do these backgrounds describe the same Narain lattice $\Narain'=\Narain$? 
This happens when for each point $P \in\Narain$ there is a unique point $P'\in\Narain'$ which is 
identical to it: In the parametrization \eqref{NarainElement} this amounts to
\equ{\label{eqn:NarainEqual}
U\, E\, N ~=~  U\, P ~=~ P' ~=~ E'\, N'~,
}
such that the integer vectors $N$ and $N'$ are mapped to each other one-to-one, i.e.\ 
$N = \widehat M\, N'$ with $\widehat M = \left(E^{-1}\, U^T E'\right)$. Note that we added 
in eqn.~\eqref{eqn:NarainEqual} a rotation matrix $U\in\text{O}(D;\Real)\times\text{O}(D+16;\Real)$, 
which is unphysical as discussed above. Hence, the Narain lattices $\Narain$ and $\Narain'$ are the 
same if there exists a rotation matrix $U$ such that $\widehat M \in \text{GL}(2D+16; \Intr)$. 
Moreover, we assumed that both $E$ and $E'$ give the same Narain metric 
$\widehat{\get}$, see~\eqref{Minkowski}. This implies that the matrix $\widehat M$ is actually an element 
of the so-called $T$-duality group $\text{O}_{\widehat{\get}}(D,D+16;\Intr)$, i.e. 
\equ{ \label{TdualityMatrix} 
\widehat{M}^T\, \widehat{\get}\, \widehat M ~=~ \widehat{\get}~.
}
(More details on the $T$-duality group $\text{O}_{\widehat{\get}}(D,D+16;\Intr)$ are given in 
Section~\ref{SecTDualityGroup}.) Therefore, Narain compactifications based on the vielbeins 
$E = R\, \widehat{E}$ and $E' = U\,E\,\widehat M$ are physically equivalent, i.e. 
\equ{ \label{eqn:EquivGeneralizedVielbein}
E ~=~ R\, \widehat{E} ~\sim~ E' ~=~ U\, R\, \widehat{E}\, \widehat{M}~,
}
if $U\in \text{O}(D;\Real)\times\text{O}(D+16;\Real)$ and 
$\widehat M \in \text{O}_{\widehat{\get}}(D,D+16;\Intr)$. In terms of $\widehat{E}$ this equivalence 
relation reads 
\equ{ \label{eqn:EquivGeneralizedVielbeinEHat}
\widehat{E} ~\sim~ \widehat U\, \widehat{E}\, \widehat{M}~,
}
where $\widehat U =  R^{-1} U\, R$. This equivalence relation can be used to define a quotient 
space. Consequently, the moduli space of Narain compactifications is uniquely parametrized by an 
element $\widehat E$ in the coset
\equ{ \label{ModuliSpace} 
\mathcal{M} ~=~ 
{\text{O}(D;\Real)\!\times\!\text{O}(D+16;\Real) \backslash 
\text{O}_{\widehat{\get}}(D,D+16;\Real)} / 
 \text{O}_{\widehat{\get}}(D,D+16;\Intr)~.
}
Here, it is understood that the first two factors in the denominator act from the left (via 
$\widehat U$) while the last factor acts from the right (via $\widehat{M}$), see 
eqn.~\eqref{eqn:EquivGeneralizedVielbeinEHat}. The $T$-duality transformations $\widehat{M}$ are 
said to change the duality frame. 

An explicit parametrization of the matrix $U\in \text{O}(D;\Real)\times\text{O}(D+16;\Real)$, 
satisfying 
\equ{ \label{ConstraintsSODD+16} 
U^T\,U=\Id~, \qquad U^T\,\get\, U = \get~, 
}
is given by
\equ{ \label{Uexplicit} 
U ~=~ \pmtrx{ u_\text{r} & 0 & 0 \\ 0 & u_\text{l} & u_\text{lL} \\ 0 & u_\text{Ll} & u_\text{L} }~, 
}
provided that the constraints $u_\text{r}^T u_\text{r} = u_\text{l}^Tu_\text{l} +  u^T_\text{Ll}u_\text{Ll} = \Id_D$, 
$u_\text{lL}^Tu_\text{lL}+u_\text{L}^Tu_\text{L}=\Id_{16}$ and $u_\text{l}^Tu_\text{lL}+u_\text{Ll}^Tu_\text{L}=0$ 
are fulfilled. As we have already seen above, often the closely related matrix
\equ{ \label{eqn:Defintionupm} 
\widehat{U} ~=~ R^{-1} U\, R ~=~ \pmtrx{
u_+ & u_- & \sfrac1{\sqrt 2} u_\text{lL} \ga_\text{g} \\ 
u_- & u_+ & \sfrac1{\sqrt 2} u_\text{lL} \ga_\text{g} \\ 
\sfrac 1{\sqrt 2}\ga_\text{g}^{-1}u_\text{Ll} & \sfrac 1{\sqrt 2}\ga_\text{g}^{-1}u_\text{Ll} & \ga_\text{g}^{-1} u_\text{L} \ga_\text{g}
}~,
\quad\text{where}\quad  
u_\pm ~=~ \frac{1}{2} \left(u_\text{l} \pm u_\text{r}\right)~,
}
is more convenient.

Modulo the transformations $\widehat{U}$ and $\widehat{M}$, the general solution to 
eqn.~\eqref{ConditionEhat} can be represented as 
\equ{ \label{NarainModuli} 
\widehat{E} ~=~ \pmtrx{
e & 0 & 0 
\\[1ex]
-e^{-T} C^T & e^{-T} & -e^{-T} A^T \ga_\text{g}
\\[1ex] 
\ga_\text{g}^{-1} A & 0 & \Id_{16} 
}~, 
\qquad 
C ~=~ B + \sfrac 12\, A^T A~.
}
Hence, $\widehat{E} = \widehat{E}(e,B,A)$ is parametrized by the Narain moduli $e$, $B$ and $A$, 
where $e$ is the $D$-dimensional vielbein of the $D$-torus with metric $G = e^T e$. $A$ is a 
$16 \times D$ matrix, whose $i$-th column contains the Wilson line which is associated to the 
$i$-th basis vector in $e$ and, finally, $B$ denotes the anti-symmetric Kalb-Ramond $B$-field.

In summary, we can specify the most general form of the generalized vielbein $E$ with Narain metric 
$\widehat{\get} = E^T \get\, E$ as given in eqn.~\eqref{eqn:etahat}. It reads
\equ{ \label{eqn:MostGeneralGeneralizedVielbein}
E ~=~ U\, R\, \widehat{E}\, \widehat{M}~, 
}
with $U\in \text{O}(D;\Real)\times\text{O}(D+16;\Real)$ and $\widehat M \in \text{O}_{\widehat{\get}}(D,D+16;\Intr)$. 
The matrix $R$ is given in eqn.~\eqref{eqn:vielbeinR} and the moduli dependent part 
$\widehat{E} = \widehat{E}(e,B,A)$ is specified in eqn.~\eqref{NarainModuli}. In fact, we may take 
$\widehat{M} = \Id$ without loss of generality as we show in Section~\ref{sec:CosetDecomposition}.

\subsubsection*{Equivalent Narain metrics}

One may encounter different Narain metrics, say $\widehat{\get}$ and $\widehat{\get}'$ from 
$\text{GL}(2D+16; \Intr)$, such that 
\equ{ 
E^T\, \get\, E ~=~ \widehat{\get}~, 
\qquad 
E^{\prime T}\, \get\, E' ~=~ \widehat{\get}^{\,\prime}~. 
}
In this case one cannot immediately compare the moduli in $E$ and $E'$, because their hatted 
versions $\widehat{E}$ and $\widehat{E}'$ lie in two different moduli spaces. Since we are talking 
about two representations of the same Narain lattice we have 
\equ{
E\, N ~=~ E'\, N'~, 
\qquad \text{with} \qquad
N ~=~ \widehat M\, N'~, 
}
where $\widehat M \in \text{GL}(2D+16; \Intr)$. Consequently, $E' = E\, \widehat M$ so that 
\equ{ \label{EquivNarainMetrics} 
\widehat{M}^T\, \widehat{\get}\, \widehat{M} ~=~ \widehat{\get}^{\,\prime}~.  
}
Obviously, only those $\widehat M \not\in O_{\widehat{\get}}(D, D+16; \Intr)$ can change the form of the 
Narain metric. Importantly, all Narain metrics can be reached from $\widehat{\get}$ given 
in eqn.~\eqref{eqn:etahat} by transformations $\widehat M \not\in O_{\widehat{\get}}(D, D+16; \Intr)$. Hence, 
we assume in the following that the Narain metric $\widehat{\get}$ is given by eqn.~\eqref{eqn:etahat}.

\subsection[Coordinate fields and momenta]{Coordinate fields and momenta} 

Consider the generalized vielbein in its most general form, i.e.\ 
$E = U\, R\, \widehat E\, \widehat M$, and choose $U=\Id$ and $\widehat{M}=\Id$, see 
eqn.~\eqref{eqn:MostGeneralGeneralizedVielbein}. Then, a Narain lattice vector $P$ is represented as 
\equ{ \label{eqn:NarainEN}
P ~=~ \pmtrx{ p_\text{r} \\ P_\text{L} } ~=~ E N ~=~
\pmtrx{
\frac 1{\sqrt 2} e^{-T} 
\big( (G+C^T)\, m - n + A^T \ga_\text{g}\, q \big) 
\\[1ex] 
\frac 1{\sqrt 2} e^{-T} 
\big( (G-C^T)\, m + n - A^T \ga_\text{g}\, q \big) 
\\[1ex] 
\ga_\text{g}\, q + A\, m
}~.
}
It can be thought of to describe both: On the one hand, $L \in \Narain$ defines the periodicity
for the compactification on a Narain lattice, see eqn.~\eqref{NarainBoundaryConditions}. On the 
other hand, $P \in\ \Narain$ gives the conjugate momentum, see eqn.~\eqref{eqn:ConjugateMomentum}.

The matrix $R$ induces the change of right- and left-moving coordinate fields, $y_\text{r}$, 
$y_\text{l}$ and $y_\text{L}$, to $D$ mixed fields $X$, $\tilde{X}$ and the remaining 16 
left-moving gauge coordinates $X_\text{g}$ 
\equ{ \label{CoordFieldBasisChange}
\widehat{Y} ~=~ R^{-1} \, Y ~=~ \pmtrx{ 
\sfrac   1{\sqrt 2} \Id_D & \sfrac 1{\sqrt 2} \Id_D & 0 \\[1ex]
\sfrac{-1}{\sqrt 2} \Id_D & \sfrac 1{\sqrt 2} \Id_D & 0 \\[1ex]  
0 & 0 & \ga_\text{g}^{-1}
}
\pmtrx{ y_\text{r} \\[1ex] y_\text{l} \\[1ex]  y_\text{L} } ~=~
\pmtrx{ {X} \\ \tilde{X} \\ X_\text{g} } 
~,
}
see eqn.~\eqref{LRmoversX}. This relation thus defines which combination of right- and left-moving 
degrees of freedom are interpreted as the physical coordinates $X$ and which as the dual 
coordinates $\tilde{X}$. The torus periodicities, 
\equ{ \label{UntwBoundaryCondition}
Y ~\sim~ Y + E\, N~, 
}
read in terms of the coordinates $X$, their duals $\tilde{X}$ and gauge coordinates $X_\text{g}$
\equ{ \label{UntwistedBoundaryCondition}
\pmtrx{ {X} \\[1ex] \widetilde{X} \\[1ex] X_\text{g} } 
~\sim~ 
\pmtrx{ {X} \\[1ex] \widetilde{X} \\[1ex] X_\text{g} } 
+ \widehat{E}\, N
~\sim~ 
\pmtrx{ {X} \\[1ex] \widetilde{X} \\[1ex] X_\text{g} } 
+
\pmtrx{ 
e\, m 
\\[1ex] 
e^{-T}\big( n - C^T m - A^T \ga_\text{g}\, q\big)
\\[1ex] 
q + \ga_\text{g}^{-1}A\, m
 }~.
}

On-shell the right- and left-moving coordinate fields, $y_\text{r}, y_\text{l}$, have 
anti-holomorphic and holomorphic mode expansions for a string with boundary 
condition~\eqref{UntwBoundaryCondition} given by
\equ{\label{eqn:YModeExpansions}
y_\text{r}(\bgs) ~=~ y_{\text{r}_0} + p_\text{r}\, \bgs + \sum_{n\neq 0} \bga_n\, e^{2\pi i\, n \bgs}~,
\qquad 
Y_\text{L}(\gs)  ~=~ Y_{\text{L}_0} + P_\text{L}\, \gs  + \sum_{n\neq 0} \ga_n\,  e^{2\pi i\, n \gs}~, 
}
respectively. Using the change of coordinate field basis~\eqref{CoordFieldBasisChange}, we see 
that the conventional coordinate field $X$ and its dual $\widetilde{X}$ have the expansions
\begin{subequations} 
\equa{
X(\gs,\bgs) ~=~ \frac{1}{\sqrt{2}} \left( y_{\text{l}_0} + y_{\text{r}_0} \right) + \frac{1}{\sqrt{2}} \left( p_\text{l} + p_\text{r} \right) \sigma_1 + \frac{1}{\sqrt{2}} \left( p_\text{l} - p_\text{r} \right) \sigma_0 + \text{oscillators}~, 
\\[2ex]
\widetilde{X}(\gs,\bgs) ~=~ \frac{1}{\sqrt{2}} \left( y_{\text{l}_0} -y_{\text{r}_0} \right) + \frac{1}{\sqrt{2}} \left( p_\text{l} - p_\text{r} \right) \sigma_1 + \frac{1}{\sqrt{2}} \left( p_\text{l} + p_\text{r} \right) \sigma_0 + \text{oscillators}~.
}
\end{subequations} 
The term linear in the worldsheet space variable $\sigma_1$ of $X$ gives the $D$-dimensional 
winding modes, i.e.
\equ{
\frac{1}{\sqrt{2}}\left( p_\text{l} + p_\text{r} \right) ~=~ e\,m~.
}
The term linear in the worldsheet time variable $\gs_0$ of $X$ corresponds to the $D$-dimensional 
momentum which is given by
\equ{
\frac{1}{\sqrt{2}}\left( p_\text{l} - p_\text{r} \right) ~=~ e^{-T}\left(n -C^T m - A^T \ga_\text{g}\, q\right)~.
}
As expected, for the dual coordinate $\widetilde{X}$ the roles of momentum and winding are interchanged.


\afterpage{
 \begin{table}[p!]
\begin{center} 
\scalebox{1}{ \tabulinesep=.4ex
\tabul{| c || l | l |}{
\hline 
& \multicolumn{2}{|c|}{\textbf{Parametrizations of subgroups of} $\boldsymbol{\text{O}_{\widehat{\get}}(D,D+16;\Real)}$}
\\ \hline\hline 
\multirow{2}{*}{\rotatebox{90}{\textbf{geometric}}} &
$\widehat M_{e}(\gD K) ~=~ \left(
\arry{ccc}{
\gD K & 0 & 0 \\[1ex]
0 & \gD K^{-T} & 0 \\[1ex]
0 & 0 & \Id_{16} 
}
\right)$
& 
$\widehat M_{W}(\gD W) ~=~ \left(
\arry{ccc}{
\Id_D & 0 & 0 \\[1ex]
0 & \Id_D & 0 \\[1ex]
0 & 0 & \ga_\text{g}^{-1}\, \gD W\, \ga_\text{g} 
}
\right)$
\\ 
& where $\gD K \in \text{GL}(D;\Real)$ & where $\gD W \in \text{O}(16;\Real)$
\\ \cline{2-3}
&
$\widehat M_{B}(\gD B) ~=~ \left(
\arry{ccc}{
\Id_D & 0 & 0 \\[1ex]
\Delta B & \Id_D & 0 \\[1ex]
0 & 0 & \Id_{16} \\[1ex]
}
\right)$
& 
$\widehat M_{A}(\gD A) ~=~ \pmtrx{
\Id_D & 0 & 0 \\[1ex]
-\frac{1}{2}\Delta A^T \Delta A  & \Id_D & -\Delta A^T \ga_\text{g} \\[1ex]
\ga_\text{g}^{-1}\Delta A & 0 & \Id_{16} 
}$
\\
& where $\gD B^T = -\gD B \in \text{M}_{D\times D}(\Real)$ & $
\gD A \in M_{16\times D}(\Real)$
\\ \hline\hline 
\multirow{2}{*}{\rotatebox{90}{$\textbf{non-geometric}$}}  &
$\widehat I_{(\pm i)} ~=~ \left(
\arry{ccc}{
\Id_D -\ge_i \ge_i^T  & \mp \ge_i \ge_i^T & 0 \\[1ex]
\mp \ge_i \ge_i^T  & \Id_D - \ge_i \ge_i^T & 0 \\[1ex]
0 & 0 & \Id_{16} \\[1ex]
}
\right)$
&
$\widehat{I} 
~=~
R^{-1} \get\, R = 
\left(
\arry{ccc}{
0 &  \Id_D & 0 \\[1ex]
\Id_D & 0 & 0 \\[1ex]
0 & 0 & \Id_{16} \\[1ex]
}
\right)$
\\ \cline{2-3}
&
$\widehat M_{\beta}(\gD \beta)
~=~ \left(
\arry{ccc}{
\Id_D & \gD \beta & 0 \\[1ex]
0 & \Id_D & 0 \\[1ex]
0 & 0 & \Id_{16} \\[1ex]
}
\right)$
& 
$\widehat M_\ga(\gD \ga)
~=~ 
\pmtrx{
\Id_D & -\frac 12 \gD \ga^T \gD \ga  & - \gD \ga^T \ga _\text{g}
\\[1ex] 
0 & \Id_D & 0 
\\[1ex] 
0 & \ga_\text{g}^{-1}\gD \ga & \Id_{16}
}$
\\
& where $\gD \beta^T = -\gD \beta \in \text{M}_{D\times D}(\Real)$ & $
\gD\ga \in M_{16\times D}(\Real)$
\\ \hline 
}
}
\end{center}
\caption{ \label{tb:DualitySubgroups} 
This table lists various subgroup elements of the duality group $\text{O}_{\widehat{\get}}(D,D+16;\Real)$. 
They are normalized such that if the parameters are taken out of $\Intr$ they represent subgroups 
of $\text{O}_{\widehat{\get}}(D,D+16;\Intr)$ (with the additional requirement that 
$\frac{1}{2}\Delta A^T \Delta A$ and $\frac 12 \gD \ga^T \gD \ga$ are integer matrices). The 
elements listed in the first two rows generate the geometric subgroup $G_\text{geom}$ of the 
duality group. The elements on the third row correspond to true $T$-duality elements that invert 
one or all radii. Note the difference between $\ga_\text{g}$ and $\gD \ga$: $\ga_\text{g}$ contains 
the simple roots of $\E{8}\times\E{8}$ and is used in the definitions of $\widehat M_{W}(\gD W)$, 
$\widehat M_{A}(\gD A)$ and $\widehat M_\ga(\gD \ga)$, while $\gD \ga$ is the parameter of 
$\widehat M_\ga(\gD \ga)$.
}
\end{table}

\begin{table}[t]
\begin{center} 
\scalebox{.95}{\tabulinesep=1ex
\tabul{| l || l |}{
\hline 
\multicolumn{2}{|c|}{\textbf{Multiplication table of duality subgroup elements}}
\\ \hline\hline  
$\widehat M_{e}(\gD K')\, \widehat M_{e}(\gD K) = \widehat M_{e}(\gD K'\gD K)$
&
$\widehat M_{e}(\gD K^{-T}) = \widehat I\, \widehat M_{e}(\gD K)\, \widehat I$
\\
$\widehat M_{W}(\gD W')\, \widehat M_{W}(\gD W) = \widehat M_{W}(\gD W'\gD W)$
&
$\widehat M_{W}(\gD W) = \widehat I\, \widehat M_{W}(\gD W)\, \widehat I$
\\
$\widehat M_{B}(\gD B')\, \widehat M_{B}(\gD B) = \widehat M_{B}(\gD B'+\gD B) $
& 
$\widehat M_{\beta}(\gD \beta) = \widehat I\, \widehat M_{B}(\gD \beta)\, \widehat I  = \widehat{M}_B(-\gD \beta)^T$
\\
$\widehat M_{A}(\gD A')\, \widehat M_{A}(\gD A)  = \widehat M_{B}(\gD B_A)\, \widehat M_{A}(\gD A'+ \gD A)$
&
$\widehat M_\ga(\gD \ga) = \widehat I\, \widehat M_{A}(\gD \ga) \,\widehat I 
= (R^TR)^{-1} \widehat M_A(-\gD \ga)^T R^TR$
\\
\text{with } 
$\gD B_A =  \frac{1}{2}\left(\gD A^T \gD A' - \gD A'{}^T \gD A\right)$
&
\\ \hline 
$\widehat M_{W}(\gD W)\,\widehat M_{e}(\gD K) = \widehat M_{e}(\gD K)\,\widehat M_{W}(\gD W)$
& 
$\widehat M_{W}(\gD W)\,\widehat M_{B}(\gD B) = \widehat M_{B}(\gD B)\,\widehat M_{W}(\gD W)$
\\
$\widehat M_{B}(\gD B)\,\widehat M_{e}(\gD K) = \widehat M_{e}(\gD K)\,\widehat M_{B}(\gD K^T \gD B\,\gD K)$
&
$\widehat M_{W}(\gD W) \, \widehat M_{A}(\gD A) = \widehat M_{A}(\gD W \gD A) \,  \widehat M_{W}(\gD W)$
\\
$\widehat M_{A}(\gD A)\,\widehat M_{e}(\gD K) =  \widehat M_{e}(\gD K)\,\widehat M_{A}(\gD A\,\gD K)$
&
$\widehat M_{A}(\gD A)\,\widehat M_{B}(\gD B) = \widehat M_{B}(\gD B)\,\widehat M_{A}(\gD A)$
\\ \hline 
}
}
\end{center} 
\caption{ \label{tb:DualityMultiplications}
Multiplication table for the generators of the duality group $\text{O}_{\widehat{\get}}(D,D+16;\Real)$.}
\end{table}

}

\section[The $\boldsymbol{T}$-duality group]{The $\boldsymbol{T}$-duality group}
\label{SecTDualityGroup}

This section is devoted to exhibit a number of properties of the $T$-duality group. In particular, 
we develop a convenient basis for this group and parametrize its maximal compact subgroup. In 
addition, we show that the non-linear transformations of the Narain moduli is a consequence of the 
coset structure in which the generalized vielbein $\widehat{E}$ lives.

\subsection[Decomposition of the generalized vielbein]{Decomposition of the generalized vielbein}
\label{sec:SubgroupsIntro}

A general $T$-duality transformation is described by an element 
$\widehat M \in \text{O}_{\widehat{\get}}(D,D+16; \Intr)$. In addition, in eqn.~\eqref{NarainModuli} we 
parametrized the Narain moduli by the generalized vielbein 
$\widehat{E} \in \text{O}_{\widehat{\get}}(D,D+16; \Real)$. Therefore, it is very convenient to describe 
the properties of matrices $\widehat M \in \text{O}_{\widehat{\get}}(D,D+16; \Real)$ first in general, 
based on the field of real numbers $\Real$. To do so we define a number of specific matrix elements 
of this group in Table~\ref{tb:DualitySubgroups}. These matrices are chosen such that if we 
restrict the parameters to be from $\Intr$ rather than $\Real$, these matrices have only integral 
entries. 

As a first application of the matrices of Table~\ref{tb:DualitySubgroups}, we decompose the 
generalized vielbein~\eqref{NarainModuli} as a product
\equ{\label{GenVielbeinDecomposition}
\widehat{E} = 
\widehat E(e,B,A) = \widehat M_{e}(e)\, \widehat M_B(B)\, \widehat M_A(A)~, 
}
of basis matrices $\widehat M_i \in \text{O}_{\widehat{\get}}(D,D+16; \Real)$ as given in 
Table~\ref{tb:DualitySubgroups}. Here, the index $i=e,B,A$ labels the matrix $\widehat M_i$ and 
each matrix $\widehat M_i$ depends on the corresponding kind of Narain moduli $e,$ $B$ and $A$. 
This parametrization will turn out to be very useful throughout this paper.

\subsection[Coset decomposition the $T$-duality group]{Coset decomposition of the $\boldsymbol{T}$-duality group}
\label{sec:CosetDecomposition}

In Section~\ref{SecNarainVielbein} we recalled that the moduli space of Narain compactifications 
can be described geometrically as a coset space~\eqref{ModuliSpace}. This already shows the central 
role that the coset space plays in our discussion and therefore we expand on this property in some 
detail here.

The generalized vielbein $\widehat{E}$ is an element of the coset 
\equ{ \label{Coset}
\text{O}(D;\Real)\!\times\!\text{O}(D+16;\Real) \backslash 
\text{O}_{\widehat\get}(D,D+16;\Real)~.  
}
This means that any element $\widehat{H} \in \text{O}_{\widehat\get}(D,D+16;\Real)$ can be 
decomposed as 
\equ{ \label{CosetDecomposition} 
\widehat{H} ~=~ \widehat{U} \, \widehat{E}~, 
}
where the specific standard form~\eqref{NarainModuli} of the generalized vielbein 
$\widehat{E}$ lies inside the coset~\eqref{Coset} and $\widehat{U} = R^{-1}\,U\,R$ with 
$U \in \text{O}(D;\Real)\!\times\!\text{O}(D+16;\Real)$ is given in eqns.~\eqref{Uexplicit} 
and~\eqref{eqn:Defintionupm}.

As this applies to {\em any} element of the $T$-duality group, it applies in particular to 
$\widehat{E} \widehat{M}$ with $\widehat{M} \in \text{O}_{\widehat{\get}}(D,D+16;\Real)$, i.e.
\equ{ \label{CosetDecompEM} 
\widehat{U}_{\widehat{M}}\, \widehat{E}(e',B',A') ~=~\widehat{E}(e,B,A)\, \widehat{M}~. 
}
The subscript $\widehat{M}$ of $\widehat{U}_{\widehat{M}}$ emphasizes that the 
$\text{O}(D;\Real)\!\times\!\text{O}(D+16;\Real)$ group element on the left hand side depends on 
the $T$-duality group element $\widehat{M}$ under consideration. Both $\widehat E(e,B,A)$ and 
$\widehat E(e',B',A')$ are given here in the standard form~\eqref{NarainModuli}. This 
equation~\eqref{CosetDecompEM} will be used frequently throughout this paper, for example, when we 
discuss $T$-duality transformations of Narain moduli in Section~\ref{sec:NarainModuliTrafo} and 
when we analyze the stabilization of Narain moduli in generalized orbifolds in 
Section~\ref{sec:ModuliStabilization}.

\subsubsection*{Simplified standard form of the generalized vielbein}

Eqn.~\eqref{CosetDecompEM} can be used to further simplify the generalized 
vielbein~\eqref{eqn:MostGeneralGeneralizedVielbein}: For any discrete $T$-duality element 
$\widehat{M} \in \text{O}_{\widehat{\get}}(D,D+16; \Intr) \subset \text{O}_{\widehat{\get}}(D,D+16; \Real)$ 
there is a matrix $U_{\widehat{M}} \in \text{O}(D;\Real) \times \text{O}(D+16;\Real)$ 
such that eqn.~\eqref{CosetDecompEM} holds. Consequently, we find
\equ{
E ~=~ U\,R\,\widehat{E}(e,B,A)\,\widehat{M} ~=~ \left(U\,U_{\widehat{M}}\right)\,R\,\widehat{E}(e',B',A') ~=~ U'\, R\,\widehat{E}(e',B',A')~, 
}
where $U' = U\, U_{\widehat{M}} \in \text{O}(D;\Real) \times \text{O}(D+16;\Real)$ is arbitrary 
since $U$ is arbitrary. Relabelling our expression by removing the primes we obtain the most 
general from of the generalized vielbein as 
\equ{\label{eqn:verymostgeneralvielbein}
E ~=~ U\, R\, \widehat{E}(e,B,A)~,
}
where $\widehat{E}(e,B,A)$ is given in eqns.~\eqref{NarainModuli} 
and~\eqref{GenVielbeinDecomposition} and 
$U \in \text{O}(D;\Real) \times \text{O}(D+16;\Real)$ may be chosen freely.

\subsubsection*{Compact subgroup in the coset decomposition}

In what follows, we consider eqn.~\eqref{CosetDecompEM} and first compute the explicit matrix expression of 
$U_{\widehat{M}} \in \text{O}(D;\Real) \times \text{O}(D+16;\Real)$, and determine the transformed 
moduli, $e',B',A'$, in terms of $\widehat{M}$ and the initial moduli $e$, $B$ and $A$.

To do so, we decompose $\widehat M$ into its $3\times 3$-block structure, i.e.\ 
\equ{
\widehat M ~=~ \left(
\arry{ccc}{
\widehat M_{11} & \widehat M_{12} & \widehat M_{13} \\[1ex]
\widehat M_{21} & \widehat M_{22} & \widehat M_{23} \\[1ex]
\widehat M_{31} & \widehat M_{32} & \widehat M_{33} \\[1ex]
}
\right)~, 
}
where $\widehat M_{11}$, $\widehat M_{12}$, $\widehat M_{21}$ and $\widehat M_{22}$ are 
$D\times D$-matrices, $\widehat M_{13}$ as well as $\widehat M_{23}$ are $D\times 16$-matrices, 
$\widehat M_{31}$ as well as $\widehat M_{32}$ are $16 \times D$-matrices, while $\widehat M_{33}$ is 
a $16\times 16$-matrix, respectively. 
Furthermore, in order to avoid lengthy formulae, we introduce short-hand notations
\begin{subequations}
\label{GammaDeltaBA}
\begin{eqnarray}
\widehat M_1 & = & -\widehat M_{21} + (G + C^T) \widehat M_{11} + A^T \ga_\text{g}\, \widehat M_{31}~, 
\\[1ex] 
\widehat M_2 & = & -\widehat M_{22} + (G + C^T) \widehat M_{12} + A^T \ga_\text{g}\, \widehat M_{32}~, 
\\[1ex] 
\widehat M_3 & = & -\widehat M_{23} + (G + C^T) \widehat M_{13} + A^T \ga_\text{g}\, \widehat M_{33}~,
\end{eqnarray}
\end{subequations} 
which will recur frequently throughout the rest of this work. Next, we compute the products of 
matrices contained in eqn.~\eqref{CosetDecompEM}, i.e.
\equ{
\widehat E(e,B,A)\, \widehat{M} \quad\text{and }\quad \widehat U_{\widehat{M}}\,\widehat E(e',B',A')~,
}
where each matrix is given in its $3 \times 3$-block structure, e.g.~$\widehat{U}_{\widehat{M}}$ is 
given in eqn.~\eqref{eqn:Defintionupm}. The result is set equal which yields $3 \times 3 = 9$ equations 
from eqn.~\eqref{CosetDecompEM}. By doing 
so, we can solve for the blocks of $U_{\widehat{M}} = R\,\widehat{U}_{\widehat{M}}\,R^{-1}$ 
as defined in eqn.~\eqref{eqn:Defintionupm} and obtain
\begin{subequations}\label{eqn:UBlocksFromM}
\begin{eqnarray}
u_\text{l}  & = & \left( \Id_D - 2\,e\,\widehat{M}_{12}\,\widehat{M}_2^{-1}\,e^T\right) u_\text{r}~,\label{eqn:ulBlocksFromM}
\\[1ex] 
u_\text{lL} & = & \sqrt{2}\,e\,\left(\widehat{M}_{13}-\widehat{M}_{12}\,\widehat{M}_2^{-1}\,\widehat{M}_3\right)\,\ga_\text{g}^{-1}~,
\\[1ex] 
u_\text{Ll} & = & -\sqrt{2}\,\left(\ga_\text{g}\,\widehat{M}_{32}+A\,\widehat{M}_{12}\right)\widehat{M}_2^{-1}\,e^T\,u_\text{r}~,
\\[1ex] 
u_\text{L}  & = & A\,\widehat{M}_{13}\,\ga_\text{g}^{-1}+\ga_\text{g}\,\widehat{M}_{33}\,\ga_\text{g}^{-1}-\left(A\,\widehat{M}_{12}+\ga_\text{g}\,\widehat{M}_{32}\right)\widehat{M}_2^{-1}\,\widehat{M}_3\,\ga_\text{g}^{-1}~,
\end{eqnarray}
\end{subequations} 
for arbitrary $u_\text{r} \in \text{O}(D;\Real)$. We have checked explicitly that these equations 
give a matrix $U$ such that the conditions~\eqref{ConstraintsSODD+16} are satisfied. Let us remark 
one observations from eqn.~\eqref{eqn:ulBlocksFromM}: $\widehat{M}_{12} \neq 0$ is a 
necessary condition for $u_\text{r}\neq u_\text{l}$. In other words, if $\widehat{M}_{12} = 0$ then 
$u_\text{r}= u_\text{l}$. In addition, let us stress that these equations~\eqref{eqn:UBlocksFromM} 
will become very important later in the context of Narain orbifolds where $U$ becomes the orbifold 
twist $\gTh$, for example in Section~\ref{sec:ExistenceOfNarainOrbifolds}. Furthermore, we identify 
the following three expressions 
\label{TransformationeBAEasy} 
\equ{ 
\widehat{M}_2^T ~=~ -\left(e'^{-1} u_\text{r}^{-1}\,e\right)~, 
\quad 
G' + {C'}^T ~=~ \left(e'^{-1} u_\text{r}^{-1}\, e\right)^{-T} \widehat{M}_1~, 
\quad 
A'\left(e'^{-1} u_\text{r}^{-1}\,e\right) ~=~ \ga_\text{g}^{-T}\widehat{M}_3^T~,
} 
from eqn.~\eqref{CosetDecompEM}, which we use in the following discussion.

\subsection[Transformation of Narain moduli]{Transformation of Narain moduli} 
\label{sec:NarainModuliTrafo}

Using the coset decomposition discussed above, we can derive the transformation properties of the 
Narain moduli $G$, $B$ and $A$ under general $T$-duality transformations $\widehat{M}\in\text{O}_{\widehat{\get}}(D,D+16; \Real)$. 
Using the results of Section~\ref{SecNarainVielbein} we see that the generalized 
vielbein~\eqref{GenVielbeinDecomposition} transforms under $\widehat M$ as
\equ{\label{eqn:GenarlCaseModuliTrafo}
\widehat E(e,B,A) ~\mapsto~ \widehat E(e',B',A') ~=~ \widehat{U}_{\widehat{M}}^{-1}\, \widehat E(e,B,A)\, \widehat{M}~,
}
where $\widehat{U}_{\widehat{M}} = R^{-1}\,U_{\widehat{M}}\,R$ and 
$U_{\widehat{M}} \in \text{O}(D;\Real) \times \text{O}(D+16;\Real)$. In other words, assume we have 
given a $T$-duality transformation $\widehat{M} \in \text{O}_{\widehat{\get}}(D,D+16; \Real)$. Then, 
there exists a matrix $U_{\widehat{M}}$ as given in eqn.~\eqref{eqn:UBlocksFromM} such that 
$\widehat E(e',B',A')$ is in the standard form~\eqref{GenVielbeinDecomposition}.

Hence, we are able to identify the transformation properties of $e$, $G+C^T$ and $A$ 
under general $T$-duality transformations from eqn.~\eqref{TransformationeBAEasy}. We find
\equa{ \label{TransformationeBA}
e' ~=~ -u_\text{r}^{-1}\,e\,\widehat{M}_2^{-T} \quad, \quad G' + {C'}^T ~=~ -\widehat{M}_2^{-1}\widehat{M}_1 \quad\text{and}\quad A' ~=~ -\ga_\text{g}^{-T}\widehat{M}_3^T\widehat{M}_2^{-T}~,
}
where $u_\text{r}^T u_\text{r} = \Id_D$. These transformations can be expanded out (by 
taking the anti-symmetric part of $G' + {C'}^T$ to solve for $B'$) and we obtain the 
transformations of the moduli $G,B,A$, i.e.
\equ{ \label{TransformationGBA} 
\hspace{-1ex} 
G \mapsto G' = \widehat M_2^{-1} G\, \widehat M_2^{-T}~,
\quad  
B  \mapsto  B' =
 \frac{1}{2}\Big( \widehat M_2^{-1} \widehat M_1 - \widehat M_1^T \widehat M_2^{-T} \Big)~,
 \quad 
A \mapsto A' = -\ga_\text{g}^{-T} \widehat M_3^T \widehat M_2^{-T}~,
} 
using the short-hands defined in eqn.~\eqref{GammaDeltaBA}. This generalizes the results for 
$\text{O}(D,D)$ (see e.g.\ \cite{Blumenhagen:2013aia}) to the heterotic 
case~\cite{Blumenhagen:2014iua}. As a cross-check, using 
$\widehat{M}\,\widehat{\get}^{-1}\widehat{M}^T = \widehat{\get}^{-1}$ one can show that 
eqn.~\eqref{TransformationGBA} yields $G' + {C'}^T = -\widehat{M}_2^{-1}\widehat{M}_1$ as given in 
eqn.~\eqref{TransformationeBA}.

\subsection[Specific elements of the $T$-duality group]{Specific elements of the $\boldsymbol{T}$-duality group}
\label{sec:DecompositionTGroup}

Next, we discuss various elements and subgroups of the group $\text{O}_{\widehat{\get}}(D,D+16; \Real)$ 
in detail and analyze their actions on the Narain moduli $G, B, A$. The parametrizations of these 
subgroups can be found in Table~\ref{tb:DualitySubgroups} and their most important products are 
given in Table~\ref{tb:DualityMultiplications}.


\subsubsection[The geometric subgroup]{The geometric subgroup}

The elements $\widehat M_{e}$, $\widehat M_{W}$, $\widehat M_{A}$ and $\widehat M_{B}$ as listed in 
Table~\ref{tb:DualitySubgroups} generate a subgroup of $\text{O}_{\widehat{\get}}(D,D+16; \Real)$ which 
we denote by $G_\text{geom}(\Real)$. This is the largest $T$-duality subgroup, that still admits a 
standard geometrical interpretation, hence the name: geometric subgroup. In more detail, all 
elements $\widehat M_\text{geom} \in G_\text{geom}(\Real)$ can be cast to the form
\equ{\label{eqn:ElementOfGGeom}
\widehat M_\text{geom} ~=~ \widehat{M}_W(\gD W)\, \widehat M_{e}(\gD K)\, \widehat M_B(\gD B)\, \widehat M_A(\gD A)~. 
}
Then, using the results of Section~\ref{SecNarainVielbein} we see that the generalized 
vielbein~\eqref{GenVielbeinDecomposition} transforms under $\widehat M_\text{geom}$ as
\begin{subequations}\label{GeomVielTrans}
\equ{ 
\widehat E(e,B,A) ~\mapsto~ \widehat E(e',B',A') ~=~ \widehat{U}_\text{geom}^{-1}\, \widehat E(e,B,A)\, \widehat{M}_\text{geom}~, 
}
where
\begin{eqnarray}
e' & = & (u_\text{r}^\text{geom})^{-1}\,e\, \gD K~, \label{GeomVielTranse} \\[1ex] 
B' & = & \gD K^T B\, \gD K + \gD B + \sfrac 12\big(\gD A^T \, \gD W^T A\, \gD K - \gD K^T A^T\gD W\, \gD A\big)~, \\[1ex] 
A' & = & \gD W^T A \,\gD K + \gD A~.
\end{eqnarray}
\end{subequations}
Here $\widehat U_\text{geom} = R^{-1}\,U_\text{geom}\,R$ and 
$U_\text{geom} \in \text{O}(D;\Real) \times \text{O}(D+16;\Real)$ must be 
chosen such that $\widehat E(e',B',A')$ is given in the standard form~\eqref{GenVielbeinDecomposition}.
Furthermore, in eqn.~\eqref{GeomVielTrans} we have used various group multiplication properties as 
given in Table~\ref{tb:DualityMultiplications} to compute the product 
$\widehat E(e,B,A)\, \widehat{M}_\text{geom}$ (analogously, one could have used the general 
transformations~\eqref{TransformationeBA} and~\eqref{TransformationGBA} for 
$\widehat M = \widehat M_\text{geom}$ to derive eqn.~\eqref{GeomVielTrans}). Notice that under a 
$\widehat M_{W}(\gD W)$-transformation the form of the generalized vielbein is not strictly 
preserved. Nevertheless, it is of the correct form such that it can be absorbed by the choice of 
$\widehat U_\text{geom} = \widehat{M}_W(\gD W) \widehat{M}_e(u_\text{r}^\text{geom})$, i.e.
\equ{
U_\text{geom} ~=~ R\, \widehat{M}_W(\gD W)\, \widehat{M}_e(u_\text{r}^\text{geom}) \, R^{-1} ~=~ 
\pmtrx{ u_\text{r}^\text{geom} \\ & u_\text{r}^\text{geom} \\ & & \gD W}~,
}
since $U_\text{geom}$ is an element of $\text{O}(D;\Real) \times \text{O}(D+16;\Real)$ because  
$\gD W \in \text{O}(16;\Real)$ and $u_\text{r}^\text{geom}\in \text{O}(D;\Real)$.

In the following, we give details for various elements of the $T$-duality group. We start with the 
four generators $\widehat M_{e}$, $\widehat M_{W}$, $\widehat M_{A}$ and $\widehat M_{B}$ of the 
geometric subgroup $G_\text{geom}(\Real)$ and use eqns.~\eqref{GeomVielTrans} in order to compute 
the transformation of moduli.

\subsubsection*{Change of geometrical basis $\boldsymbol{\widehat M_{e}(\gD K)}$}

Changes of the geometrical basis $e$ are given by $\widehat M_{e}(\gD K)$ with 
$\gD K \in \text{GL}(D;\Real)$. The unit element $\widehat M_{e}(\gD K) = \Id$ has $\gD K = \Id_D$. 
From eqns.~\eqref{GeomVielTrans} we identify the transform of the background fields $G$, $B$ and 
$A$: $\widehat M_{e}(\gD K)$ leads to a change of basis of the $D$-dimensional torus, 
$e\mapsto e' = (u_\text{r}^{\text{geom}})^{-1} e\, \gD K$, and 
\equ{ \label{ChangeOfBasisT6} 
G ~\mapsto~ G' ~=~ \gD K^T\, G\, \gD K~, 
\qquad 
B ~\mapsto~ B' ~=~ \gD K^T\, B\, \gD K~, 
\qquad 
A ~\mapsto~ A' ~=~ A\, \gD K~. 
} 

\subsubsection*{Change of basis in the gauge degrees of freedom $\boldsymbol{\widehat M_{W}(\gD W)}$}

In addition, we may change the basis in the gauge degrees of freedom by $\widehat M_{W}(\gD W)$ with 
$\gD W\in \text{O}(16;\Real)$. The unit element $\widehat M_{W}(\gD W) = \Id$ has 
$\gD W = \Id_{16}$. $\widehat M_W(\gD W)$ induces a transformation
\equ{ \label{ChangeOfBasisGauge}
A  ~\mapsto~  A' ~=~ \gD W^T A~
} 
of the Wilson lines, while $G$ and $B$ remain invariant. 

In the case of the discrete $T$-duality group we define 
$\rho_W = \ga_\text{g}^{-1}\, \gD W\, \ga_\text{g}$. Then, 
$\widehat M_{W}(\gD W) \in \text{O}_{\widehat{\get}}(D,D+16;\Intr)$ if 
$\rho_W \in \text{O}_{g}(16;\Intr)$, i.e.\ $\rho_W^T g\, \rho_W = g$ using 
$g = \ga_\text{g}^T \ga_\text{g}$. Hence, $\rho_W$ is an automorphism of the $\E{8}\times\E{8}$ 
root  lattice spanned by $\ga_\text{g}$.

\subsubsection*{$\boldsymbol{B}$-field shifts $\boldsymbol{\widehat M_{B}(\gD B)}$} 

Matrices of the form $\widehat M_{B}(\gD B)$ with 
$\Delta B^T = -\Delta B \in \text{M}_{D\times D}(\Real)$ leave $G$ and $A$ invariant and only 
induce $B$-field shifts, i.e
\equ{ \label{BShift} 
B ~\mapsto~ B' ~=~ B + \Delta B~. 
} 
$B$-field shifts generate a subgroup $G_B(\Real) \subset \text{O}_{\widehat{\get}}(D,D+16; \Real)$. The 
unit element $\widehat M_{B}(\gD B) = \Id$ is given by $\gD B = 0$.

\subsubsection*{Wilson line shifts $\boldsymbol{\widehat M_{A}(\gD A)}$} 

Wilson line shifts are generated by $\widehat M_{A}(\gD A)$ with 
$\ga_\text{g}^{-1} \gD A \in M_{16\times D}(\Real)$. Indeed, we obtain
\equ{ 
A ~\mapsto~ A' ~=~ A + \Delta A~, 
\qquad 
B ~\mapsto~ B '~=~ B + \frac{1}{2}\Big(\Delta A^T  A - A^T \Delta A \Big)~.
} 
Hence, transformations of the Wilson lines $A$ are accompanied by a $B$-field transformation, while 
the metric $G$ is kept invariant. Furthermore, we find
\equ{ \label{WilsonLineGroup}
\widehat M_{A}(\gD A')\, \widehat M_{A}(\gD A)  = \widehat M_{B}(\gD B_A)\, \widehat M_{A}(\gD A'+ \gD A)~, 
}
with $\gD B_A = \frac{1}{2}\left(\gD A^T \gD A' - \gD A'{}^T \gD A\right)$, where we remark that 
Wilson line shifts and $B$-field shifts commute, see Table~\ref{tb:DualityMultiplications}.

Due to eqn.~\eqref{WilsonLineGroup}, Wilson line shifts do not generate a subgroup of 
$\text{O}_{\widehat{\get}}(D,D+16; \Real)$ on their own, but only when combined with $B$-field shifts 
$\widehat M_{B}(\gD B)$. We denote this subgroup by $G_\text{WL}(\Real)$.  
The subgroup $G_B(\Real)$ of $B$-field shifts and the subgroup $G_\text{WL}(\Real)$ of combined 
Wilson line and $B$-field shifts are both normal subgroups of the geometric subgroup 
$G_\text{geom}(\Real)$. In particular, it follows that 
\equ{ 
G_\text{geom}/ G_\text{WL} = \text{GL}(D;\Real) \times \text{O}(16; \Real)~.
}
Note that $\widehat M_{B}(\gD B) \widehat M_{A}(\gD A)$ with $\gD B \not\in \text{M}_{D\times D}(\Intr)$ 
can be an element of the discrete $T$-duality group, i.e.\ 
$\widehat M_{B}(\gD B) \widehat M_{A}(\gD A) \in \text{O}_{\widehat{\get}}(D,D+16;\Intr)$, if 
$\ga_\text{g}^{-1} \gD A \in M_{16\times D}(\Intr)$ and
\equ{ \label{DiscreteWilsonLineBShift}
-\frac{1}{2}\Delta A^T \Delta A + \gD B ~\in~ M_{D\times D}(\Intr)~.
}

\subsubsection[Non-geometric elements]{Non-geometric elements} 
\label{sec:nongeomericTduality}

In the following, we give details for non-geometric elements of the $T$-duality group. We use 
eqns.~\eqref{TransformationGBA} in order to compute the transformation of moduli.

\subsubsection*{$\boldsymbol{T}$-duality inversions}

We can define $\Intr_2$ involutions
\equ{
\label{TDualityTrafo_single}
\widehat I_{(\pm i)} ~=~ \left(
\arry{ccc}{
\Id_D -\ge_i \ge_i^T  & \mp \ge_i \ge_i^T & 0 \\[1ex]
\mp \ge_i \ge_i^T  & \Id_D - \ge_i \ge_i^T & 0 \\[1ex]
0 & 0 & \Id_{16} \\[1ex]
}
\right) \quad\text{for}\quad i = 1,\ldots, D~,
}
where $\ge_i$ denotes the standard basis vector in the $i$-th torus direction. The element 
$\widehat{I}_{(\pm i)}$ can be written as conjugation of a reflection in the $i$-th left- or 
right-moving direction as $\widehat{I}_{(\pm i)} = R^{-1} I_{(\pm i)}R$ using
\equ{
I_{(+i)} ~=~ 
\left(
\arry{ccc}{
\Id_D & 0 & 0 \\[1ex] 
0 & \Id_D - 2\, \ge_i \ge_i^T &  0 \\[1ex]
0 & 0 & \Id_{16}
}
\right)~, 
\quad 
I_{(-i)} ~=~ 
\left(
\arry{ccc}{
\Id_D - 2\, \ge_i \ge_i^T & 0 & 0 \\[1ex]
0 & \Id_D & 0 \\[1ex] 
0 & 0 & \Id_{16}
}
\right)~.
}
Therefore, all the elements $\widehat{I}_{(\pm i)}$ can be obtained from $\widehat{I}_{(\pm 1)}$ by 
conjugation with an appropriate change of basis element $\widehat M_{e}(\gD K)$.

The element $\widehat{I}_{(-i)}$ induces a $T$-duality inversion along the $i$-th torus direction. 
We can preform the $T$-duality inversion in all torus directions simultaneously by 
\equ{
\label{TDualityTrafo}
\widehat I ~=~
\widehat I_{(-1)} \cdot \cdots \cdot 
\widehat I_{(-D)}~,
}
as given in Table~\ref{tb:DualitySubgroups}. Using the general 
results~\eqref{TransformationGBA} we find for this element 
\equ{ \label{eqn:modulitrafoI}
G ~\mapsto~ G' ~=~ \widehat M_2^{-1} G\, \widehat M_2^{-T}~, 
\quad 
B ~\mapsto~ B' ~=~ -\widehat M_2^{-1} B\, \widehat M_2^{-T}~,
\quad 
A ~\mapsto~ A' ~=~ - A\, \widehat M_2^{-T}~, 
} 
where $\widehat M_2 = G +C^T$. For $A=0$ we get $\widehat M_2 = G-B$. Hence, 
eqn.~\eqref{eqn:modulitrafoI} yields the famous transformation $(G+B) \mapsto (G+B)^{-1}$.

\subsubsection*{Maximal subgroup of $\boldsymbol{\text{O}_{\widehat{\get}}(D,D+16;\Real)}$ connected to the identity}

As an application of the special duality elements $\widehat{I}_{(\pm i)}$ we discuss the maximal 
non-compact subgroup $\text{SO}^+_{\widehat{\get}}(D,D+16;\Real)$ of the $T$-duality group 
$\text{O}_{\widehat{\get}}(D,D+16;\Real)$ that is connected to the identity. 
The quotient group 
\equ{
\text{O}_{\widehat{\get}}(D,D+16;\Real)/\text{SO}^+_{\widehat{\get}}(D,D+16;\Real)
~\cong~ \Intr_2 \times \Intr_2~
}
is of order four and, hence, corresponds to four disconnected components of 
$O_{\widehat{\get}}(D,D+16;\Real)$: One can choose the two $\Intr_2$-generators as the elements 
$\widehat{I}_{(-1)}$ and $\widehat{I}_{(+1)}$. The matrix representations of the four 
disconnected components are obtained by multiplying $\Id, \widehat{I}_{(-1)}, \widehat{I}_{(+1)}$ 
or $\widehat{I}_{(-1)}\widehat{I}_{(+1)}$ by arbitrary matrices of 
$\text{SO}^+_{\widehat{\get}}(D,D+16;\Real)$.

\subsubsection*{Inverted $\boldsymbol{B}$-field shifts $\boldsymbol{\widehat M_{\beta}(\gD \gb)}$}

Even though the following two elements $\widehat M_{\beta}(\gD \gb)$ and $\widehat M_{\ga}(\gD \ga)$ 
can be obtained by combining the $B$- and $A$-shifts with the inversion element $\widehat I$, we 
list them explicitly as they are important in the context of non-geometry.

Inverted $B$-field shifts, often referred to as $\beta$-transformations, are generated by 
\equ{
\label{betaShift} 
\widehat M_{\beta}(\gD \beta) = \widehat I\, \widehat M_{B}(\gD \beta)\, \widehat I~, 
}
with $\gD \beta^T = -\gD \beta \in M_{D\times D}(\Real)$. The $\beta$-transformations of the 
metric, $B$-field and gauge backgrounds take the form 
\begin{subequations} 
\equ{ 
G ~\mapsto~ G' ~=~ \widehat M_2^{-1}\, G \, \widehat M_2^{-T}~, 
\quad 
A ~\mapsto~ A' ~=~ -A\, \widehat M_2^{-T}~,
\\[1ex] 
B ~\mapsto~ B' ~=~ \widehat M_2^{-1}\, \Big( B - (G+C^T)\gD\beta(G+C)\Big) \, \widehat M_2^{-T}~, 
}
\end{subequations} 
using $\widehat M_2 = -\Id_D + (G+C^T)\gD \beta$ in eqn.~\eqref{TransformationGBA}.

\subsubsection*{Inverted Wilson line shifts $\boldsymbol{\widehat M_\ga(\gD \ga)}$}
 
Finally, by inverting the Wilson line shifts $\widehat M_{A}$ we obtain 
\equ{ 
\widehat M_\ga(\gD \ga) = \widehat I \, \widehat M_{A}(\gD \ga) \,\widehat I~,
}
with $\ga_\text{g}^{-1} \gD \ga \in M_{16\times D}(\Real)$. 

The inversion of changes of bases, i.e.\ $\widehat I\,\widehat M_{e}(\gD K)\widehat I$ and 
$\widehat I\,\widehat M_{W}(\gD W)\widehat I$, 
just become changes of bases again. Hence, they do not give us novel transformations. For completeness 
we nevertheless list them in Table~\ref{tb:DualityMultiplications}. Indeed, counting the number of 
generators shows that this list contains all possible $\text{O}_{\widehat{\get}}(D,D+16, \Real)$ 
transformations.

\subsection[The maximal compact subgroup of $\text{O}_{\widehat{\get}}(D,D+16; \Real)$]{The maximal compact subgroup of $\boldsymbol{\text{O}_{\widehat{\get}}(D,D+16; \Real)}$}
\label{sec:maximalcompactsubgroup}

Next, we discuss the maximal compact subgroup of $\text{O}_{\widehat{\get}}(D,D+16; \Real)$. 
To do so, we note that the maximal compact subgroup of $\text{O}_{\get}(D,D+16; \Real)$ is 
$\text{O}(D;\Real)\times\text{O}(D+16;\Real)$. By conjugation with $R$ one maps elements 
$U \in \text{O}_{\get}(D,D+16; \Real)$ one-to-one to elements 
$\widehat{U} \in \text{O}_{\widehat{\get}}(D,D+16; \Real)$, i.e.~$\widehat{U} = R^{-1} U R$. Thus, the maximal 
compact subgroup of $\text{O}_{\widehat{\get}}(D,D+16; \Real)$ is also 
$\text{O}(D;\Real)\times\text{O}(D+16;\Real)$. An explicit parametrization of this subgroup is 
given by $\widehat{U}$ in eqn.~\eqref{eqn:Defintionupm}. Note that, as discussed in Section~\ref{SecNarainVielbein}, 
elements $U\in \text{O}(D;\Real)\times\text{O}(D+16;\Real) \subset \text{O}_{\get}(D,D+16; \Real)$ 
map physically identical Narain configurations to each other.

Using the generators of the $\text{O}_{\widehat{\get}}(D,D+16; \Real)$ listed in 
Table~\ref{tb:DualitySubgroups} an element $U$ from the identity component of 
$\text{O}(D;\Real)\times\text{O}(D+16;\Real)$ defined by eqn.~\eqref{ConstraintsSODD+16} can be 
expressed as follows
\equ{ \label{Udecomposition}
U = R\, \widehat{U} R^{-1}~; 
\quad 
\widehat{U} ~=~ \widehat M_e(\gD\gth) \widehat M_W(\gD W)  \, 
\widehat M_\ga(\gD A) \widehat M_\gb(\gD B) 
\widehat M_e(\gD K) 
\widehat M_A(\gD A) \widehat M_B(\gD B)~,  
}
if $u_+$ is invertible, see eqn.~\eqref{eqn:Defintionupm}, and we defined
\begin{subequations}
\equ{
\gD K ~=~ \Id_D + \gD C~,
\qquad \gD C ~=~ \gD B + \sfrac 12 \gD A^T \gD A~,
\qquad 
\gD B^T ~=~ -\gD B~, 
\\[1ex] 
\gD\gth^T \gD\gth ~=~ \Id_D 
\quad\text{and}\quad 
\gD W^T \gD W ~=~ \Id_{16}~. 
}
\end{subequations}
The first two factors $\widehat M_e(\gD\gth) \widehat M_W(\gD W)$ in eqn.~\eqref{Udecomposition} 
define the subgroup $\text{O}(D;\Real) \times \text{O}(16;\Real)$, where the $\text{O}(D;\Real)$-factor 
lies diagonally in both the left- and right-moving directions. This can be seen from 
eqn.~\eqref{Udecomposition} by using the expressions for the duality group elements given in 
Table~\ref{tb:DualitySubgroups} and the matrix $R$ defined in eqn.~\eqref{eqn:vielbeinR}. Then, we 
obtain
\equ{  \label{UexplicitParameterization} 
U = 
\pmtrx{ 
\gD \gth & 0 & 0 \\[1ex]  
0 & \gD \gth & 0 \\[1ex]  
0 &  0 & \gD W 
} 
\pmtrx{ 
\Id_D & 0 & 0 \\[1ex]  
0 & (\Id_D-\gD C)^T\big(\Id_D+\gD C\big)^{-T} &  - \sqrt{2} \big(\Id_D+\gD C\big)^{-T} \gD A^T \\[1ex]  
0 & \sqrt{2} \gD A \big(\Id_D+\gD C\big)^{-T} & \Id_{16} - \gD A\big(\Id_D+\gD C\big)^{-T} \gD A^T }. 
}
By comparing this with eqn.~\eqref{Uexplicit} one can read off the expressions for the submatrices 
\equ{\label{eqn:thetafromU}
\arry{ll}{ 
 u_\text{l}  = \gD \gth  \big(\Id_D-\gD C\big)^T\big(\Id_D+\gD C\big)^{-T}~, \qquad & u_\text{lL} =  - \sqrt{2} \gD \gth \big(\Id_D+\gD C\big)^{-T} \gD A^T~,  \\[1ex] 
 u_\text{Ll} =  \sqrt{2} \gD W \gD A \big(\Id_D+\gD C\big)^{-T}~,            \qquad & u_\text{L} = \gD W\Big( \Id_{16} - \gD A\big(\Id_D+\gD C\big)^{-T} \gD A^T \Big)
}
}
and $u_\text{r} = \gD \gth$. One can verify that these expressions satisfy the 
constraints~\eqref{ConstraintsSODD+16}. 

In addition, for a given element 
$U\in \text{O}(D;\Real)\times\text{O}(D+16;\Real)$ one can use eqn.~\eqref{eqn:thetafromU} to 
decompose $\widehat{U}=R^{-1}U\,R$ according to eqn.~\eqref{Udecomposition}, i.e.
\equ{ 
\arry{ll}{ 
\gD \gth ~=~ u_\text{r}~,\qquad &
\gD C ~=~ -u_-^T\, u_+^{-T}~, \\[1ex]
\gD A ~=~ - \frac{1}{\sqrt{2}}  u_\text{lL}^T\, u_+^{-T}~,\qquad  & 
\gD W ~=~ u_\text{L} \Big( \Id_{16} - \frac{1}{2} u_\text{lL}^T\, u_+^{-T} u_\text{r}^{-1} u_\text{lL} \Big)^{-1}~,
} 
}
where we assumed that $u_+$ is invertible.


\section[Generalized space groups of Narain orbifolds]{Generalized space groups of Narain orbifolds}
\label{SecOrbifolds} 

In this section we introduce the generalized space group for heterotic Narain orbifolds and discuss 
some of its properties. In particular, we define orbifold projections to characterize quantization 
conditions of the generalized shift vectors and state the conditions to preserve $\cN=1$ 
supersymmetry.

\subsection[Heterotic Narain orbifolds]{Heterotic Narain orbifolds}
\label{SecHetNarainOrbifolds} 

Next, we consider orbifolds of the heterotic Narain lattice construction denoted by
\equ{
T^{2D+16}_\Narain/\mathbf{P}~.
}
Here, the $2D+16$-dimensional torus $T^{2D+16}_\Narain$ is specified by a Narain lattice $\Narain$. 
In addition, the Narain point group $\mathbf{P}$ is defined as a (sub-)group of the rotational symmetries 
of $\Narain$, as we will see later in eqn.~\eqref{LatticeCompatibility}. Hence, the Narain point group 
$\mathbf{P}$ is finite. The generators of $\mathbf{P}$ are $(2D+16)\times(2D+16)$ matrices and they 
are denoted by $\gTh_\ga$, for $\ga = 1,\ldots, N_\mathbf{P}$. $K_\ga$ is the order of 
$\gTh_\ga$. In more detail, for each generator $\gTh_\ga$, the order $K_\ga$ is the 
smallest non-negative integer such that $\gTh_\ga^{K_\ga} = \Id$. Elements of $\mathbf{P}$ 
are often called twists. In the following, a generic twist will be denoted by $\gTh$ and $K$ gives 
its order.

To define the compactification of the heterotic string on a Narain orbifold\cite{Narain:1986qm,Narain:1990mw}, 
the main idea is to generalize the boundary conditions~\eqref{NarainBoundaryConditions} of the 
$2D+16$-dimensional right- and left-moving coordinate-vector $Y$ to
\equ{\label{YOrbiBoundaryConditions} 
Y(\gs+1,\bgs+1) =  \gTh \, Y(\gs,\bgs) + V_\gTh + L~,
}
for all elements $\gTh\in\mathbf{P}$ and $L \in \Narain$. In general, for each twist $\gTh$ there 
is a so-called generalized shift $V_\gTh$ associated to it, which will be discussed in detail later. 
Importantly, the twists $\gTh$ are not allowed to mix right- and left-moving fields in 
eqn.~\eqref{YOrbiBoundaryConditions}. Hence, for all $\gTh \in \mathbf{P}$ we demand
\equ{\label{eqn:TwistBlockform}
\gTh ~=~ \pmtrx{\gth_\text{r} & 0 \\ 0 & \gTh_\text{L}} ~\in~ O(D;\Real) \times O(D+16;\Real)~.
}
Consequently, we find the conditions

\equ{\label{eqn:ThetaProperties}
\gTh_\ga^T \gTh_\ga ~=~ \Id~, 
\qquad  
\gTh_\ga^T \get\, \gTh_\ga ~=~ \get
\quad\text{and}\quad
\gTh_\ga^{K_\ga} ~=~ 0~,
}
for all generators $\gTh_\ga$ of the Narain point group.

Furthermore, we call a Narain orbifold symmetric~\cite{Dixon:1985jw, Dixon:1986jc}, if there is a 
basis such that all generators $\gTh_\ga \in \mathbf{P}$ are simultaneously of the form
\equ{\label{eqn:SymmetricOrbifoldTwist}
\gTh_\ga ~=~ \pmtrx{\gth_\ga & 0 & 0 \\ 0 & \gth_\ga & 0\\ 0 & 0 & \Id_{16}} ~\in~ O(D;\Real) ~\subset~ O(D;\Real) \times O(D+16;\Real)~. 
}
If such a basis does not exist, then the corresponding Narain orbifold is asymmetric. 
Even though this definition of symmetric orbifolds involves a choice of basis, this property is in 
fact basis independent. Nevertheless, in a given basis it might be difficult to see whether a 
Narain orbifold is symmetric or asymmetric: One can bring a symmetric twist $\gTh_\text{sym}$ into 
a seemingly asymmetric twist $\gTh_\text{asym}=U^{-1}\, \gTh_\text{sym}\,U$ by the choice of 
$U \in \text{O}(D;\Real) \times \text{O}(D+16;\Real)$, see also the example in 
Section~\ref{sec:O22Z12Orbifolds}. However, the conjugation with $U$ can neither change 
the orders of $\gth_\text{r}$ and $\gTh_\text{L}$, nor the two finite groups which are generated by 
either $\gth_{\ga\,\text{r}}$ or $\gTh_{\ga\,\text{L}}$.

\subsection[Generalized space group]{Generalized space group}


It has been proven to be very convenient to employ a space group formulation of the heterotic 
string on symmetric orbifolds, especially in the context of classifications~\cite{Fischer:2012qj}. This language 
can be extended to Narain orbifolds naturally. The generalized space group $\mathbf{S}$ associated to 
a Narain orbifold is defined as being generated by the elements 
\equ{\label{eqn:NarainSpaceGroupElement}
\big( \Id, L \big) 
\quad\text{and}\quad 
\big(\gTh_\ga, V_\ga\big) 
\quad\text{for all}\quad L \in \Narain \quad\text{and}\quad \gTh_\ga \in \mathbf{P}~,
}
where $V_\ga$, a vector with $2D+16$ components, is the so-called generalized shift which is 
associated to the twist $\gTh_\ga$. Conversely, we demand that for 
all space group elements of the form $(\Id,L') \in \mathbf{S}$ it follows that $L' \in \Narain$. 
So, the Narain lattice $\Narain$ is the subgroup of $\mathbf{S}$ that contains all pure 
translations of $\mathbf{S}$. Note that a generator $(\gTh_\ga, V_\ga)$ is a generalized 
roto-translation if $V_\ga \neq 0$, see~\cite{Fischer:2012qj}. These generators build the so-called 
Narain orbifold group $\mathbf{O}$, which is defined modulo lattice translations. Hence, just as 
$\mathbf{P}$, the Narain orbifold group $\mathbf{O}$ is a finite group.

A general space group element $g = (\gTh, \lambda) \in \mathbf{S}$ is defined to act on $Y$ as
\equ{ \label{SpaceGroupAction} 
Y ~\mapsto~ g [Y] ~=~ \big(\gTh, \lambda\big) [Y] ~=~ \gTh\, Y + \lambda~. 
}
Consequently, the unit element of $\mathbf{S}$ is given by
\equ{
\big(\Id, 0\big) ~\in~ \mathbf{S}~.
}
The inverse element $g^{-1}$ of $g = (\gTh, \lambda) \in \mathbf{S}$ reads
\equ{
g^{-1} ~=~ \big(\gTh^{-1}, -\gTh^{-1}\lambda\big) ~\in~ \mathbf{S}~.
}
Furthermore, two elements $g = (\gTh, \gl)$ and $g' = (\gTh',\gl')$ are multiplied as
\equ{
g\, g' ~=~ \big(\gTh, \lambda\big)\, \big(\gTh', \lambda'\big) ~=~ \big(\gTh\, \gTh', \gTh\,\lambda' + \lambda\big) ~\in~ \mathbf{S}~.
}
Hence, the generalized space group $\mathbf{S}$ is in general non-Abelian even if the Narain point 
group $\mathbf{P}$ is Abelian.

For orbifolds, each sector of string states is characterized by a boundary 
condition~\eqref{YOrbiBoundaryConditions} and, thus, by the so-called constructing element 
$g=(\gTh, \gl) \in \mathbf{S}$, where $\gl = V_\gTh + L$ and $L \in \Narain$. Only 
those elements $g' \in \mathbf{S}$ that commute with the constructing element $g$ yield projections 
and, hence, give rise to non-vanishing contributions to the twisted sector partition function. This 
only happens when 
\equ{
\gTh \gTh' ~=~ \gTh' \gTh
\quad\text{and}\quad
(\Id - \gTh) \gl' ~=~  (\Id - \gTh') \gl~. 
}

\subsection[Conditions on the twists $\gTh_\ga$]{Conditions on the twists $\boldsymbol{\gTh_\ga}$}
\label{sec:ConditionsOnTwists}

Furthermore, we choose $L \in \Narain$ and consider
\equ{
\big(\gTh_\ga, V_\ga\big)\, \big(\Id, L\big)\, \big(\gTh_\ga, V_\ga\big)^{-1} ~=~ \big(\Id, \gTh_\ga L\big) ~\stackrel{!}{\in}~ \mathbf{S} \quad\Rightarrow\quad \gTh_\ga L ~\stackrel{!}{\in}~ \Narain~.
}
Thus, the lattice $\Narain$ is a normal subgroup of $\mathbf{S}$ and the Narain point group 
$\mathbf{P}$ has to consist of automorphisms of the Narain lattice, i.e.
\equ{ \label{LatticeCompatibility} 
\gTh_\ga\, \Narain ~=~ \Narain~. 
}
In addition, we have to impose eqn.~\eqref{eqn:ThetaProperties} on the twist generators $\gTh_\ga$.

It is is interesting to pause here and reflect on the possible orders of twists for a given number 
of dimensions $D_\Gamma$ for general orbifolds associated to a lattice $\gG$. As is 
well-known~\cite{Vaidyanathaswamy:1928:IRU}, if the order $K$ satisfies
\equ{\label{eqn:EulerPhiBoundary}
\phi(K) ~\leq~ D_\Gamma~,
}
then there exists at least one lattice $\Gamma$ with rotational symmetry of order $K$. Here, 
$\phi(K)$ is the Euler $\phi$-function and this bound does not take into account that one can build 
point groups as direct sums of lower dimensional cases. However, in the current paper we are not 
working with a general lattice $\Gamma$ in $D_\Gamma$ dimensions, but with Narain lattices 
$\Gamma=\Narain$ with $D_\Narain=2D+16$. Hence, contrary to the Euclidean case, it is not 
guaranteed that there exists a Narain lattice for each order $K$ satisfying the 
bound~\eqref{eqn:EulerPhiBoundary}.

\subsection[Orbifold projections of $\Narain$]{Orbifold projections of $\boldsymbol{\Narain}$}

In general, a twist $\gTh \in \mathbf{P}$ of order $K$ acts as the identity in some directions of 
$Y$ while it acts as a $\Intr_K$ twist on others. To identify these directions, we define 
projection operators for each twist $\gTh \in \mathbf{P}$: The projection operators 
$\ProjP{\mathcal{P}}{\gTh}$ and $\ProjO{\mathcal{P}}{\gTh}$ project a vector onto the directions in 
which $\gTh$ acts trivially and non-trivially, respectively. In detail, we define the projectors
\equ{ \label{eqn:Projector}
\ProjP{\mathcal{P}}{\gTh} ~=~ \frac{1}{K}\, \sum_{j=0}^{K-1} \gTh^j
\qquad\text{and}\qquad
\ProjO{\mathcal{P}}{\gTh} ~=~ \Id - \ProjP{\mathcal{P}}{\gTh}~, 
}
with the properties 
\equ{
\big(\ProjP{\mathcal{P}}{\gTh}\big)^2 = \ProjP{\mathcal{P}}{\gTh}~, 
\qquad 
\big(\ProjO{\mathcal{P}}{\gTh}\big)^2  = \ProjO{\mathcal{P}}{\gTh}~, 
\qquad 
\gTh\, \ProjP{\mathcal{P}}{\gTh} = \ProjP{\mathcal{P}}{\gTh}~,
\qquad 
\ProjO{\mathcal{P}}{\gTh} \, \ProjP{\mathcal{P}}{\gTh} =
\ProjP{\mathcal{P}}{\gTh} \, \ProjO{\mathcal{P}}{\gTh} = 0~.
}
Then, any vector $\lambda \in \Real^{2D+16}$ can be decomposed into two vectors 
$\ProjP{\lambda}{\gTh}$ and $\ProjO{\lambda}{\gTh}$ according to
\equ{
\ProjP{\lambda}{\gTh} ~=~ \ProjP{\mathcal{P}}{\gTh}\, \lambda~, \quad \ProjO{\lambda}{\gTh} ~=~ \ProjO{\mathcal{P}}{\gTh}\, \lambda \quad\text{so that}\quad
\lambda ~=~ \ProjP{\lambda}{\gTh} + \ProjO{\lambda}{\gTh}~,
}
and $\gTh\, \ProjP{\lambda}{\gTh} = \ProjP{\lambda}{\gTh}$. The final relation clarifies the use of 
the subscript $\parallel$: It defines the directions which are left invariant by $\gTh$.

Moreover, it is important to realize that the projected Narain lattice 
$\ProjP{\Narain}{\gTh} = \ProjP{\mathcal{P}}{\gTh} \Narain$ is in general not Narain. In detail, 
even if $\Narain$ and $\gTh\,\Narain$ are Narain lattices, see 
eqn.~\eqref{LatticeCompatibility}, the normalisation $1/K$ in the projection operator 
$\ProjP{\mathcal{P}}{\gTh}$ in eqn.~\eqref{eqn:Projector} can make $\ProjP{\Narain}{\gTh}$ non-Narain. 
A Narain lattice is said to be factorized w.r.t.\ the orbifold twists when 
\equ{
\ProjP{\Narain}{\gTh} ~\subset~ \Narain \,
}
for all twists $\gTh \in \mathbf{P}$. In this case, obviously, all projected Narain lattices are 
themselves Narain again.

\subsection[Quantization of the generalized shifts ${V_\ga}$]{Quantization of the generalized shifts $\boldsymbol{V_\ga}$}
\label{sec:QuantizationOfShift}

For each Narain point group generator $\gTh_\ga$ of order $K_\ga$ we consider the generator 
$(\gTh_\ga, V_\ga)$ of the generalized space group $\mathbf{S}$. Then, its $K_\ga$-th 
power reads
\equ{
\big(\gTh_\ga, V_\ga\big)^{K_\ga} ~=~ \big(\gTh^{K_\ga}, \sum_{j=0}^{K_\ga-1}\, \gTh_\ga^j\, V_\ga \big) ~=~ \big(\Id, K_\ga \ProjP{\mathcal{P}}{\ga} V_\ga \big) ~\stackrel{!}{\in}~ \mathbf{S}~,
}
(where $\ProjP{\mathcal{P}}{\ga} = \ProjP{\mathcal{P}}{\gTh_\ga}$) without summation over 
$\ga$. Consequently, we have to demand the condition
\equ{ \label{eqn:ProjVFromNarain}
K_\ga \ProjP{V_\ga}{} ~=~
K_\ga \ProjP{\mathcal{P}}{\ga}  V_\ga
~=~ \sum_{j=0}^{K_\ga-1}\, \gTh_\ga^j\, V_\ga ~=~ L_\ga
~\stackrel{!}{\in}~ \Narain~.
}
That is, the shift $V_\ga$ needs to be quantized in units of $K_\ga$ in the directions where $\gTh_\ga$ 
acts trivially, i.e.\ $V_\ga$ is given by
\equ{\label{eqn:ShiftQuantization}
V_\ga ~=~ \frac{L_\ga}{K_\ga} +  \lambda_\ga  \quad\text{with}\quad \gTh_\ga\, L_\ga = L_\ga~,~ \ProjP{\mathcal{P}}{\ga} \lambda_\ga ~=~ 0 \quad\text{and}\quad L_\ga ~\in~ \Narain~.
}
The same procedure can be applied to some arbitrary element $\gTh\in \mathbf{P}$ of order $K$ 
with associated element $(\gTh, V_\gTh) \in \mathbf{S}$. This yields
\equ{\label{eqn:ShiftQuantizationInGeneral}
V_\gTh ~=~ \frac{L_\gTh}{K} +  \lambda_\gTh  \quad\text{with}\quad \gTh\, L_\gTh = L_\gTh~,~ \ProjP{\mathcal{P}}{\gTh} \lambda_\gTh ~=~ 0 \quad\text{and}\quad L_\gTh \in \Narain~.
}
As a remark, for example in the case when $\gTh_\ga^k$ has a fixed torus for $0 < k < K_\ga$ 
(i.e.~when $\gTh_\ga^k$ has more invariant directions than $\gTh_\ga$) 
eqn.~\eqref{eqn:ShiftQuantizationInGeneral} gives stronger quantization conditions on the shift 
$V_\ga$ than eqn.~\eqref{eqn:ShiftQuantization}. 

Various choices for $V_\ga$ correspond to the same Narain orbifold. Indeed, one can shift the origin, i.e.
\equ{\label{eqn:OriginShift}
Y(\gs,\bgs) ~\mapsto~ Y(\gs,\bgs) + Y_0~,
}
and hence transform the generalized shifts $V_\gTh \mapsto V_\gTh - (\Id -  \gTh)\, Y_0$ for 
$Y_0 \in \Real^{2D+16}$. (In light of the equivalence~\eqref{LRgaugeTrans}, only the (lower) $D+16$ 
components of $Y_0$ actually modify the description.) By doing so, one can set the components of 
$\lambda_\gTh$ either to zero or to some quantized value for each element 
$(\gTh, \ProjP{V_\gTh}{} + \lambda_\gTh)$ of the Narain orbifold group $\mathbf{O}$. Especially, if 
the Narain point group is isomorphic to $\Z{K}$ (with one generator $\gTh$ of order $K$) the 
generalized shift can be chosen as $V_\gTh=L_\gTh/K$ with $\gTh L_\gTh = L_\gTh \in \Narain$ 
without loss of generality.

\subsection[Preserving at least ${\cN=1}$ target-space supersymmetry]{Preserving at least $\boldsymbol{\cN=1}$ target-space supersymmetry}
\label{sc:SUSYpreserving} 

To enable the discussion on target-space supersymmetry we first need to recall a few facts about 
supersymmetry on the worldsheet. By construction the heterotic string has $(1,0)$ worldsheet 
supersymmetry. Hence, we can identify the worldsheet supercurrent
\equ{ \label{eqn:worldsheetsupercurrent}
T_\text{F} ~=~ \gps_\text{R}^\gm\, \bder x_\gm + 
\gps_\text{R}^T\, u_\text{Rr}\, \bder y_\text{r}~, 
}
where $\gps_\text{R} =(\gps_\text{R}^i)$ are the real worldsheet fermions of the $D$ compactified 
dimensions and $u_\text{Rr}$ is a $D \times D$ matrix. For each twist $\gTh_\ga$, the space group 
action~\eqref{SpaceGroupAction} is defined to be accompanied by a transformation of $\gps_\text{R}$ 
as
\equ{ 
\gps_\text{R} ~\mapsto~ g [\gps_\text{R}] ~=~ \gth_{\ga\,\text{R}}\, \gps_\text{R}~, 
}
where $\gth_{\ga\,\text{R}} \in \text{O}(D;\Real)$. Since the first term 
$\gps_\text{R}^\gm\, \bder x_\gm$ in eqn.~\eqref{eqn:worldsheetsupercurrent} is orbifold invariant 
the worldsheet supercurrent $T_\text{F}$ has to be orbifold invariant as well. Consequently, we 
need to require that the twists on the 
right-moving coordinates $y_\text{r}$ and on the right-moving fermions $\gps_\text{R}$ are 
identified: $\gth_{\ga\,\text{R}} = u_\text{Rr}\, \gth_{\ga\,\text{r}}\, u_\text{Rr}^{-1}$. 

Given that the properties of target-space fermions are determined by the right-moving momentum 
$p_\text{R}$ associated to these right-moving fermions, as given eqn.~\eqref{FermiLatticeSum}, the 
question of target-space supersymmetry is only affected by the transformations generated by 
$\gth_{\ga\, \text{R}}$ in the right-moving sector. In particular, target-space supersymmetry is 
independent of the choice one makes for $\gTh_{\ga\,\text{L}}$. Only if one restricts oneself to 
symmetric orbifolds, for which $\gth_{\ga\,\text{L}} = \gth_{\ga\,\text{l}} \oplus \Id_{16}$ and 
$\gth_{\ga\,\text{l}} = \gth_{\ga\,\text{r}} = \gth_{\ga\,\text{R}}$ with $u_\text{Rr}=\Id_D$, see 
eqn.~\eqref{eqn:SymmetricOrbifoldTwist}, this connection is made.

Consequently, in order to preserve at least $\cN=1$ supersymmetry in the $d$-dimensional 
target-space, the generators $\gth_{\ga\,\text{R}} \in \text{SO}(D;\Real)$ have 
to lie inside the appropriate special holonomy subgroup of $\text{SO}(D;\Real)$. For $D=4, 6, 7, 8$ 
these subgroups are $\text{SU}(2)$, $\text{SU}(3)$, $\text{G}_2$ and $\text{Spin}(7)$, 
respectively, see e.g.~\cite{Polchinski:1998rr}. 
For example, assume $D=6$ and an Abelian Narain point group, i.e.
\equ{
\mathbf{P} ~\cong~ \Z{K_1} \times \ldots \Z{K_{N_\mathbf{P}}}~.
}
Then, the four-dimensional effective low energy theory possesses at least $\mathcal{N}=1$ 
supersymmetry if 
\equ{\label{eqn:N1SUSYin4d}
\gf_{\ga\,\text{R}}^m = 0~, 
\qquad 
K_\ga\, \gf_{\ga\,\text{R}}^{a} \equiv 0~, 
\qquad 
\sfrac 12\, \sum_a \gf_{\ga\,\text{R}}^a = 0~.
}
Here, we introduced the so-called twist vector 
$\gf_{\ga\, \text{R}} = (\gf_{\ga\, \text{R}}^m, \gf_{\ga\,\text{R}}^a)$ as the vector of phases 
corresponding to $\gth_{\ga\,\text{R}}$, such that $\gth_{\ga\,\text{R}}$ acts as
\equ{
\gps_\text{R}^m ~\mapsto~ e^{2\pi i\, \gf_{\ga\, \text{R}}^m} \, \gps_\text{R}^m~, 
\qquad 
\gps_\text{R}^a ~\mapsto~ e^{2\pi i\, \gf_{\ga\, \text{R}}^a} \, \gps_\text{R}^a~, 
}
using the complex indices defined below eqn.~\eqref{FermiLatticeSum}. In fact, the last condition 
of eqn.~\eqref{eqn:N1SUSYin4d} only needs to be imposed mod integers (i.e.~$\equiv$) and this 
specific choice fixes the unbroken supercharges for $d=4$ and $\gf_{\ga\, \text{R}}^a\neq 0$ 
to be represented as $\pm(\sfrac 12,\sfrac 12,\sfrac 12,\sfrac 12)$.

\newpage

\section[Moduli stabilization in Narain orbifolds]{Moduli stabilization in Narain orbifolds}
\label{sec:ModuliStabilization}

\noindent
As we have seen in the previous section, the space group description of Narain orbifolds is 
naturally formulated using the twist $\gTh$ and the generalized vielbein $E$. On the other hand, 
the question about moduli stabilization and classification, in particular, are more conveniently 
discussed in the so-called lattice basis in which the twist is encoded by an integral matrix $\widehat{\gr}$. 
Therefore, we begin this section with a discussion of Narain orbifolds in the lattice basis. Beside 
the integral twist matrices $\widehat{\gr}$, we introduce the generalized metric $\cH$ and a 
closely related $\Intr_2$-grading $\cZ$. After that we investigate under which conditions Narain 
orbifolds exist and derive restrictions on the Narain moduli that have to be imposed in order to 
be compatible with the orbifold action. In particular, we derive a character formula that counts 
the dimension of the orbifold Narain moduli space.

\subsection[Narain orbifolds in the lattice basis]{Narain orbifolds in the lattice basis}
\label{sec:TwistInLatticeBasis}

\subsubsection*{Twists and shifts in the lattice basis}

We have seen in eqn.~\eqref{LatticeCompatibility} that each point group generator $\gTh_\ga$ has to 
map a Narain vector $E\,N$ to another Narain vector $E\,N' = \gTh_\ga\, E\,N$, see 
eqn.~\eqref{NarainElement}. It follows that $N' = \widehat\rho_\ga N$, where we define 
$\widehat{\rho}_\ga$ as
\equ{ \label{IntroRho} 
\widehat{\rho}_\ga ~=~ E^{-1} \gTh_\ga\, E ~=~ \widehat E^{-1} \widehat\gTh_\ga\, \widehat E ~\in~ \text{GL}(2D+16; \Intr)~.
}
Here, we used $E = U\,R\, \widehat E$ and we absorbed $U$ in the definition of 
$\widehat\gTh_\ga= R^{-1}\, U^{-1}\,\gTh_\ga\, U\,R$.

The matrices $\widehat{\gr}_\ga$ represent the generating twists $\gTh_\ga$ in the so-called 
lattice basis. They have to be invertible over the integers 
(i.e.~$\widehat{\gr}_\ga\in\text{GL}(2D+16; \Intr)$) because each $\widehat\rho_\ga$ has to 
map an integer vector $N$ one-to-one to another integer vector $N'$. Furthermore, they 
inherit the following conditions 
\equ{ \label{RhoProperties1}
\widehat \rho_\ga^{\,T} \widehat{\get}\, \widehat \rho_\ga ~=~ \widehat{\get}
\qquad\text{and}\qquad
\widehat \rho_\ga^{\,K_\ga} ~=~ \Id~,
}
since the generating twists $\gTh_\ga$ are elements of $\text{O}_{\get}(D,D+16;\Real)$ of finite 
order $K_\ga$. The integral matrices $\widehat \rho_\ga$ generate the so-called Narain point group 
in the lattice basis $\widehat{\mathbf{P}} \subset \text{O}_{\widehat{\get}}(D,D+16;\Intr)$, while 
twists $\gTh_\ga \in \mathbf{P}$ are given in the so-called coordinate basis. The lattice basis 
will be of special importance for the classification of Narain orbifolds later in 
Section~\ref{sec:NarainQandZClass}. Moreover, the space group generators $(\gTh_\ga, V_\ga)$ and 
$(\Id, L) \in \mathbf{S}$ can be represented in the lattice basis as 
\equ{ \label{SpaceGroupGenLatticeBasis} 
(\widehat{\gr}_\ga, \frac 1{K_\ga} N_\ga) ~\in~ \widehat{\mathbf{S}} \quad\text{and}\quad (\Id, N) ~\in~ \widehat{\mathbf{S}}~,
}
where $V_\ga = \frac 1{K_\ga}\, E\, N_\ga$ and $L = E\, N$ for 
$N, N_\ga \in \Intr^{2D+16}$.

\subsubsection*{Generalized metric}

Eqn.~\eqref{RhoProperties1} represents two out of the three properties~\eqref{eqn:ThetaProperties} 
of the generators $\gTh_\ga$ in the lattice basis. The remaining one, $\gTh_\ga^T\gTh_\ga = \Id$, 
can be cast in the form  
\equ{\label{eqn:RhoOrthogonal}
\widehat \rho_\ga^T \cH\, \widehat \rho_\ga ~=~ \cH~, 
}
where we have introduced the so-called generalized metric $\cH$ defined as 
\equ{ \label{cHDefinition} 
\cH ~=~ E^T E ~=~ \widehat{E}^T(e,B,A) \,R^T R\, \widehat{E}(e,B,A)~.
}
In other words, condition~\eqref{eqn:RhoOrthogonal} states that the generators $\widehat \rho_\ga$ 
and the generalized metric $\cH$ have to be compatible.

The generalized metric is given explicitly by
\equ{
\label{GeneralizedMetric}
\cH(G,B,A) ~=~ 
\pmtrx{
G + A^T A + C G^{-1} C^T & -C G^{-1} & (\Id_D + C G^{-1}) A^T \ga_\text{g}
\\[1ex] 
-G^{-1} C^T & G^{-1} & - G^{-1} A^T \ga_\text{g}
\\[1ex] 
\ga_\text{g}^T A (\Id_D+ G^{-1} C^T) & -\ga_\text{g}^T A G^{-1} & \ga_\text{g}^T \big(\Id_{16}+ A G^{-1} A^T \big) \ga_\text{g}
}~, 
}
using eqns.~\eqref{eqn:vielbeinR} and~\eqref{NarainModuli}. It is an interesting object in its own 
right: Assume one is given a generic Narain lattice (with moduli-independent Narain metric 
$\widehat{\get} = E^T \get\, E$ as given in eqn.~\eqref{eqn:etahat}) by specifying the generalized 
vielbein $E$, then it might be rather awkward to determine the matrix $U$ from $E = U R \widehat E$ 
such that we can read off the moduli contained in the matrix $\widehat E$. As the generalized metric 
$\cH$ is independent of $U$, it can be used to read off the metric $G$ of the $D$-dimensional torus, 
the $B$-field and the Wilson line matrix $A$. As the explicit expression of the generalized 
metric~\eqref{GeneralizedMetric} shows, not all its components are independent, i.e.\ $\cH$ is not 
a generic $(2D+16) \times (2D+16)$ matrix. Indeed, $\cH$ satisfies the constraints
\equ{ \label{ConstraintGenMetric} 
\cH\, \widehat\get^{\,-1} \cH ~=~ \widehat{\get}
\qquad\text{and}\qquad 
\cH^T = \cH~, 
}
as follows from its definition~\eqref{cHDefinition} .

\subsubsection*{A $\boldsymbol{\Intr_2}$ grading}
\label{sec:Z2Grading}

The compatibility condition~\eqref{eqn:RhoOrthogonal} of the orbifold twists in the lattice basis 
can also be represented as 
\equ{ \label{eqn:ZCommutes}
\cZ\, \widehat \rho_\ga ~=~ \widehat \rho_\ga\,\cZ~,
}
where we have defined 
\equ{\label{eqn:GenMetricParam}
\cZ ~=~ \widehat{\get}^{-1}\cH ~=~ E^{-1} \get E~=~  \widehat E(e,B,A)^{-1}\, \widehat I\, \widehat E(e,B,A)
 ~\in~ \text{O}_{\widehat{\get}}(D,D+16, \Real)~. 
}
The second expression in this equation is obtained using  $\widehat I = R^{-1} \get\, R$, as given 
in Table~\ref{tb:DualitySubgroups}, and the relation $R^TR = \widehat{\get}\, \widehat I$. 
Explicitly, $\cZ$ is given by
\equ{
\label{GradingZ}
\cZ(G,B,A) ~=~ 
\pmtrx{
-G^{-1} C^T & G^{-1} & - G^{-1} A^T \ga_\text{g}
\\[1ex] 
G + A^T A + C G^{-1} C^T & -C G^{-1} & (\Id_D + C G^{-1}) A^T \ga_\text{g}
\\[1ex] 
\ga_\text{g}^{-1} A (\Id_D+ G^{-1} C^T) & -\ga_\text{g}^{-1} A G^{-1} & \ga_\text{g}^{-1} \big(\Id_{16}+ A G^{-1} A^T \big) \ga_\text{g}
}~.
}
The constraints~\eqref{ConstraintGenMetric}, which the generalized metric satisfies, translate to 
the following conditions on $\cZ$: 
\equ{ \label{eqn:ZConditions}
\cZ^T \,\widehat{\get}\, \cZ ~=~ \widehat{\get}
\qquad\text{and}\qquad 
\cZ^2 ~=~ \Id~.
}
This can be confirmed by using eqn.~\eqref{ConstraintGenMetric} and the fact that 
$\widehat{I}^{\, 2} =  \Id$. Given its definition~\eqref{eqn:GenMetricParam}, the matrix $\cZ$ has 
signature $(D,D+16)$, just as $\get$ (and $\widehat{I}$). This leads to a grading of the 
Narain lattice: It characterizes the distinction between $D$ right- and $D+16$ left-moving 
directions of the Narain lattice.

\subsection[On the existence of Narain orbifolds for a given point group]{On the existence of Narain orbifolds for a given point group}
\label{sec:ExistenceOfNarainOrbifolds}

Assume a given finite point group 
$\widehat{\mathbf{P}} \subset \text{O}_{\widehat{\get}}(D,D+16;\Intr)$ with generators 
$\widehat{\gr}_\alpha$ in the lattice basis. We want to understand these generators 
$\widehat{\gr}_\ga$ as the crucial ingredient in the definition of a Narain orbifold. Therefore, we 
have to address the following question: Under which condition does a corresponding Narain orbifold 
exist? In terms of the terminology introduced in Section~\ref{SecOrbifolds} this can be phrased as 
follows: When does a Narain lattice exist, such that all generators $\gTh_\alpha$ of the 
corresponding group $\mathbf{P}$ in the coordinate basis satisfy~\eqref{eqn:ThetaProperties} and 
are symmetries of this lattice~\eqref{LatticeCompatibility}? 

In the following, we will answer this question in the lattice basis. Then, the conditions on 
$\gTh_\ga$ translate to conditions~\eqref{RhoProperties1} and~\eqref{eqn:RhoOrthogonal} on 
$\widehat{\gr}_\ga \in \widehat{\mathbf{P}}$. In fact, eqn.~\eqref{RhoProperties1} is fulfilled 
by assumption (i.e~$\widehat{\mathbf{P}} \subset \text{O}_{\widehat{\get}}(D,D+16;\Intr)$ and 
finite). Thus, it remains to show that eqn.~\eqref{eqn:RhoOrthogonal} is fulfilled, i.e.~we have to 
find a generalized metric that is compatible with all generators $\widehat{\gr}_\ga$. 
Consequently, a Narain orbifold with given point group $\widehat{\mathbf{P}}$ exists if 
one finds Narain moduli $G$, $B$ and $A$ that are invariant under 
$\widehat{\gr}_\alpha \in \widehat{\mathbf{P}}$.

If such a generalized vielbein exists, then generically, not all the moduli of the Narain torus 
compactification are still free; some Narain moduli are stabilized.  Thus, we can use our 
discussion on the transformation properties of Narain moduli under general $T$-duality 
transformations in Section~\ref{sec:NarainModuliTrafo} in order to derive conditions for moduli 
stabilization.

To address these questions, we study the existence of both a twist 
$\gTh_\ga  \in \text{O}(D;\Real) \times \text{O}(D+16;\Real)$ for each $\widehat{\rho}_\ga$ and a 
compatible generalized vielbein $\widehat{E}(e,B,A)$, i.e. 
\equ{\label{eqn:OrbiCompatible}
\gTh_\ga\, R\,\widehat{E}(e,B,A) ~=~  R\,\widehat{E}(e,B,A)\, \widehat{\rho}_\ga~,
}
which is equivalent to eqn.~\eqref{IntroRho} by absorbing $U$ in the definition of $\gTh_\ga$. 
Eqn.~\eqref{eqn:OrbiCompatible} constitutes nine coupled matrix equations for the $D(D+16)$ Narain 
moduli $G, B, A$ and the $D(D-1)/2$ and $(D+16)(D+15)/2$ parameters inside each of the generators 
$\gTh_\ga$. 

Instead of trying to solve all nine coupled matrix equations, we first focus on a subset of 
only three matrix equations
\equ{ \label{ReducedOrbiCompatible} 
W\, \widehat{\gr}_\ga ~=~ \gth_{\ga\,\text{r}}\, W
\quad\text{with}\quad 
W ~=~ \sqrt 2\, (R\, \widehat E)_\text{r} ~=~ e^{-T} \pmtrx{G+C^T, & -\Id_D, & A^T\ga_\text{g}}~, 
}
(where $\gth_{\ga\,\text{r}}$ is the $u_\text{r}$ part of the matrix $\gTh_\ga$ as defined in 
eqn.~\eqref{eqn:Defintionupm}) that determine the Narain moduli uniquely already. Expanding out 
eqn.~\eqref{ReducedOrbiCompatible}, we obtain 
\begin{subequations}
\label{eqn:FixedModuliEasyWrittenOut}
\begin{eqnarray}
 -(\widehat \gr_\ga)_{21} + (G + C^T) (\widehat \gr_\ga)_{11} + A^T \ga_\text{g}\, (\widehat \gr_\ga)_{31} &=& \gr_{\ga\,\text{r}}^{-T} (G+C^T)~, 
\\[1ex] 
 -(\widehat \gr_\ga)_{22} + (G + C^T) (\widehat \gr_\ga)_{12} + A^T \ga_\text{g}\, (\widehat \gr_\ga)_{32} &=& - \gr_{\ga\,\text{r}}^{-T}~, \label{FixedModuliMiddle}
\\[1ex] 
-(\widehat \gr_\ga)_{23} + (G + C^T) (\widehat \gr_\ga)_{13} + A^T \ga_\text{g}\, (\widehat \gr_\ga)_{33} &=& \gr_{\ga\,\text{r}}^{-T} A^T \ga_\text{g}~,
\end{eqnarray}
\end{subequations} 
where $\rho_{\ga\,\text{r}} := e^{-1} \theta_{\ga\,\text{r}}\,e$. (Note that there is a redundancy 
between $e$ and $\gth_{\ga\,\text{r}}$,  which reflects the fact that the vielbein $e$ is not 
uniquely determined by the metric $G$.) 

It is sufficient to solve only these three matrix equations~\eqref{eqn:FixedModuliEasyWrittenOut} 
in order to find a solution of all nine equations~\eqref{eqn:OrbiCompatible} because of the coset 
decomposition~\eqref{CosetDecompEM}: Indeed, we can alternatively obtain the set of coupled 
equations~\eqref{eqn:FixedModuliEasyWrittenOut} by comparing eqn.~\eqref{eqn:OrbiCompatible} to 
eqn.~\eqref{CosetDecompEM}. They are identical if we determine each twist $\gTh_\ga$ from 
eqn.~\eqref{eqn:UBlocksFromM} using $U_{\widehat{M}} = \gTh_\ga$ (hence, in particular 
$u_\text{r} = \gth_{\ga\,\text{r}}$) and $\widehat{M} = \widehat\rho_\ga$. Furthermore, we have to 
set $G' = G$, $B' = B$ and $A' = A$, where the primed objects are determined by the transformation 
of the generalized metric  
\equ{
\cH(G',B',A') ~=~ \widehat{\gr}_\ga^T\, \cH(G,B,A)\, \widehat{\gr}_\ga ~\stackrel{!}{=}~ \cH(G,B,A)~,
}
using eqn.~\eqref{cHDefinition}. Therefore, using eqn.~\eqref{TransformationeBAEasy} the moduli of 
the Narain lattice are constrained according to
\equ{\label{eqn:FixedModuliEasy}
\widehat{M}_{\ga\,1}^T \rho_{\ga\, \text{r}} ~=~ G + C \quad,\quad \widehat{M}_{\ga\,2}^T \rho_{\ga\,\text{r}} ~=~ -\Id \quad\text{and}\quad \widehat{M}_{\ga\,3}^T \rho_{\ga\,\text{r}} ~=~ \ga_\text{g}^T A~,
}
for each generator of the point group $\widehat{\gr}_\ga$.  Inserting the moduli-dependent 
short-hands $\widehat{M}_{\ga\,i}$ from eqn.~\eqref{GammaDeltaBA} the resulting equations are again 
eqns.~\eqref{eqn:FixedModuliEasyWrittenOut}. 
In summary, for a given finite group $\widehat{\mathbf{P}} \subset \text{O}_{\widehat{\get}}(D,D+16;\Intr)$ 
there exists a Narain lattice such that $\widehat{\mathbf{P}}$ is a point group of this lattice if 
the Narain moduli can be chosen such that they are invariant under the orbifold action, i.e.~$G'=G$, 
$B'=B$ and $A' = A$, see Section~\ref{sec:ModuliStabilization}.

Eqn.~\eqref{FixedModuliMiddle} can be used to constrain $\gr_{\ga\,\text{r}}$. Inserting this in 
the other two equations of eqns.~\eqref{eqn:FixedModuliEasyWrittenOut} leads to two coupled 
quadratic matrix equations 
\begin{subequations}
\label{eqn:AlgbraicRiccati}
\equ{ 
(G+C^T) \widehat\gr_{12} (G+C^T) + A^T\ga_\text{g}(\widehat\gr_{32} (G+C^T) + \widehat\gr_{31})
- \widehat\gr_{22} (G+C^T) + (G+C^T) \widehat\gr_{11}  ~=~ \widehat\gr_{21}~, 
\\[1ex] 
A^T\ga_\text{g} \widehat\gr_{32} A^T \ga_\text{g} + (G+C^T) (\widehat\gr_{13} + \widehat\gr_{12} A^T\ga_\text{g} ) - \widehat\gr_{22} A^T \ga_\text{g} + A^T\ga_\text{g} \widehat\gr_{33} ~=~ \widehat\gr_{23}~. 
}
\end{subequations}
for each generator $\widehat{\gr} = \widehat{\gr}_\ga$ of the point group $\widehat{\mathbf{P}}$. 
These conditions can be thought of as algebraic Riccati equations (see e.g.~\cite{bittanti2012riccati})
which constrain some and 
sometimes even all the moduli $G$, $B$ and $A$. Hence we have reduced the existence question of 
Narain orbifolds to the question whether these Riccati equations admit real solutions.

\subsection[Mapping from the lattice basis to the coordinate basis]{Mapping from the lattice basis to the coordinate basis}
\label{SecLatticeBtoCoordB} 

Assume we are given a finite point group 
$\widehat{\mathbf{P}} \subset \text{O}_{\widehat{\get}}(D,D+16;\Intr)$ with generators 
$\widehat{\gr}_\alpha$ in the lattice basis and we want to know a compatible Narain lattice as well 
as the twists $\gTh_\ga$ in the coordinate basis. To obtain this data we can perform the following 
steps: First, we find a solution to eqns.~\eqref{eqn:AlgbraicRiccati}, i.e.\ find orbifold invariant 
moduli $G$, $B$ and $A$. After that we {make a choice for a geometrical vielbein $e$ such that 
$e^T\, e = G$. By doing so, we have obtained a generalized vielbein $E=R\,\widehat{E}(e,B,A)$, 
which is compatible with $\widehat{\mathbf{P}}$ in the sense of eqn.~\eqref{eqn:OrbiCompatible}. 
Finally, we compute the twists in the lattice basis: Using the geometrical vielbein $e$ we can 
determine the right-moving twists $\theta_{\ga\,\text{r}} = e\, \rho_{\ga\,\text{r}}\, e^{-1}$, 
where $\rho_{\ga\,\text{r}}$ is given by eqn.~\eqref{FixedModuliMiddle}. Consequently, we can 
compute the blocks of $\gTh_\ga$ from eqn.~\eqref{eqn:UBlocksFromM}, i.e.
\begin{subequations}\label{eqn:ThetaBlocksFromM}
\begin{eqnarray}
\gth_{\ga\,\text{l}}  & = & \left( \Id_D - 2\,e\,(\widehat\gr_\ga)_{12}\,\widehat{M}_{\ga\,2}^{-1}\,e^T\right) \gth_{\ga\,\text{r}}~,\label{eqn:ThetalBlocksFromM}
\\[1ex] 
\gth_{\ga\,\text{lL}} & = & \sqrt{2}\,e\,\left((\widehat\gr_\ga)_{13}-(\widehat\gr_\ga)_{12}\,\widehat{M}_{\ga\,2}^{-1}\,\widehat{M}_{\ga\,3}\right)\,\ga_\text{g}^{-1}~,
\\[1ex] 
\gth_{\ga\,\text{Ll}} & = & -\sqrt{2}\,\Big(\ga_\text{g}\,(\widehat\gr_\ga)_{32}+A\,(\widehat\gr_\ga)_{12}\Big)\widehat{M}_{\ga\,2}^{-1}\,e^T\,\gth_{\ga\,\text{r}}~,
\\[1ex] 
\gth_{\ga\,\text{L}}  & = & A\,(\widehat\gr_\ga)_{13}\,\ga_\text{g}^{-1}+\ga_\text{g}\,(\widehat\gr_\ga)_{33}\,\ga_\text{g}^{-1}-\Big(A\,(\widehat\gr_\ga)_{12}+\ga_\text{g}\,(\widehat\gr_\ga)_{32}\Big)\widehat{M}_{\ga\,2}^{-1}\,\widehat{M}_{\ga\,3}\,\ga_\text{g}^{-1}~,
\end{eqnarray}
\end{subequations}
where $\widehat{M}_{\ga\,i}$ for $i=1,2,3$ are defined in eqn.~\eqref{GammaDeltaBA} setting 
$\widehat{M}=\widehat{\gr}_\ga$. This method we will be exemplified in 
Section~\ref{sec:spacegroupexamples} where we discuss a number of two-dimensional Narain orbifolds.

An important characterization of heterotic Narain orbifolds is whether they are symmetric or 
asymmetric. In Section~\ref{SecHetNarainOrbifolds} we defined a Narain orbifold to be 
symmetric if there is a coordinate basis such that eqn.~\eqref{eqn:SymmetricOrbifoldTwist} holds. 
In the lattice basis, a sufficient but not necessary condition for a Narain orbifold to be 
symmetric is $(\widehat\gr_\ga)_{12} = 0$: First of all notice that  $(\widehat\gr_\ga)_{12} = 0$ 
implies $(\widehat\gr_\ga)_{13} = 0$ and $(\widehat\gr_\ga)_{32} = 0$ since 
$\widehat\gr_\ga^T\,\widehat\get\,\widehat\gr_\ga=\widehat\get$. Consequently, the 
conditions~\eqref{eqn:AlgbraicRiccati} become linear in the moduli and, hence, not all Narain 
moduli are frozen. Furthermore, using eqns.~\eqref{eqn:ThetaBlocksFromM} we obtain 
\equ{
\gth_{\ga\,\text{l}} = \gth_{\ga\,\text{r}}~,
\qquad  
\gth_{\ga\,\text{lL}} = \gth_{\ga\,\text{Ll}} = 0~, 
\qquad 
\gth_{\ga\,\text{L}} = \ga_\text{g}\,(\widehat\gr_\ga)_{33}\,\ga_\text{g}^{-1}~. 
}
Hence, any generator $\widehat{\gr}_\alpha \in \widehat{\mathbf{P}}$ with 
$(\widehat\gr_\ga)_{12} = 0$ and $(\widehat\gr_\ga)_{33} = \Id_{16}$ corresponds to a symmetric 
twist. However, the converse is in general not true. In Section~\ref{sec:spacegroupexamples} 
we provide examples for both cases: In Section~\ref{SecZN2DNarainOrbifolds} we list several 
Narain orbifolds that are necessarily symmetric because $(\widehat\gr_\ga)_{12} = 0$ and in 
Section~\ref{sec:O22Z3SymmetricOrbifolds} we give one Narain orbifold that is symmetric even though 
$(\widehat\gr_\ga)_{12} \neq 0$.

\subsection{Dimensionality of the Narain orbifold moduli space}
\label{sec:IdentifyModuli}

Assuming that a Narain orbifold exists, i.e.\ assuming that we have found a generalized vielbein 
$\widehat{E}_0$ that satisfies eqn.~\eqref{eqn:OrbiCompatible}, we want to determine the number of 
unconstrained Narain moduli. In other words, we want to count the number of moduli 
perturbations $\gd \cH$ that can deform the associated generalized metric $\cH_0$ such that 
$\cH_0 + \gd \cH$ remains invariant under the Narain orbifold action. 

To address this question, we make use of the results from Appendix~\ref{sec:ModuliDeformations} 
and set $\widehat{\mathbf{H}} = \widehat{\mathbf{P}}$. Then, the tangent space to the 
orbifold-invariant moduli space is given by
\equ{
\cM_{\widehat{\mathbf{P}}} ~=~ \Big\{ \gd\frak{m}_{\widehat{\mathbf{P}}} ~=~ \cP_{\widehat{\mathbf{P}}}\,  \gd\frak{m} \Big\}~,
}
where the projection operator $\cP_{\widehat{\mathbf{P}}}$ is defined in eqn.~\eqref{HinvProjection}. 
The moduli deformations $\gd \cH$, can be parametrized as follows  
\equ{ \label{DeltaFrakh} 
\gd \cH ~=~ E_0^T \gd \frak{h}\, E_0
~, 
\quad 
\gd \frak{h} ~=~  
\pmtrx{ 0 & \gd \frak{m} \\ 
\gd \frak{m}^T & 0 }~
\quad 
\gd \frak{m} ~=~ 
 \pmtrx{e_0^{-T}(\gd G - \gd B')\,e_0^{-1}~, & \sqrt 2\, e_0^{-T}\, \gd A^T}~, 
}
where $\gd B' = \gd B +\frac 12\, \gd A^T\, A_0 - \frac 12\, A_0^T\, \gd A$, 
$\gd G = \gd e^T e_0 + e_0^T \gd e$.

According to eqn.~\eqref{eqn:NumberOfHModuli} the dimension of the orbifold-invariant Narain moduli 
space, i.e.~the number of moduli, is determined by 
\equ{\label{eqn:orbifoldmoduli}
\dim(\cM_{\widehat{\mathbf{P}}}) ~=~ 
 \langle \chi_\text{r}, \chi_\text{L} \rangle
~=~ 
 \frac1{|\mathbf{P}|} \sum_{\gTh \in \mathbf{P}} \chi_\text{r}(\gTh)\, \chi_\text{L}(\gTh)^*~,  
}
where we have introduce the right- and left-characters 
\equa{ \label{rLcharacters} 
\chi_\text{r}(\gTh)  ~=~ \text{tr}[ \gth_\text{r}] ~=~ \text{tr}\Big[ \frac{\Id- \get}2\, \gTh\Big]~,  
\qquad 
\chi^{\;}_\text{L}(\gTh) ~=~  \text{tr}[ \gTh_\text{L}] ~=~ \text{tr}\Big[ \frac{\Id+ \get}2\, \gTh\Big]~,
}
respectively. Because of this character formula~\eqref{eqn:orbifoldmoduli}, the number of moduli 
$\dim(\cM_{\widehat{\mathbf{P}}})$ for Narain orbifolds only depends on the representations of 
$\gth_\text{r}$ and $\gTh_\text{L}$ of the point group $\mathbf{P}$, but not on conjugation of 
$\gTh$ with $U \in \text{O}(D;\Real) \times \text{O}(D+16;\Real)$.

The number of fixed moduli is given by $D(D+16) - \dim(\cM_{\widehat{\mathbf{P}}})$. In particular, 
all Narain moduli are frozen if $\dim(\cM_{\widehat{\mathbf{P}}}) = 0$. In this case, the 
Narain orbifold moduli space $\cM_{\widehat{\mathbf{P}}}$ is a point (or a set of disjoint points). 
This happens when the right- and left-characters~\eqref{rLcharacters} are orthogonal. In light of 
this, we can use the property that characters of irreducible representations form an orthonormal 
basis to analyze eqn.~\eqref{eqn:orbifoldmoduli}. In detail, for two (complex) irreducible 
representations $\boldsymbol{\mu}$ and $\boldsymbol{\nu}$ of the finite point group $\mathbf{P}$ we have
\equ{ 
\langle \chi_{\boldsymbol{\gm}}, \chi_{\boldsymbol{\gn}} \rangle 
~=~ 
\bigg\{ \begin{array}{ll}1 & \text{ if } \boldsymbol{\gm} = \boldsymbol{\gn}\\[1ex] 0 & \text{ else} \end{array}~.
}
This can be used to construct some situations with all moduli fixed, i.e.\ $\dim(\cM_{\widehat{\mathbf{H}}}) = 0$:  
\items{ 
\item 
If the matrix representations of $\gth_\text{r}$ and $\gTh_\text{L}$ are 
both irreducible, they have to be different, since the former is 
$D$-dimensional while the latter is $(D+16)$-dimensional, and hence, their characters are orthogonal. 
\item 
If the representations of $\gth_\text{r}$ and $\gTh_\text{L}$ are reducible, one can decompose them 
into irreducible ones as 
\equ{
\gth_\text{r} ~=~ \bigoplus_{\boldsymbol{\gm}} \gth_{\text{r}\, \boldsymbol{\gm}}~, 
\quad 
 \gTh_\text{L} ~=~ \bigoplus_{\boldsymbol{\gn}} \gTh_{\text{L}\, \boldsymbol{\gn}}~, 
\quad\Rightarrow\quad 
\chi_\text{r} ~=~ \sum_{\boldsymbol{\gm}} \chi_{\text{r}\, \boldsymbol{\gm}}~,
\quad 
\chi_\text{L} ~=~ \sum_{\boldsymbol{\gn}} \chi_{\text{r}\, \boldsymbol{\gn}}~,
}
where the irreducible representations $\gth_{\text{r}\, \boldsymbol{\gm}}$ and 
$\gTh_{\text{L}\, \boldsymbol{\gn}}$ are in 
general complex. Hence, if and only if $\gth_\text{r}$ and $\gTh_\text{L}$ do not contain any 
irreducible representation in common, again the characters $\gch_\text{r}$ and $\gch_\text{L}$ are 
orthogonal. An particular example of this is obtained, when $\gTh_\text{L}=\Id$ and $\gth_\text{r}$ 
does not contain any trivial one-dimensional representations of $\mathbf{P}$.
}

\subsection[A $T$-fold constructed as an asymmetric $\Z{2}$ Narain orbifold]{A $\boldsymbol{T}$-fold constructed as an asymmetric $\boldsymbol{\Z{2}}$ Narain orbifold}
\label{sec:AsymmetricZ2Example}

To illustrate the various results, we conclude this section by considering a simple but instructive 
construction of a $T$-fold: We define an asymmetric $\Z{2}$ Narain orbifold by choosing
\equ{
\widehat{\rho} ~=~ \widehat{I} ~=~ \pmtrx{ 0 & \Id_D & 0 \\[1ex] \Id_D & 0 & 0 \\[1ex] 0 & 0 & \Id_{16}}~,
}
see Table~\ref{tb:DualitySubgroups}. First, we identify a specific example of a compatible Narain 
lattice using the $\Intr_2$ grading $\cZ$. Then, we will use the discussion from 
Section~\ref{sec:ExistenceOfNarainOrbifolds} to see that this is actually the most general solution. 
Finally, we confirm this by counting the number of unstabilized Narain moduli using 
Section~\ref{sec:IdentifyModuli}.

To find a compatible Narain lattice, we notice that $\cZ=\widehat{I}$ is a valid $\Z{2}$ grading 
satisfying eqn.~\eqref{eqn:ZCommutes}. Hence, we can easily read off 
\equ{ \label{Z2FixedModuli}
e ~=~ G ~=~ \Id_D \quad,\quad B ~=~ 0 \quad\text{and}\quad A ~=~ 0~.
}
from eqn.~\eqref{GradingZ} as a possible choice for the Narain moduli. Alternatively, we can study 
the solutions of eqns.~\eqref{eqn:AlgbraicRiccati}. In this case these equations read: 
\equ{
(G+C^T) (G+C) = \Id_D~, 
\qquad 
A^T \ga_\text{g} = 0~. 
}
Again, it is not difficult to confirm that eqns.~\eqref{Z2FixedModuli} constitute a solution. 

Consequently, we find $\widehat{E}(e,B,A)=\Id$ and we obtain the twist $\gTh$ in the coordinate 
basis from eqn.~\eqref{eqn:OrbiCompatible} as
\equ{
\gTh ~=~ R\,\widehat{E}(e,B,A)\, \widehat{\rho}\, \widehat{E}(e,B,A)^{-1}\,R^{-1} ~=~ R\, \widehat{I}\, R^{-1} ~=~ \get ~=~ \left( \begin{array}{ccc} -\Id_D & 0 & 0\\ 0 & \Id_D & 0 \\ 0 & 0 & \Id_{16}\end{array}\right)~,
}
i.e.~$\theta_\text{r} = -\Id_D$, $\theta_\text{l} = \Id_D$ and $\gth_\text{L}=\Id_{16}$.

In fact, all Narain moduli are stabilized in this case as we are going to show next. We use 
eqn.~\eqref{GammaDeltaBA} with $\widehat{M} = \widehat{I}$, which yields
\equ{
\widehat M_1 ~=~ -\Id_D \quad, \quad \widehat M_2 ~=~ G + C^T \quad\text{and}\quad \widehat M_3 ~=~ A^T \ga_\text{g}~.
}
Then, the Narain moduli are subject to the constraints~\eqref{eqn:FixedModuliEasy}. In this example, 
they read 
\equ{ 
\Id_D ~=~ G + C \quad\text{and}\quad A ~=~ -A~.
}
using $\rho_\text{r} = e^{-1} \theta_\text{r}\,e = -\Id_D$. Consequently, all Narain moduli are 
stabilized and their values are given by eqns.~\eqref{Z2FixedModuli}.

The fact that all Narain moduli are stabilized in this example is also easy to understand 
using the number of unstabilized Narain moduli $\dim(\cM_{\widehat{\mathbf{P}}})$, see 
eqn.~\eqref{eqn:orbifoldmoduli}: $\gth_\text{r}$ consists of $D$ non-trivial irreducible 
representations of the $\Z{2}$ point group, while $\gTh_\text{L}$ consists of $D+16$ trivial 
irreducible representations. As the characters of different irreducible representations are 
orthogonal, we easily find $\dim(\cM_{\widehat{\mathbf{P}}}) = 0$.

\section{Towards a classification of Narain orbifolds}
\label{SecClassification}

In this section we would like to lay the foundations for a classification of inequivalent Narain 
orbifolds. In general, the key to a classification of any structure is to identify those 
transformations that relate (or even define) equivalent structures. These transformations can be 
used to define equivalence relations {that consequently give rise to equivalence classes. For the 
classification of $D$-dimensional -- geometrical -- orbifolds the structure turns out to be the 
space group and the equivalence relations are based on the notions of $\Ratl$-, $\Intr$- and 
affine-classes~\cite{Fischer:2012qj}. In this section we show that extending these notions to 
generalized space groups is the key for a classification of Narain orbifolds.

In more detail, for the classification of Narain orbifolds we identify three main structures: 
(i) the integral Narain point group $\widehat{\mathbf{P}}$ of finite lattice automorphisms, (ii) an 
associated Narain lattice $\Narain$ (given by a geometrical torus with metric $G$, a $B$-field and 
Wilson lines $A$) that is compatible with the point group and, finally, (iii) the full generalized 
space group $\mathbf{S}$, which fully specifies a Narain orbifold as we have seen in 
Section~\ref{SecOrbifolds}. The main purposes of this section are to define equivalences for these 
three structures, namely Narain $\Ratl$-, $\Intr$- and Poincar\'e-equivalences, together with 
their associated equivalence-classes and to analyze their interpretations.

\subsection[Narain $\Ratl$- and $\Intr$-classes]{Narain $\boldsymbol{\Ratl}$- and $\boldsymbol{\Intr}$-classes}
\label{sec:NarainQandZClass}

For the definition of Narain $\Ratl$- and $\Intr$-classes we need to describe the Narain point group 
in the lattice basis, where $\widehat{\mathbf{P}} \subset \text{O}_{\widehat{\get}}(D,D+16;\Intr)$, 
see Section~\ref{sec:TwistInLatticeBasis}. Then, one only has to consider integral finite order 
elements $\widehat{\gr}_\alpha \in \widehat{\mathbf{P}}$. Since Narain $\Ratl$- and $\Intr$-classes 
are analogously defined, we take the field $\mathbbm{F}$ to be either $\Ratl$- and $\Intr$ and 
begin with the definition of $\mathbbm{F}$-equivalence: Two matrices 
$\widehat{\gr} \in \text{O}_{\widehat{\get}}(D,D+16;\Intr)$ and 
$\widehat{\gr}' \in \text{O}_{\widehat{\get}'}(D,D+16;\Intr)$ are defined to be 
$\mathbbm{F}$-equivalent if there exists a matrix $\widehat{M} \in \text{GL}(2D+16;\mathbbm{F})$ 
such that
\equ{\label{eqn:QZequiv}
\widehat{\gr}' ~=~ \widehat{M}^{-1}\, \widehat{\gr}\, \widehat{M}
\quad\text{and}\quad
\widehat{\get}' ~=~ \widehat{M}^T\, \widehat{\get}\, \widehat{M}~.
}

Two Narain points groups $\widehat{\mathbf{P}} \subset \text{O}_{\widehat{\get}}(D,D+16;\Intr)$ 
and $\widehat{\mathbf{P}}' \subset \text{O}_{\widehat{\get}'}(D,D+16;\Intr)$ are said to be 
$\mathbbm{F}$-equivalent if there exists a single matrix $\widehat{M} \in \text{GL}(2D+16;\mathbbm{F})$ such 
that 
\equ{
\widehat{\mathbf{P}}' ~=~  \widehat{M}^{-1}\, \widehat{\mathbf{P}}\, \widehat{M}
\quad\text{and}\quad
\widehat{\get}' ~=~ \widehat{M}^T\, \widehat{\get}\, \widehat{M}~.
}
Note that if two point groups are from the same $\Intr$-class they are also from the same 
$\Ratl$-class, because if $\widehat{M} \in \text{GL}(2D+16;\Intr)$ then $ \widehat{M} \in \text{GL}(2D+16;\Ratl)$. 
But the converse is not true, i.e.~two point groups from the same $\Ratl$-class can 
be in inequivalent $\Intr$-classes.

\subsection[Interpretation of Narain $\Ratl$- and $\Intr$-classes]{Interpretation of Narain $\boldsymbol{\Ratl}$- and $\boldsymbol{\Intr}$-classes}  

To prepare the interpretation of the Narain $\Ratl$- and $\Intr$-classes, let us assume that two 
Narain point groups $\widehat{\mathbf{P}}$ and $\widehat{\mathbf{P}}'$ are from the same 
$\mathbbm{F}$-class, where the field $\mathbbm{F}$ is either $\Ratl$ or $\Intr$. Then, 
there exists a matrix $\widehat{M} \in \text{GL}(2D+16;\mathbbm{F})$ such that for each generator 
$\widehat{\gr}_\ga \in \widehat{\mathbf{P}}$ there is a generator 
$\widehat{\gr}_\ga^{\,\prime} \in \widehat{\mathbf{P}}'$ with
\equ{ \label{eqn:rhoTrafoFclass}
\widehat{\gr}_\ga^{\,\prime} ~=~ \widehat{M}^{-1}\, \widehat{\gr}_\ga\, \widehat{M}~.
}
Now, consider a Narain lattice spanned by a generalized vielbein $E$, such that $E$ is compatible 
with all generators $\widehat{\gr}_\ga$ and insert 
eqn.~\eqref{eqn:rhoTrafoFclass}, i.e.
\equ{\label{eqn:QZequivEVielbein}
\gTh_\ga\, E ~=~ E\, \widehat{\gr}_\ga ~=~ E\, \widehat{M}\,\widehat{\gr}_\ga^{\,\prime}\,\widehat{M}^{-1}~.
}
Consequently, we find
\equ{\label{eqn:QZequivVielbeins}
\gTh_\ga\, E' ~=~ E'\, \widehat{\gr}_\ga^{\,\prime} \quad\text{where}\quad E' ~=~ E\, \widehat{M}~.
}
Hence, we can interpret eqn.~\eqref{eqn:QZequivVielbeins} as follows: 

If $\widehat{\mathbf{P}}$ is a symmetry of a Narain lattice with generalized vielbein $E$ and 
Narain metric $\widehat{\get}$ then $\widehat{\mathbf{P}}'$ is a symmetry of a Narain lattice with 
generalized vielbein $E'=E\widehat{M}$ and Narain metric 
$\widehat{\get}' = \widehat{M}^T\, \widehat{\get}\, \widehat{M}$. Furthermore, we notice that both 
point groups have the same geometrical action $\gTh_\ga$ which corresponds to both $\widehat{\gr}_\ga$ 
and $\widehat{\gr}_\ga^{\,\prime}$. In other words, the corresponding point groups $\mathbf{P}$ and 
$\mathbf{P}'$ in the coordinate basis are identical (up to a trivial basis change) for point groups 
from the same $\mathbbm{F}$-class. Consequently, the question of symmetric or asymmetric 
orbifolds, the number of unbroken supersymmetries in $d$ 
uncompactified dimensions and the number of invariant Narain moduli 
eqn.~\eqref{eqn:orbifoldmoduli} are also equal. This is independent of the choice for the field 
$\mathbbm{F}$ to be $\Ratl$ or $\Intr$.

Next, we have to distinguish between these two Narain classes: Let us first consider the case 
$\mathbbm{F} = \Ratl$. The Narain lattices spanned by $E$ and $E'=E\widehat{M}$, are in general 
physically inequivalent, because if $\widehat{M} \in \text{GL}(2D+16;\Ratl)$ then in general 
$\widehat{M} \not\in \text{GL}(2D+16;\Intr)$. A representation of a $\Ratl$-class only gives one 
example of a compatible Narain lattice. To characterize all inequivalent lattices for a given 
$\Ratl$-class one needs to consider $\Intr$-classes. That is, if $\mathbbm{F} = \Intr$ the 
generalized vielbeins $E$ and $E'=E\widehat{M}$ span identical Narain lattices.

Finally, if $\widehat{M}$ additionally preserves the Narain metric $\widehat{\get}$, 
i.e.~if
\equ{
\widehat{M} ~\in~ \text{O}_{\widehat{\get}}(D,D+16; \mathbbm{F}) ~\subset~ \text{GL}(2D+16;\mathbbm{F})~, 
}
which means that $\widehat{M}$ is a $T$-duality transformation, we can analyze the consequences of 
eqn.~\eqref{eqn:QZequivVielbeins} for the Narain moduli $G$, $B$ and $A$. In this case, we take the 
most general vielbein $E=U\,R\,\widehat{E}(e,B,A)$ from eqn.~\eqref{eqn:verymostgeneralvielbein} 
and use eqn.~\eqref{CosetDecompEM} in order to transfer $\widehat{M}$ into 
$U_{\widehat{M}}$ for the generalized vielbein $E' = E\, \widehat{M}$. Consequently, one can show 
that $E'$ is given by
\equ{\label{eqn:latticeEprime}
E' ~=~ \left(U\,R\,\widehat{E}(e,B,A)\right)\, \widehat{M} ~=~ U_\text{B}\,U\,R\,\widehat{E}(e',B',A') \quad\text{where}\quad U_\text{B} ~=~ U\,U_{\widehat{M}}\,U^{-1}~,
}
and the $\widehat{M}$-transformed Narain moduli are given in eqn.~\eqref{TransformationGBA}. Hence, 
if two Narain point groups $\widehat{\mathbf{P}}$ and $\widehat{\mathbf{P}}'$ are 
$\mathbbm{F}$-equivalent and defined with respect to the same Narain metric $\widehat{\get}$ then the 
lattice $E=U\,R\,\widehat{E}(e,B,A)$ of $\widehat{\mathbf{P}}$ corresponds to the lattice $E'$ of 
$\widehat{\mathbf{P}}'$ as given in eqn.~\eqref{eqn:latticeEprime}. This change of lattices from 
$E$ to $E'$ involves a transformation of moduli from $G$, $B$ and $A$ to $G'$, $B'$ and $A'$ using 
the $T$-duality transformation $\widehat{M}$ and, in addition, a rotation in the coordinate basis with 
$U_\text{B} \in \text{O}(D;\Real) \times \text{O}(D+16;\Real)$. Moreover, from 
eqns.~\eqref{eqn:QZequivEVielbein} and~\eqref{eqn:QZequivVielbeins} we obtain
\begin{subequations}\label{eqn:QZequivChangeOfTheta}
\begin{eqnarray}
\gTh_\ga\, \left(U\,R\,\widehat{E}(e,B,A)\right) & = & \left(U\,R\,\widehat{E}(e,B,A)\right)\,\widehat{\gr}_\ga~, \\[1ex] 
\left(U_\text{B}^{-1}\,\gTh_\ga\,U_\text{B}\right)\, \left(U\,R\,\widehat{E}(e',B',A')\right) & = & \left(U\,R\,\widehat{E}(e',B',A')\right)\,\widehat{\gr}_\ga'~.
\end{eqnarray}
\end{subequations}
That is, even though we have seen in eqns.~\eqref{eqn:QZequivEVielbein} 
and~\eqref{eqn:QZequivVielbeins} that the Narain point groups $\mathbf{P}$ and 
$\mathbf{P}'$ are identical in the same $\mathbbm{F}$-class, their generators $\gTh_\ga$ and 
$\gTh_\ga' = U_\text{B}^{-1}\,\gTh_\ga\,U_\text{B}$ can look different, for example, one is 
symmetric and the other looks asymmetric. This is the case if one 
chooses the corresponding Narain lattices as different points, specified by $(e,B,A)$ and 
$(e',B',A')$, in the same representation of the Narain moduli space, i.e.~with the same $U$ in 
eqn.~\eqref{eqn:QZequivChangeOfTheta}. As an example for eqn.~\eqref{eqn:QZequivChangeOfTheta}, 
we will discus two $\mathbbm{F}$-equivalent $\Z{3}$ point groups $\widehat{\mathbf{P}}_{(1)}$ and 
$\widehat{\mathbf{P}}_{(2)}$ in Section~\ref{sec:O22Z3SymmetricOrbifolds}, where the point group 
$\widehat{\mathbf{P}}_{(1)}$ is symmetric while $\widehat{\mathbf{P}}_{(2)}$ looks asymmetric due to a 
non-trivial transformation $U_\text{B}$.

\subsection[Narain Poincar\'e-classes]{Narain Poincar\'e-classes} 

As final type of equivalence transformations, we want to generalize affine transformations 
$(F,\lambda)$ of Euclidean $D$-dimensional orbifolds (with linear mapping $F\in\text{GL}(D; \Real)$ 
and translation $\lambda \in \Real^{D}$) to the Narain case. Importantly, the $(2D+16)$-dimensional 
Narain lattice is equipped with a metric $\get$ with signature $(D,D+16)$, which has to be 
preserved by any transformation. Hence, it is essential for the Narain case to restrict affine 
transformations in $2D+16$ dimensions to Poincar\'e transformations $(F,\lambda)$ of the Narain 
lattice, where $F \in \text{O}_\get(D,D+16;\Real)$ and $\lambda \in \Real^{2D+16}$. Therefore, we 
need to introduce Poincar\'e-classes instead of affine classes in order to describe Narain orbifolds.

This might give the impression that Poincar\'e transformations of Narain orbifolds are more 
restrictive than affine transformations of ordinary Euclidean orbifolds. This is not the case 
since $\text{O}_\get(D,D+16;\Real)$ transformations contain $\text{GL}(D;\Real)$ transformations. 
This can be made explicit by the parametrization 
$E \widehat{M}_e(\gD K) E^{-1} \in \text{O}_\get(D,D+16;\Real)$, where $\widehat{M}_e(\gD K)$ is is 
given in Table~\ref{tb:DualitySubgroups} with $\gD K \in \text{GL}(D;\Real)$. Consequently, 
Poincar\'e-classes generalize the notion of affine classes to Narain orbifolds.

In light of this, we define the following equivalence relation: Consider two Narain orbifolds, 
i.e.\ two space groups $\mathbf{S}_{(1)}$ and $\mathbf{S}_{(2)}$ with point groups in the same 
$\Intr$-class. Two such Narain space groups are defined to be equivalent if there exists a Poincar\'e 
transformation $(F,\lambda)$ with $F \in \text{O}_\get(D,D+16;\Real)$ and $\lambda \in \Real^{2D+16}$ 
such that 
\equ{
\label{eqn:AffineTrafo}
\mathbf{S}_{(2)} =  (F,\lambda)^{-1}\, \mathbf{S}_{(1)}\, (F,\lambda)~. 
}
More explicitly, in terms of the generators $(\gTh_{(\kappa)\ga}, V_{(\kappa)\ga})$ and $(\Id, L_{(\kappa)})$ of the space groups $\mathbf{S}_{(\kappa)}$ for $\kappa=1,2$ this reads 
\equ{ \label{eqn:AffineTrafoExplicit}
L_{(2)} = F^{-1}\, L_{(1)}~, 
\qquad 
\gTh_{(2)\ga} = F^{-1} \,\gTh_{(1)\ga}\, F~, 
\qquad 
V_{(2)\ga} = F^{-1} \big(V_{(1)\ga} - (\Id - \gTh_{(1)\ga}) \gl \big)~,
}
see eqn.~\eqref{eqn:NarainSpaceGroupElement}.
Notice that Narain $\Ratl$- and $\Intr$-classes involve transformations in the lattice basis, while 
Narain Poincar\'e classes involve transformations in the coordinate basis. 
Since Narain Poincar\'e transformations act on all defining quantities of the space group, see 
eqn.~\eqref{eqn:AffineTrafoExplicit}, their interpretation is more involved.

\subsection[Interpretation of Narain Poincar\'e-classes]{Interpretation of Narain Poincar\'e-classes} 
\label{SecInterpretationPoincareClasses} 

First of all, we show that two generalized space groups from the same affine class correspond 
to the same Narain orbifold but possibly at different points in the moduli space.
To see this, let us denote the generalized vielbeins that specify the Narain lattices from the respective 
generalized space groups $\mathbf{S}_{(\kappa)}$ by $E_{(\kappa)}= U_{(\kappa)}\,R\,\widehat{E}(e_{(\kappa)},B_{(\kappa)},A_{(\kappa)})$ for ${\kappa}=1,2$, 
where $U_{(\kappa)} \in \text{O}(D;\Real) \times \text{O}(D+16;\Real)$ 
and $\widehat{E}(e_{(\kappa)},B_{(\kappa)},A_{(\kappa)})$ is given in eqn.~\eqref{GenVielbeinDecomposition}. 
Since $L_{(\kappa)} = E_{(\kappa)} N_{(\kappa)}$ are related by the transformation~\eqref{eqn:AffineTrafoExplicit}, 
a Poincar\'e transformation $(F,\lambda)$ of the corresponding generalized vielbeins $E_{(1)}$ and 
$E_{(2)}$ is given by 
\equ{\label{eqn:AffineTrafoVielbein}
U_{(2)}\,R\,\widehat{E}(e_{(2)},B_{(2)},A_{(2)}) ~=~ E_{(2)} ~=~ F^{-1}\, E_{(1)} ~=~  F^{-1}\,U_{(1)}\,R\,\widehat{E}(e_{(1)},B_{(1)},A_{(1)})~,
}
where we assume without loss of generality that we do not perform a discrete $T$-duality 
transformation (i.e.\ $N_{(2)} = N_{(1)}$). This can be rewritten as
\equ{
\widehat{U}_{(2)}\, \widehat{E}(e_{(2)},B_{(2)},A_{(2)}) ~=~ \widehat{E}(e_{(1)},B_{(1)},A_{(1)})\, \widehat{M}_F~,
}
where  
\equ{
\widehat{U}_{(2)} ~=~ R^{-1}\,U_{(2)}\,R
\quad\text{and}\quad 
\widehat{M}_F ~=~ 
E_{(1)}^{-1} \, U_{(1)}\,  F^{-1}\, E_{(1)} 
~\in~ \text{O}_{\widehat{\get}}(D,D+16,\Real)~.
}
Since $\widehat{M}_F$ parametrizes a general $T$-duality transformation, we can make use of 
eqn.~\eqref{CosetDecompEM} to determine the transformation of the moduli by setting 
$\widehat{M} = \widehat{M}_F$, i.e. 
\equ{
\widehat{U}_{\widehat{M}_F}\, \widehat{E}(e'_{(1)},B'_{(1)},A'_{(1)}) ~=~\widehat{E}(e_{(1)},B_{(1)},A_{(1)})\, \widehat{M}_F~. 
}
Since the generalized vielbein is uniquely defined up to $\text{O}(D;\Real)\times\text{O}(D+16;\Real)$ 
transformations, we conclude that 
\equ{ 
e_{(2)} ~=~ e'_{(1)}~,
\qquad 
B_{(2)} ~=~ B'_{(1)}~, 
\qquad 
A_{(2)} ~=~ A'_{(1)}~, 
}
where the prime denotes the resulting moduli under the $T$-duality transformation $\widehat{M}_F$. 
This tells us that two generalized space groups from the same Poincar\'e-class can correspond to the same 
Narain orbifold but at different points in the moduli space.
In fact only if $F \in \text{O}(D;\Real) \times \text{O}(D+16;\Real) \backslash \text{O}_\get(D,D+16;\Real)$ 
we get a proper moduli transformation. Indeed, if $F \in \text{O}(D;\Real) \times \text{O}(D+16;\Real)$ 
we find that $U_{(2)} = F^{-1}\,U_{(1)}$ as well as $e_{(2)}=e_{(1)}$, $B_{(2)}=B_{(1)}$ and 
$A_{(2)}=A_{(1)}$.
In this case, also the left- and right-moving mass formulae of the heterotic string stay the same.

So far we only gave an interpretation of the first equivalence relation in 
eqns.~\eqref{eqn:AffineTrafoExplicit}. The second relation tells us that the orbifold twists can 
take various guises by conjugation with $F \in \text{O}_\get(D,D+16;\Real)$. 
The third equivalence relation in 
eqns.~\eqref{eqn:AffineTrafoExplicit} can be interpreted by resorting to the decomposition 
mentioned in Section~\ref{sec:QuantizationOfShift}.

\section[Symmetric orbifolds as Narain orbifolds]{Symmetric orbifolds as Narain orbifolds}
\label{sec:SymmetricZKExample}

The main objective of our study in this paper is to set up a framework to investigate asymmetric orbifolds. 
Nevertheless, it is very instructive to apply the Narain formalism also to symmetric 
orbifolds~\cite{Dixon:1985jw, Dixon:1986jc}: 
It provides us with a unified view on both, geometric moduli and Wilson lines~\cite{Ibanez:1986tp}. Moreover, this case 
can be used to illustrate the power of the $T$-duality group approach in the investigation of 
moduli stabilization. For concreteness and simplicity, we only consider symmetric $\Z{K}$ orbifolds 
in this section. Extending the discussion is straightforward, yet beyond the scope of the present 
paper.

\subsection[Symmetric $\Z{K}$ orbifolds]{Symmetric $\boldsymbol{\Z{K}}$ orbifolds}

The Narain point group of a symmetric $\Intr_K$ orbifold is generated by a single twist $\gTh$ of 
order $K$ and the associated generator of the generalized space group is given by $(\gTh, V)$. For 
the orbifold to be symmetric, we choose the twist $\gTh$ to be of the form given in 
eqn.~\eqref{eqn:SymmetricOrbifoldTwist}. Thus, we obtain for $\gTh^k$, $k=1,\ldots,K$,
\equ{
\widehat\gTh^k ~=~ R^{-1} \gTh^k\, R ~=~ \left( \begin{array}{ccc} \theta^k & 0 & 0\\ 0 & \theta^k & 0 \\ 0 & 0 & \Id_{16}\end{array}\right) ~=~ \widehat M_{e}(\theta^k)~,
}
see Table~\ref{tb:DualitySubgroups} and using $\theta^T\, \theta = \Id_D$. Using the 
definition~\eqref{IntroRho} of the integral matrix $\widehat{\gr}$ we can subsequently obtain an 
expression for $\widehat\rho^{\,k}$, which can be further evaluated with the help of the 
multiplication Table~\ref{tb:DualityMultiplications} for $T$-duality group elements. This yields
\equ{ \label{eqn:SymmetricRho}
\widehat\rho^k  ~=~ \widehat{E}(e,B,A)^{-1}\,\widehat{\gTh}^k\,\widehat{E}(e,B,A) ~=~ \widehat M_{e}(\hat\theta^k)\, \widehat M_B(\Delta B_k)\, \widehat M_A(\Delta A_k)~, 
}
where we defined
\begin{subequations}\label{eqn:SymmetricRhoModuli}
\begin{eqnarray}
\hat\theta & = & e^{-1}\theta\, e\, \label{eqn:SymmetricRhoModuliG} \\[1ex]
\Delta B_k & = & B - \hat\theta^{kT} B\, \hat\theta^k + \frac{1}{2}\left(\hat\theta^{kT} A^TA - A^TA\,\hat\theta^k\right) \quad\text{with}\quad \Delta B_k^T = -\Delta B_k \label{eqn:SymmetricRhoModuliB}\\[1ex]
\Delta A_k & = & A\, \left(\Id_D - \hat\theta^k\right)~. \label{eqn:SymmetricRhoModuliA}
\end{eqnarray}
\end{subequations}
Since $\widehat\gr$ is an integral matrix, $\hat\theta$, $\Delta B_k$ and $\Delta A_k$ all have to 
be constant, i.e.~moduli-independent, matrices. As a cross-check, let us confirm that for $k=K$ 
we obtain $\widehat{\gr}^{\,K}=\Id$: Indeed, in this case we get $\hat\gth^K = \Id_D$, 
$\Delta A_K = 0$ and $\Delta B_K = 0$ and consequently, $\widehat\rho^K = \widehat M_{e}(\Id_D) = \Id$, as required. 
Furthermore, we find from eqn.~\eqref{eqn:SymmetricRho} that $\widehat\rho$ is an element of the 
discrete geometric subgroup $G_\text{geom}(\Intr) \subset \text{O}_{\hat\get}(D,D+16; \Intr)$, see 
eqn.~\eqref{eqn:ElementOfGGeom} with $\gD W = \Id_{16}$.

The twist $\gTh$ is in general accompanied by a shift 
$V^T = (V_\text{r}^T, V_\text{l}^T, V_\text{L}^T)$, see eqn.~\eqref{eqn:NarainSpaceGroupElement}. 
As we have seen in Section~\ref{sec:QuantizationOfShift}, the shift is quantized, i.e.\ 
$K \ProjP{V}{\gTh} = E\, N_V \in \Narain$. It is instructive to analyze this in more 
detail for the case that $\gth$ rotates in all $D$ compact dimensions. Then, the projection 
operator eqn.~\eqref{eqn:Projector} reads
\equ{
\ProjP{\mathcal{P}}{\gTh} ~=~ \left( \begin{array}{ccc} 0 & 0 & 0\\ 0 & 0 & 0 \\ 0 & 0 & \Id_{16}\end{array}\right)~,
}
and we obtain the condition
\equ{
K \ProjP{V}{\gTh} ~=~ \left(\begin{array}{c} 0 \\ 0\\ K\, V_\text{L}\end{array}\right) ~\stackrel{!}{=}~ E\, N_V ~\in~ \Narain \qquad\text{with}\qquad N_V ~=~ \left(\begin{array}{c} m_V \\ n_V\\ q_V\end{array}\right) ~\in~ \Z{}^{2D+16}~.
}
This is solved by
\equ{
K\, V_\text{L} ~=~ \ga_\text{g}\, q_V ~\in~\Lambda_{\E{8}\times\E{8}}~, \quad K A^T V_\text{L} ~=~ n_V \in \Z{}^6 \quad\text{and}\quad m_V = 0~,
}
where $\Lambda_{\E{8}\times\E{8}}$ denotes the root lattice of $\E{8}\times\E{8}$ and we used 
eqn.~\eqref{eqn:NarainEN}. Hence, $V_\text{L}$ is the gauge shift vector of order $K$ known to the 
symmetric orbifold literature, e.g.~\cite{Ibanez:1987pj,Bailin:1999nk}. Furthermore, we can set 
$V_\text{r} = V_\text{l} = 0$ by shifting the origin using the transformation~\eqref{eqn:OriginShift}.

\subsection[Moduli stabilization in symmetric $\Z{K}$ orbifolds]{Moduli stabilization in symmetric $\boldsymbol{\Z{K}}$ orbifolds}

The fact that even for symmetric $\Intr_K$ orbifolds a certain number of moduli, $G$, $B$ and $A$, 
become constrained, can be inferred in two ways: First of all, the 
conditions~\eqref{eqn:SymmetricRhoModuli} can be obtained from eqns.~\eqref{eqn:SymmetricRho}, as 
shown above by using the fact that for symmetric orbifolds the twist $\widehat\rho$ is an element of 
the geometric subgroup $G_\text{geom}(\Intr) \subset \text{O}_{\hat\get}(D,D+16; \Intr)$. A second 
derivation of eqn.~\eqref{eqn:SymmetricRhoModuli} follows from the general discussion in 
Section~\ref{sec:ExistenceOfNarainOrbifolds}, which is valid for both, symmetric and asymmetric 
orbifolds: To see this, we use 
\equ{\label{eqn:SymmtericOrbifoldGeneralRho}
\widehat\rho ~=~ \widehat M_{e}(\hat\theta)\, \widehat M_B(\Delta B_1)\, \widehat M_A(\Delta A_1) ~=~ \widehat{E}(\hat\theta, \Delta B_1, \Delta A_1)~,
}
see eqn.~\eqref{eqn:SymmetricRho} and eqn.~\eqref{GenVielbeinDecomposition}. Then, we set 
$\widehat{M} = \widehat\rho$ in eqn.~\eqref{GammaDeltaBA} and obtain
\begin{subequations}
\begin{eqnarray}
\widehat M_1 & = & \hat\theta^{-T} \left(-\Delta B_1 + \frac{1}{2} \Delta A_1^T \Delta A_1\right)  + (G + C^T) \hat\theta + A^T \Delta A_1~, 
\\[1ex] 
\widehat M_2 & = & -\hat\theta^{-T}~, \qquad \widehat M_3 ~=~ \left(\Delta A_1\, \hat\theta^{-1} + A\right)^T\ga_\text{g}~, 
\end{eqnarray}
\end{subequations}
and in addition we have $\rho_\text{r} = e^{-1} \theta_\text{r}\,e = \hat\theta$. Consequently, 
the Narain moduli are constrained by eqns.~\eqref{eqn:FixedModuliEasy}, which are equivalent to 
eqns.~\eqref{eqn:SymmetricRhoModuli}. Thus, we found two equivalent ways to derive the 
conditions~\eqref{eqn:SymmetricRhoModuli} for Narain moduli stabilization in the case of symmetric 
orbifolds.

Let us now discuss the consequences of eqns.~\eqref{eqn:SymmetricRhoModuli} for Narain moduli 
stabilization. Since $\widehat\rho$ in eqn.~\eqref{eqn:SymmtericOrbifoldGeneralRho} has to be an 
integer matrix, i.e.~$\widehat\rho \in \text{O}_{\hat\get}(D,D+16; \Intr)$, we have to demand that
\equ{\label{eqn:SymmtricOrbifoldConditionOnAB}
 \hat\theta ~\in~ \text{GL}(D;\Intr)~,\qquad 
 \alpha_\text{g}^{-1}\Delta A_k ~\in~ M_{16\times D}(\Intr)~, \qquad 
 -\frac{1}{2}\Delta A_k^T \Delta A_k + \gD B_k ~\in~ M_{D\times D}(\Intr)~, 
}
as can be inferred from eqns.~\eqref{NarainModuli} and~\eqref{DiscreteWilsonLineBShift}.

We start with fixing moduli in the metric $G$. From 
eqn.~\eqref{eqn:SymmetricRhoModuliG} and $\gTh^T \gTh = \Id$ we obtain the condition
\equ{\label{eqn:SymmtricOrbifoldConditionOnG}
\hat\theta^T G\, \hat\theta ~\stackrel{!}{=}~ G \qquad\Leftrightarrow\qquad \hat\theta ~\in~ O_G(D;\Intr)~,
}
which fixes some of the moduli, as is well-known. The general solution to 
eqn.~\eqref{eqn:SymmtricOrbifoldConditionOnG} for a given $\hat\theta$ can be parametrized as
\equ{\label{eqn:SymmetricOrbifoldG}
G ~=~ \frac 1K\, \sum_{k = 0}^{K-1} \hat\theta^{kT}\,G_0\, \hat\theta^k~, 
}
where $G_0$ is some symmetric positive definite matrix, for example $G_0 = \Id_D$. Now, it 
is easy to demonstrate that some metric moduli remain unconstrained for symmetric orbifolds: at 
least we can scale $G_0$ with an arbitrary positive factor, while 
eqn.\eqref{eqn:SymmtricOrbifoldConditionOnG} stays fulfilled.

Next, we consider the Wilson lines. If $\gth$ rotates in all $D$ compact dimensions 
$(\Id_D - \hat\theta)$ is invertible, i.e.
\begin{equation} 
(\Id_D - \hat\theta)^{-1} ~=~ -\frac{1}{K}\sum_{n=1}^{K-1} n \hat\theta^n\;,
\end{equation}
and the Wilson lines are uniquely determined from $\Delta A_k$ in eqn.~\eqref{eqn:SymmetricRhoModuliA}, 
e.g.\ from $k=1$
\begin{equation} \label{eqn:SymmetricOrbifoldA}
A ~\stackrel{!}{=}~ -\frac{1}{K}\sum_{n=1}^{K-1} n\, \Delta A_1\, \hat\theta^n\;.
\end{equation}
Consequently, the Wilson lines $A$ are completely frozen as they have to be discrete, i.e.\ quantized 
in units of $1/K$ in the directions where $\gth$ acts non-trivially. As a further consequence 
of eqn.~\eqref{eqn:SymmetricRhoModuliA} we see that two Wilson lines (i.e.\ two columns of $A$) 
have to be identical up to some trivial $\Delta A_k$ if the corresponding columns in the 
geometrical vielbein $e$ are mapped to each other by $\hat\theta^k$.

Finally, the $B$-field is constrained by the condition~\eqref{eqn:SymmetricRhoModuliB}
\equ{\label{eqn:SymmetricOrbifoldB}
B - \hat\theta^{\,T} B\, \hat\theta ~\stackrel{!}{=}~ \Delta B ~=~ \Delta B_1 - \frac{1}{2}\left(\hat\theta^{T} A^TA - A^TA\,\hat\theta\right)~,
}
combined with eqn.~\eqref{eqn:SymmtricOrbifoldConditionOnAB}. In analogy to 
eqn.~\eqref{eqn:SymmetricOrbifoldG} the general solution of this equation can written as 
\equ{
B = \frac 1K\, \sum_{k=0}^{K-1}  \hat\theta^{kT} B_0\, \hat\theta^k + B_P~, 
}
where $B_0$ is an arbitrary anti-symmetric matrix (for example, $B_0=0$) and $B_P$ is a 
particular solution to eqn.~\eqref{eqn:SymmetricOrbifoldB}. For example, in $D=2$ the 
anti-symmetric $2 \times 2$ matrix $B$ contains a single modulus. It is subject to 
eqn.~\eqref{eqn:SymmetricOrbifoldB}, i.e.
\equ{\label{eqn:SymmetricOrbifoldB2D}
B - \hat\theta^{T} B\, \hat\theta ~=~ (1 - \text{det}(\hat\theta) )\, B ~\stackrel{!}{=}~ \Delta B~,
}
where $\text{det}(\hat\theta) = \pm 1$. Thus, for $\text{det}(\hat\theta) = 1$ we obtain 
$\Delta B \stackrel{!}{=} 0$ and the single $B$-field modulus in $B$ is unconstrained and $B_P=0$. 
On the other hand, $B$ is stabilized at $B_P = \frac{1}{2}\Delta B$ if $\text{det}(\hat\theta) = -1$.

\subsubsection*{Number of moduli in symmetric $\boldsymbol{\Z{K}}$ orbifolds}

We can compute the number of (real) unstabilized moduli for symmetric $\Z{K}$ orbifolds for general 
$K$ using the results of Section~\ref{sec:IdentifyModuli}. To do so, we assume for simplicity $D=6$ 
and $K \neq 2$. Furthermore, we choose a $\Z{K}$ twist vector 
$\gf_\text{R} = (0,\gf_\text{R}^1,\gf_\text{R}^2,-\gf_\text{R}^1-\gf_\text{R}^2)$ such that 
$\mathcal{N}=1$ supersymmetry survives in four dimensions, see Section~\ref{sc:SUSYpreserving}. 
Hence, $K=3,4,6,7,8$ or $12$. Then, eqn.~\eqref{eqn:orbifoldmoduli} yields
\equ{
\dim(\cM_{\Z{K}}) 
\,=\, 6 + 2 \big(\delta_{\gf_\text{R}^1,\frac{1}{2}} + \delta_{\gf_\text{R}^2,\frac{1}{2}} + \delta_{\gf_\text{R}^1+\gf_\text{R}^2,\frac{1}{2}}\big) + 4 \big(\delta_{\gf_\text{R}^1,\gf_\text{R}^2} + \delta_{\gf_\text{R}^1,-\gf_\text{R}^2} + \delta_{\gf_\text{R}^1,-2\gf_\text{R}^2} + \delta_{2\gf_\text{R}^1,-\gf_\text{R}^2}\big)\,,
}
where $\delta_{a,b} = 1$ if $a \equiv b$ and $\delta_{a,b} = 0$ otherwise. For example, for $\Z{3}$ 
we take $\gf_\text{R}^1 = \gf_\text{R}^2 = \frac{1}{3}$ and obtain 
$\dim(\cM_{\Z{3}}) = 6 + 2 \times 0 + 4 \times (1+0+1+1)= 18$. As is well-known, these 18 (real) 
moduli correspond to 9 complex structure moduli, see e.g.~\cite{Cvetic:1988yw}.

\section[Two-dimensional Abelian Narain orbifolds]{Two-dimensional Abelian Narain orbifolds}
\label{sec:spacegroupexamples}

In this section, we study examples of generalized space groups of Narain orbifolds with Abelian 
Narain point groups $\Z{K}$ in two dimensions. Many of them correspond to previously unknown 
two-dimensional Narain orbifolds. We collect them in a comprehensive table. Furthermore, to 
illustrate various aspects of the theory developed in previous sections, we describe some of these 
two-dimensional $\Z{K}$ Narain orbifolds in more detail. For example, by an explicit construction 
we show that it is possible to have \Z{12} two-dimensional Narain orbifolds, while it is well-known 
that for Euclidean orbifolds in $D=2$ the largest order of a twist is $K=6$. Moreover, the 
$\Ratl$- and $\Intr$-classes are used to distinguish seemingly asymmetric from truly asymmetric 
orbifolds.

\subsection[$(D,D)$-Narain orbifold formalism]{$\boldsymbol{(D,D)}$-Narain orbifold formalism}
\label{sec:appendixODD} 

To prepare the discussion of various illustrative examples of two-dimensional Narain orbifolds, 
we briefly restrict the Narain orbifold formalism to the case where $\get$ has signature $(D,D)$:
\equ{
\widehat{\get} ~=~ R^T\, \get\, R ~=~ \pmtrx{
0     & \Id_D \\
\Id_D & 0
} \quad\text{with}\quad R ~=~ \frac{1}{\sqrt{2}}\, 
\pmtrx{
\Id_D &-\Id_D \\
\Id_D & \Id_D
}~.
}
The generalized vielbein $\widehat{E}$ is an element from $\text{O}_{\widehat{\get}}(D,D;\Real)$,
\equ{ \label{NarainModuliIn2D} 
\widehat{E}(e,B) ~=~ \pmtrx{
e        & 0      \\[1ex]
e^{-T} B & e^{-T}
}~.
}
Analogously to the discussion in Section~\ref{sec:NarainModuliTrafo}, for each element 
\equ{
\widehat{M} ~=~ \pmtrx{
\widehat{M}_{11} & \widehat{M}_{12} \\[1ex]
\widehat{M}_{21} & \widehat{M}_{22}
} ~\in~ \text{O}_{\widehat{\get}}(D,D;\Real)~, 
}
there exist a choice for a matrix $U_{\widehat{M}}\in O(D;\Real) \times O(D;\Real)$ and transformed 
moduli $e'$ and $B'$, such that
\equ{\label{eqn:ModuliTrafoIn2D}
\widehat{U}_{\widehat{M}}\, \widehat{E}(e',B') ~=~ \widehat{E}(e,B)\, \widehat{M}~,
\qquad
U_{\widehat{M}}  ~=~ R\, \widehat{U}_{\widehat{M}}\, R^{-1} ~=~ \pmtrx{
u_\text{r} & 0      \\[1ex]
0          & u_\text{l}
}~. 
}
In detail, defining 
\equ{\label{eqn:ODDDefinitionOfM}
\widehat{M}_1 ~=~ - \widehat{M}_{21} + (G-B)\widehat{M}_{11} \qquad\text{and}\qquad
\widehat{M}_2 ~=~ - \widehat{M}_{22} + (G-B)\widehat{M}_{12}~,
}
in accordance with eqn.~\eqref{GammaDeltaBA}, we obtain
\equ{\label{eqn:ODD_ul}
u_\text{l} ~=~ \left( \Id_D - 2\,e\,\widehat{M}_{12}\,\widehat{M}_2^{-1}\,e^T\right) u_\text{r} ~\in~ O(D;\Real)
}
for arbitrary $u_\text{r} \in O(D;\Real)$. This shows that $\widehat{M}_{12}\neq 0$ is a 
necessary condition for $u_\text{r} \neq u_\text{l}$. Furthermore, the Narain moduli transform as
\equ{ \label{eqn:ODDModuliTrafo}
e' ~=~ -u_\text{r}^{-1}\,e\,\widehat{M}_2^{-T}~,
\qquad 
G' ~=~ \widehat{M}_2^{-1}\, G\, \widehat{M}_2^{-T}~,
\qquad 
B' ~=~ \frac{1}{2}\left(\widehat{M}_2^{-1}\, \widehat{M}_1 - \widehat{M}_1^{T}\, \widehat{M}_2^{-T}\right)~.
}
By restricting $\widehat{M}$ to lie either inside $ \text{O}_{\widehat{\get}}(D,D;\Ratl)$ or 
$\text{O}_{\widehat{\get}}(D,D;\Intr)$, we obtain the transformations that map different 
representations within the same $\Ratl$- or $\Intr$-class to each other. 

Next, we discuss Narain orbifolds with Abelian $\Intr_K$ point groups 
$\widehat{\mathbf{P}}  \subset \text{O}_{\widehat{\get}}(D,D;\Intr)$. We use 
eqn.~\eqref{eqn:ODDDefinitionOfM} and set $\widehat{M} = \widehat{\rho}$, where $\widehat{\rho}$ is 
the generator of $\widehat{\mathbf{P}}$. Then, we find invariant moduli $G'=G$ and $B'=B$ from the 
latter two transformations in eqns.~\eqref{eqn:ODDModuliTrafo}. Moreover, we obtain the 
right-moving twist $\gth_\text{r} = u_\text{r}$ from the first relation in 
eqn.~\eqref{eqn:ODDModuliTrafo} by choosing a vielbein $e'=e$, which is in agreement with $G'=G$. 
By identifying the full twist $\widehat{\gTh} = \widehat{U}_{\widehat{\rho}}$ from 
eqn.~\eqref{eqn:ODD_ul} the Narain orbifold condition follows from 
eqn.~\eqref{eqn:ModuliTrafoIn2D}, i.e.
\equ{\label{eqn:DDOrbifoldIn2D}
\widehat{\gTh}\, \widehat{E}(e,B) ~=~ \widehat{E}(e,B)\, \widehat{\rho}~. 
}
Then, in analogy to Section~\ref{sec:ExistenceOfNarainOrbifolds} we know that the $\Z{K}$ Narain 
orbifold exists.

If the matrix-block $\widehat{\rho}_{12}$ is zero the orbifold is 
symmetric (i.e.~$\gth_{\text{r}} = \gth_{\text{l}}$) and a necessary (but not sufficient) 
condition for the orbifold to be asymmetric is $\widehat{\rho}_{12}\neq 0$, as can be seen from 
eqn.~\eqref{eqn:ODD_ul}.


\subsection[$\Ratl$- and $\Intr$-classes of two-dimensional $\Z{K}$ Narain orbifolds]{$\boldsymbol{\Ratl}$- and $\boldsymbol{\Intr}$-classes of two-dimensional $\boldsymbol{\Z{K}}$ Narain orbifolds}
\label{SecZN2DNarainOrbifolds} 

Following the discussion of the last section we focus on two-dimensional Narain 
orbifolds with point groups $\widehat{\mathbf{P}} \subset \text{O}_{\widehat{\get}}(2,2;\Intr)$, 
generated by a single twist $\widehat{\rho}$ of order $K$.

To initiate this investigation, we give a brief discussion on the possible orders following 
Section~\ref{sec:ConditionsOnTwists}: For Narain orbifolds with $D=2$ we have to set 
$D_\Gamma = 2D = 4$. Then, eqn.~\eqref{eqn:EulerPhiBoundary} yields the following list of possible 
orders 
\equ{
K ~\in~ \{\,1,2,3,4,5,6,8,10,12\,\}~.
}
In contrast, for two-dimensional symmetric orbifolds we have $D_\Gamma=D=2$ which yields only 
$K \in \{1,2,3,4,6\}$. Indeed, as we discuss in the following, we found examples for $K=12$. They are 
genuine asymmetric because twists of order 12 are not possible for $D_\Gamma=2$. On the other hand, 
we did not find any examples for $K=5, 8$ and $10$ in the scan of two-dimensional Narain orbifold 
we performed for this paper. 

In Table~\ref{TabZN2DNarainOrbifolds} we list a number of Abelian $\Z{K}$ Narain orbifolds of order 
$K$, which we constructed explicitly in our scan. For each Narain point group 
$\widehat{\mathbf{P}} \subset \text{O}_{\widehat{\get}}(2,2;\Intr)$ this table displays the 
following data in the various columns: 
\begin{enumerate}
\item column labels the inequivalent orbifolds and characterizes the orbifold as symmetric or asymmetric;
\item column gives a representation of the generating twist $\widehat{\rho}$ of order $K$ in the lattice basis;
\item column displays the corresponding right-twist $\gth_\text{r}$;
\item column displays the corresponding left-twist $\gth_\text{l}$;
\item column indicates the relation between these twists; 
\item column gives a choice of the geometrical vielbein $e$; 
\item column gives to resulting  metric as $G=e^T e$;
\item column gives the anti-symmetric $B$-field.
\end{enumerate}
A couple of further comments about the conventions of this table are in order: Our labelling 
conventions for inequivalent Narain orbifolds are as follows. The inequivalent $\Ratl$-classes of 
a given order $K$ are enumerated by a Roman number R=I,II,$\ldots$ as $\Z{K}$-R. Furthermore, when 
we give inequivalent $\Intr$-classes within a given $\Ratl$-class, we enumerate them with 
$n=1,2,3$ as $\Z{K}$-R-$n$. In fact, only the $\Ratl$-class $\Z{2}$-II is subdivided into 
three inequivalent $\Intr$-classes. Furthermore, the given right- and left-twists depend on our 
choice for the geometrical vielbein $e$ and on the Narain moduli $G$ and $B$.

To describe all these two dimensional Narain orbifolds in detail would lead to a lengthy 
discussion. Therefore, we focus in the following subsections on a number of striking features of 
some of these orbifolds instead. Before, doing so we make a couple of observations: First of all, 
we see that the number of asymmetric orbifolds greatly outweighs the number of symmetric orbifolds. 
This might imply that there exist many more asymmetric Narain orbifolds than symmetric ones. 
Most of the asymmetric orbifolds constructed in the past have twists that are trivial for either
the left- or the right-moving sectors, like the \Z{3}-II and \Z{3}-III orbifolds. In our scan we 
also encountered such examples, but again it seems that the majority of asymmetric orbifolds are 
not of this type: Most of them have non-trivial left- and right-moving twists simultaneously. In 
fact, there are even cases where the orders of the left- and right-moving twists are co-prime: 
the \Z{6}-IV and \Z{6}-VII Narain orbifolds. Since their orders are coprime, all their 
characters are orthogonal. Using the results of Section~\ref{sec:IdentifyModuli} this immediately 
implies that all moduli are stabilized for these orbifolds.

\afterpage{ 
\scriptsize 
\begin{longtable}[c]
{|>{\hspace{-2.8ex}$}c<{$\hspace{-2.8ex}}||>{\hspace{-1.5ex}$}c<{$\hspace{-1.5ex}}|>{\hspace{-1.5ex}$}c<{$\hspace{-1.5ex}}|>{\hspace{-1.5ex}$}c<{$\hspace{-1.5ex}}|>{\hspace{-1.5ex}$}c<{$\hspace{-1.5ex}}|>{\hspace{-1.5ex}$}c<{$\hspace{-1.5ex}}|>{\hspace{-1.5ex}$}c<{$\hspace{-1.5ex}}|>{\hspace{-1.5ex}$}c<{$\hspace{-1.5ex}}|}
\hline
\text{\bf label} & \text{\bf twist $\boldsymbol{\widehat{\rho}}$} & \text{\bf twist $\boldsymbol{\gth_\text{r}}$} & \text{\bf twist $\boldsymbol{\gth_\text{l}}$} & \text{\bf relation} & \text{\bf vielbein $\boldsymbol{e}$} & \text{\bf metric $\boldsymbol{G}$} & \text{\bf $\boldsymbol{B}$-field} \\
\hline\hline
\endhead
\hline
\multicolumn{8}{r@{}}{continued \ldots}
\endfoot
\endlastfoot
\begin{array}{c}   
\Z{2}\text{-I}\\
\text{sym.}
\end{array}
 &
\pmtrx{
 -1 &  0 &  0 &  0\\
  0 & -1 &  0 &  0\\
  0 &  0 & -1 &  0\\
  0 &  0 &  0 & -1}
 &
-\Id_2
 &
-\Id_2
 &
\gth_\text{r} = \gth_\text{l}
 &
\pmtrx{
 R_1 & R_2 \cos\ga\\[1ex]
 0   & R_2 \sin\ga
}
 &
\begin{array}{c}
\pmtrx{
 R_1^2 & w\\[1ex]
 w     &  R_2^2
}\\
w=R_1R_2\cos\ga
\end{array}
 &
\pmtrx{
 0 & b\\[1ex]
-b & 0
}
\\[1ex] \hline
\begin{array}{c}   
\Z{2}\text{-II-1}\\
\text{sym.}
\end{array}
 & 
\pmtrx{
  1 &  0 &  0 &  0\\
  0 & -1 &  0 &  0\\
  0 &  0 &  1 &  0\\
  0 &  0 &  0 & -1}
 &
\pmtrx{
 1 & 0\\[1ex]
 0 &-1
}
 &
\pmtrx{
 1 & 0\\[1ex]
 0 &-1
}
 &
\gth_\text{r} = \gth_\text{l}
 &
\pmtrx{
 R_1 & 0\\[1ex]
 0 & R_2
} &
\pmtrx{
 R_1^2 & 0\\[1ex]
 0 & R_2^2
} &
\pmtrx{
 0 & 0\\[1ex]
 0 & 0
}
\\[1ex]\hline
\begin{array}{c}   
\Z{2}\text{-II-2}\\
\text{sym.}
\end{array}
 &
\pmtrx{
  0 &  1 &  0 &  0\\
  1 &  0 &  0 &  0\\
  0 &  0 &  0 &  1\\
  0 &  0 &  1 &  0}
 &
\pmtrx{
 1 & 0\\[1ex]
 0 &-1
}
 &
\pmtrx{
 1 & 0\\[1ex]
 0 &-1
}
 &
\gth_\text{r} = \gth_\text{l}
 &
\pmtrx{
 R_1 & R_1\\[1ex]
-R_2 & R_2
} &
\pmtrx{
R_1^2+R_2^2 & R_1^2-R_2^2\\[1ex]
R_1^2-R_2^2 & R_1^2+R_2^2
} &
\pmtrx{
 0 & 0\\[1ex]
 0 & 0
}
\\[1ex]\hline
\begin{array}{c} 
\Z{2}\text{-II-3}\\
\text{sym.}
\end{array}
 & 
\pmtrx{
  0 &  1 &  0 &  0\\
  1 &  0 &  0 &  0\\
 -1 &  0 &  0 &  1\\
  0 &  1 &  1 &  0}
 &
\pmtrx{
 1 & 0\\[1ex]
 0 &-1
}
 &
\pmtrx{
 1 & 0\\[1ex]
 0 &-1
}
 &
\gth_\text{r} = \gth_\text{l}
 &
\pmtrx{
 R_1 & R_1\\[1ex]
-R_2 & R_2
} &
\pmtrx{
R_1^2+R_2^2 & R_1^2-R_2^2\\[1ex]
R_1^2-R_2^2 & R_1^2+R_2^2
} &
\pmtrx{
 0 & \frac{1}{2}\\[1ex]
-\frac{1}{2} & 0
}
\\[1ex]\hline\hline
\begin{array}{c}
\Z{3}\text{-I}\\
\text{sym.}
\end{array}
 &
\pmtrx{
  0 & -1 &  0 &  0\\
  1 & -1 &  0 &  0\\
  0 &  0 & -1 & -1\\
  0 &  0 &  1 &  0}
 &
\pmtrx{
-\frac{1}{2}        & -\frac{\sqrt{3}}{2}\\[1ex]
 \frac{\sqrt{3}}{2} & -\frac{1}{2}
}
 &
\pmtrx{
-\frac{1}{2}        & -\frac{\sqrt{3}}{2}\\[1ex]
 \frac{\sqrt{3}}{2} & -\frac{1}{2}
}
 &
\gth_\text{r} = \gth_\text{l}
 &
R\, \pmtrx{
 \sqrt{2} & -\frac{1}{\sqrt{2}}\\[1ex]
 0 &  \sqrt{\frac{3}{2}}
}
 &
R^2\, \pmtrx{
 2 & -1\\[1ex]
-1 &  2
}
 &
\pmtrx{
 0 & b\\[1ex]
-b & 0
} 
\\[1ex] \hline 
\begin{array}{c}
\Z{3}\text{-II}\\
\text{asym.}
\end{array}
 &
\pmtrx{
  0 &  0 &  1 &  1\\
  0 &  0 &  0 &  1\\
  1 &  0 &  0 & -1\\
 -1 &  1 &  1 &  1}
 &
\pmtrx{
-\frac{1}{2}        & -\frac{\sqrt{3}}{2}\\[1ex]
 \frac{\sqrt{3}}{2} & -\frac{1}{2}
}
 &
\Id_2
 &
\begin{array}{c}
\gth_\text{r}^3 = \Id_2\\
\gth_\text{l} = \Id_2
\end{array}
 &
\pmtrx{
 1 &-\frac{1}{2}\\[1ex]
 0 & \frac{\sqrt{3}}{2}
}
 &
\pmtrx{
 1 & -\frac{1}{2}\\[1ex]
-\frac{1}{2} &  1
}
 &
\pmtrx{
 0 & \frac{1}{2}\\[1ex]
-\frac{1}{2} & 0
} 
\\[1ex] \hline 
\begin{array}{c}
\Z{3}\text{-III}\\
\text{asym.}
\end{array}
 &
\pmtrx{
  0 &  0 & -1 &  0\\
  0 &  0 & -1 & -1\\
 -1 &  1 &  1 &  1\\
  0 & -1 & -1 &  0}
 &
\Id_2
 &
\pmtrx{
-\frac{1}{2}        &-\frac{\sqrt{3}}{2}\\[1ex]
 \frac{\sqrt{3}}{2} & -\frac{1}{2}
} 
 &
\begin{array}{c}
\gth_\text{r} = \Id_2\\
\gth_\text{l}^3 = \Id_2
\end{array}
 &
\pmtrx{
 1 &-\frac{1}{2}\\[1ex]
 0 &-\frac{\sqrt{3}}{2}
}
 &
\pmtrx{
 1 & -\frac{1}{2}\\[1ex]
-\frac{1}{2} &  1
}
 &
\pmtrx{
 0 & \frac{1}{2}\\[1ex]
-\frac{1}{2} & 0
} 
\\[1ex] \hline \hline 
\begin{array}{c}
\Z{4}\text{-I}\\
\text{sym.}
\end{array}
 &
\pmtrx{
  0 & -1 &  0 &  0\\
  1 &  0 &  0 &  0\\
  0 &  0 &  0 & -1\\
  0 &  0 &  1 &  0}
 &
\pmtrx{
 0 &-1\\[1ex]
 1 & 0
}
 &
\pmtrx{
 0 &-1\\[1ex]
 1 & 0
}
 &
\gth_\text{r}=\gth_\text{l}
 &
\pmtrx{
 R & 0\\[1ex]
 0 & R
}
 &
\pmtrx{
 R^2 & 0\\[1ex]
 0 & R^2
}
 &
\pmtrx{
 0 & b\\[1ex]
-b & 0
} 
\\[1ex] \hline
\begin{array}{c}
\Z{4}\text{-II}\\
\text{asym.}
\end{array}
 &
\pmtrx{
 -1 & -1 & -1 &  1\\
  0 &  0 & -1 &  1\\
  0 &  0 &  0 & -1\\
 -1 &  0 &  0 &  1}
 &
\pmtrx{
 0 &-1\\[1ex]
 1 & 0
}
 &
\pmtrx{
-1 & 0\\[1ex]
 0 & 1
}
 &
\begin{array}{c}
\gth_\text{r}^4=\Id_2\\
\gth_\text{l}^2=\Id_2
\end{array}
 &
\pmtrx{
 \frac{1}{\sqrt{2}} & 0\\[1ex]
 0 & \frac{1}{\sqrt{2}}
}
 &
\pmtrx{
 \frac{1}{2} & 0\\[1ex]
 0 & \frac{1}{2}
}
 &
\pmtrx{
 0 & \frac{1}{2}\\[1ex]
-\frac{1}{2} & 0
}
\\[1ex] \hline 
\begin{array}{c}
\Z{4}\text{-III}\\
\text{asym.}
\end{array}
 &
\pmtrx{
 -1 &  1 &  1 &  1\\
  0 &  0 & -1 & -1\\
  0 &  0 &  0 &  1\\
 -1 &  0 &  0 &  1}
 &
\pmtrx{
-1 & 0\\[1ex]
 0 & 1
}
 &
\pmtrx{
 0 &-1\\[1ex]
 1 & 0
}
 &
\begin{array}{c}
\gth_\text{r}^2=\Id_2\\
\gth_\text{l}^4=\Id_2
\end{array}
 &
\pmtrx{
 \frac{1}{\sqrt{2}} & 0\\[1ex]
 0 & -\frac{1}{\sqrt{2}}
}
 &
\pmtrx{
 \frac{1}{2} & 0\\[1ex]
 0 & \frac{1}{2}
}
 &
\pmtrx{
 0 & \frac{1}{2}\\[1ex]
 -\frac{1}{2} & 0
} 
\\[1ex] \hline\hline
\begin{array}{c}
\Z{6}\text{-I}\\
\text{sym.}
\end{array}
 &
\pmtrx{
  1 & -1 &  0 &  0\\
  1 &  0 &  0 &  0\\
  0 &  0 &  0 & -1\\
  0 &  0 &  1 &  1}
 &
\pmtrx{
 \frac{1}{2} & -\frac{\sqrt{3}}{2}\\[1ex]
 \frac{\sqrt{3}}{2} & \frac{1}{2}
}
 &
\pmtrx{
 \frac{1}{2} & -\frac{\sqrt{3}}{2}\\[1ex]
 \frac{\sqrt{3}}{2} & \frac{1}{2}
}
 &
\gth_\text{r}=\gth_\text{l}
 &
R\, \pmtrx{
 \sqrt{2} & -\frac{1}{\sqrt{2}}\\[1ex]
 0 &  \sqrt{\frac{3}{2}}
}
 &
R^2\,\pmtrx{
 2 & -1\\[1ex]
 -1 & 2
}
 &
\pmtrx{
 0 & b\\[1ex]
-b & 0
} 
\\[1ex] \hline
\begin{array}{c}
\Z{6}\text{-II}\\
\text{asym.}
\end{array}
 &
\pmtrx{
  0 &  0 & -1 & -1\\
  0 &  0 &  0 & -1\\
 -1 &  0 &  0 &  1\\
  1 & -1 & -1 & -1}
 &
\pmtrx{
 \frac{1}{2} & -\frac{\sqrt{3}}{2}\\[1ex]
 \frac{\sqrt{3}}{2} & \frac{1}{2}
}
 &
-\Id_2
 &
\begin{array}{c}
\gth_\text{r}^6=\Id_2\\
\gth_\text{l}^2=\Id_2
\end{array}
 &
\pmtrx{
 1 & -\frac{1}{2}\\[1ex]
 0 & -\frac{\sqrt{3}}{2}
}
 &
\pmtrx{
 1 & -\frac{1}{2}\\[1ex]
 -\frac{1}{2} & 1
}
 &
\pmtrx{
 0 & \frac{1}{2}\\[1ex]
-\frac{1}{2} & 0
} 
\\[1ex] \hline
\begin{array}{c}
\Z{6}\text{-III}\\
\text{asym.}
\end{array}
 &
\pmtrx{
  1 & -1 &  0 &  0\\
  1 & -1 & -1 & -1\\
  0 &  1 &  1 &  0\\
  0 & -1 &  0 &  0}
 &
\pmtrx{
 \frac{1}{2} & -\frac{\sqrt{3}}{2}\\[1ex]
 \frac{\sqrt{3}}{2} & \frac{1}{2}
}
 &
\pmtrx{
 1 & 0\\[1ex]
 0 & -1
}
 &
\begin{array}{c}
\gth_\text{r}^6=\Id_2\\
\gth_\text{l}^2=\Id_2
\end{array}
 &
\pmtrx{
 1 &-\frac{1}{2}\\[1ex]
 0 & \frac{\sqrt{3}}{2}
}
 &
\pmtrx{
 1 & -\frac{1}{2}\\[1ex]
 -\frac{1}{2} & 1
}
 &
\pmtrx{
 0 & \frac{1}{2}\\[1ex]
-\frac{1}{2} & 0
} 
\\[1ex] \hline
\begin{array}{c}
\Z{6}\text{-IV}\\
\text{asym.}
\end{array}
 &
\pmtrx{
 -1 &  0 &  0 &  1\\
  0 &  0 &  0 &  1\\
  0 &  0 & -1 & -1\\
 -1 &  1 &  1 &  1}
 &
\pmtrx{
-\frac{1}{2} & -\frac{\sqrt{3}}{2}\\[1ex]
 \frac{\sqrt{3}}{2} &-\frac{1}{2}
}
 &
\pmtrx{
-1 & 0\\[1ex]
 0 & 1
}
 &
\begin{array}{c}
\gth_\text{r}^3=\Id_2\\
\gth_\text{l}^2=\Id_2
\end{array}
 &
\pmtrx{
 1 &-\frac{1}{2}\\[1ex]
 0 & \frac{\sqrt{3}}{2}
}
 &
\pmtrx{
 1 & -\frac{1}{2}\\[1ex]
 -\frac{1}{2} & 1
}
 &
\pmtrx{
 0 & \frac{1}{2}\\[1ex]
-\frac{1}{2} & 0
} 
\\[1ex] \hline
\begin{array}{c}
\Z{6}\text{-V}\\
\text{asym.}
\end{array}
 &
\pmtrx{
 -1 &  1 &  1 &  1\\
 -1 &  0 &  0 &  1\\
  1 &  0 &  0 &  0\\
 -1 &  1 &  0 &  0}
 &
-\Id_2
 &
\pmtrx{
 \frac{1}{2} & -\frac{\sqrt{3}}{2}\\[1ex]
 \frac{\sqrt{3}}{2} & \frac{1}{2}
}
 &
\begin{array}{c}
\gth_\text{r}^2=\Id_2\\
\gth_\text{l}^6=\Id_2
\end{array}
 &
\pmtrx{
 1 &-\frac{1}{2}\\[1ex]
 0 &-\frac{\sqrt{3}}{2}
}
 &
\pmtrx{
 1 & -\frac{1}{2}\\[1ex]
 -\frac{1}{2} & 1
}
 &
\pmtrx{
 0 & \frac{1}{2}\\[1ex]
-\frac{1}{2} & 0
} 
\\[1ex] \hline
\begin{array}{c}
\Z{6}\text{-VI}\\
\text{asym.}
\end{array}
 &
\pmtrx{
  0 &  0 &  0 &  1\\
 -1 &  0 &  0 &  1\\
  0 &  1 &  1 &  0\\
  0 &  0 & -1 &  0}
 &
\pmtrx{
 1 & 0\\[1ex]
 0 &-1
} 
 &
\pmtrx{
 \frac{1}{2} & -\frac{\sqrt{3}}{2}\\[1ex]
 \frac{\sqrt{3}}{2} & \frac{1}{2}
}
 &
\begin{array}{c}
\gth_\text{r}^2=\Id_2\\
\gth_\text{l}^6=\Id_2
\end{array}
 &
\pmtrx{
 1 &-\frac{1}{2}\\[1ex]
 0 &-\frac{\sqrt{3}}{2}
}
 &
\pmtrx{
 1 & -\frac{1}{2}\\[1ex]
 -\frac{1}{2} & 1
}
 &
\pmtrx{
 0 & \frac{1}{2}\\[1ex]
-\frac{1}{2} & 0
} 
\\[1ex] \hline
\begin{array}{c}
\Z{6}\text{-VII}\\
\text{asym.}
\end{array}
 &
\pmtrx{
 -1 &  1 &  0 &  0\\
  0 &  0 & -1 & -1\\
  0 &  0 &  0 &  1\\
 -1 &  0 &  0 &  0}
 &
\pmtrx{
-1 & 0\\[1ex]
 0 & 1
} 
 &
\pmtrx{
-\frac{1}{2} & -\frac{\sqrt{3}}{2}\\[1ex]
 \frac{\sqrt{3}}{2} & -\frac{1}{2}
}
 &
\begin{array}{c}
\gth_\text{r}^2=\Id_2\\
\gth_\text{l}^3=\Id_2
\end{array}
 &
\pmtrx{
 1 &-\frac{1}{2}\\[1ex]
 0 &-\frac{\sqrt{3}}{2}
}
 &
\pmtrx{
 1 & -\frac{1}{2}\\[1ex]
 -\frac{1}{2} & 1
}
 &
\pmtrx{
 0 & \frac{1}{2}\\[1ex]
-\frac{1}{2} & 0
} 
\\[1ex] \hline\hline
\begin{array}{c}
\Z{12}\text{-I}\\
\text{asym.}
\end{array}
 &
\pmtrx{
  0 &  0 &  1 &  0\\
  0 &  0 &  0 &  1\\
  1 &  0 &  0 &  1\\
  0 &  1 & -1 &  0}
 &
\pmtrx{
-\frac{\sqrt{3}}{2} & \frac{1}{2} \\[1ex]
-\frac{1}{2} & -\frac{\sqrt{3}}{2}
}
 &
\pmtrx{
 \frac{\sqrt{3}}{2} & \frac{1}{2} \\[1ex]
-\frac{1}{2} & \frac{\sqrt{3}}{2}
}
 &
\gth_\text{l}=\gth_\text{r}^5
 &
\pmtrx{
 \frac{3^{\frac{1}{4}}}{\sqrt{2}} & 0\\[1ex]
 0 & \frac{3^{\frac{1}{4}}}{\sqrt{2}}
}
 &
\pmtrx{
 \frac{\sqrt{3}}{2} & 0\\[1ex]
 0 & \frac{\sqrt{3}}{2}
}
 &
\pmtrx{
 0 & -\frac{1}{2}\\[1ex]
\frac{1}{2} & 0
} 
\\\hline
\caption{This table presents a large number of examples for $\Intr_K$ Narain orbifolds in two 
dimensions. For each inequivalent orbifold it gives important data that characterizes Narain 
orbifolds, like the twists in both, the lattice and the coordinate basis and the values of the (frozen) moduli.}
\label{TabZN2DNarainOrbifolds}
\end{longtable}
\normalsize 
}


\subsection[Two equivalent asymmetric $\Z{12}$ Narain orbifolds]{Two equivalent asymmetric $\boldsymbol{\Z{12}}$ Narain orbifolds}
\label{sec:O22Z12Orbifolds}

With our first two examples we want to illustrate that we are able to construct genuine 
asymmetric orbifolds using the formalism for Narain orbifolds exposed in this paper. Concretely, we 
define two $\Z{12}$ Narain point groups $\widehat{\mathbf{P}}_{(1)}$ and $\widehat{\mathbf{P}}_{(1)}$ 
in $D=2$, each being generated by an element 
$\widehat{\rho}_{(1)}, \widehat{\gr}_{(2)} \in \text{O}_{\widehat{\get}}(2,2;\Intr)$ of order 12. 
In each case, we determine the corresponding Narain lattice and the twist $\gTh$ which is given by 
its action on right- and left-movers, $\gth_\text{r}$ and $\gth_\text{l}$, respectively. As there 
is no symmetric $\Z{12}$ orbifold in $D=2$ (i.e.~there is no two-dimensional lattice with 
rotational symmetry of order 12), these orbifolds must be genuine asymmetric\footnote{Such 
asymmetric $\Z{12}$ orbifolds were studied in the past~\cite{Harvey:1987da,Dabholkar:1998kv}.}. 
Moreover, to emphasize that the use of $\Intr$-classes is extremely powerful to investigate whether 
two orbifolds are distinct, we show that these two $\Z{12}$ point groups are in fact equivalent by 
giving an explicit $\text{O}_{\widehat{\get}}(D,D;\Intr)$ matrix that relates the two twists in the 
lattice basis.

The first asymmetric $\Z{12}$ orbifold example has a non-vanish $B$-field $B_{(1)} \neq 0$: We choose
\equ{
\widehat{\rho}_{(1)}  ~=~ \pmtrx{
  0     & \Id_2\\[1ex]
  \Id_2 & \epsilon
} ~\in~ \text{O}_{\widehat{\get}}(2,2;\Intr)\qquad\text{where}\qquad \epsilon ~=~ \pmtrx{
  0 & 1\\
 -1 & 0
}
}
and obtain
\equ{
\widehat{M}_{(1)1} ~=~ -\Id_2~, \quad \widehat{M}_{(1)2} ~=~ G_{(1)} -B_{(1)} -\epsilon~,
}
from eqn.~\eqref{eqn:ODDDefinitionOfM}. Then we follow the procedure outlined in 
Section~\ref{SecLatticeBtoCoordB} to find that all Narain moduli are stabilized and take the form 
\equ{
e_{(1)}  ~=~ \frac{3^{\frac{1}{4}}}{\sqrt{2}} \Id_2 \quad\text{and}\quad B_{(1)}  ~=~ -\frac{1}{2}\, \epsilon~, 
}
while the twist $\gTh_{(1)}$ is given by
\equ{
\gth_{(1)\text{r}} ~=~ \pmtrx{
-\frac{\sqrt{3}}{2} &  \frac{1}{2} \\[1ex]
-\frac{1}{2}        & -\frac{\sqrt{3}}{2}
} \quad\text{and}\quad \gth_{(1)\text{l}} ~=~ \gth_{(1)\text{r}}^5~.
}
This precisely corresponds to the data given for the \Z{12}-I orbifold in Table~\ref{TabZN2DNarainOrbifolds}.

An equivalent description of this asymmetric \Z{12}-I orbifold has no $B$-field at all ($B_{(2)} = 0$). For this case  we take 
\equ{
\widehat{\rho}_{(2)} ~=~ \pmtrx{
  0 &  0 &  1 &  1\\
  0 &  0 &  0 &  1\\
  1 &  0 &  0 &  0\\
 -1 &  1 &  0 &  0} ~\in~ \text{O}_{\widehat{\get}}(2,2;\Intr)~.
}
The stabilized Narain moduli are now given by 
\equ{
e_{(2)} ~=~ \pmtrx{
\frac{\sqrt{2}}{3^{\frac{1}{4}}} & -\frac{1}{\sqrt{2}\, 3^{\frac{1}{4}}} \\[1ex]
0                                & -\frac{3^{\frac{1}{4}}}{\sqrt{2}}
} \quad\text{and}\quad B_{(2)} ~=~ 0~,  
}
with the twist $\gTh_{(2)}$ is given by
\equ{
\gth_{(2)\text{r}} ~=~ \pmtrx{
-\frac{\sqrt{3}}{2} &  \frac{1}{2} \\[1ex]
-\frac{1}{2}        & -\frac{\sqrt{3}}{2}
} \quad\text{and}\quad \gth_{(2)\text{l}} ~=~ \gth_\text{r}^7 ~=~ -\gth_\text{r}~.
}

To show explicitly that these two $\Z{12}$ orbifolds are $\Intr$-equivalent (and 
consequently also $\Ratl$-equivalent), we observe that we can relate the two \Z{12} generators,
\equ{ 
\widehat{M}\, \widehat{\rho}_{(2)} ~=~ \widehat{\rho}_{(1)}\, \widehat{M}~,
}
using the matrix 
\equ{
\widehat{M} ~=~ \pmtrx{
 -1 &  0 &  0 &  0\\
  0 &  0 &  0 & -1\\
  0 &  0 & -1 & -1\\
  1 & -1 &  0 &  0} ~\in~ \text{O}_{\widehat{\get}}(2,2;\Intr)~.
}
Here, we used that both generators $\widehat{\rho}_{(1)}$ and $\widehat{\gr}_{(2)}$ are defined 
with respect to the same Narain metric $\widehat{\get}$. Hence, the corresponding Narain point 
groups $\widehat{\mathbf{P}}_{(1)}$ and $\widehat{\mathbf{P}}_{(2)}$ are identical up to the 
discrete $T$-duality transformation with $\widehat{M}$, i.e.\ these point groups lie in the same 
$\Intr$-class. In other words, we have described the same asymmetric \Z{12} orbifold in two 
different duality frames, once with and once without $B$-field.


\subsection[Exposing a seemingly asymmetric $\Z{3}$ Narain orbifold]{Exposing a seemingly asymmetric $\boldsymbol{\Z{3}}$ Narain orbifold}
\label{sec:O22Z3SymmetricOrbifolds}

It might happen that one uses a description, i.e.\ choice of duality frame, in which a given Narain 
orbifold appears to be asymmetric. Consider for example a two-dimensional $\Z{3}$ Narain orbifold 
defined by the twist 
\equ{
\widehat{\rho}_{(\text{a})} ~=~ \pmtrx{
  0 & \epsilon\\
  \epsilon & -\Id_2
} ~\in~ \text{O}_{\widehat{\get}}(2,2;\Intr)~,
}
in the lattice basis. We use the subscript (a) to refer to this seemingly asymmetric orbifold: It 
is not obviously a symmetric orbifold, as it does not meet the sufficient condition 
$(\widehat{\gr})_{12} = 0$ for being a symmetric Narain orbifold formulated in 
Section~\ref{SecLatticeBtoCoordB}. Since in this case, eqns.~\eqref{eqn:ODDDefinitionOfM} 
reduce to 
\equ{
\widehat{M}_{(\text{a})1} ~=~ -\epsilon~, \qquad \widehat{M}_{(\text{a})2} ~=~ \Id_2 + (G_{(\text{a})}-B_{(\text{a})})\epsilon~, 
}
the Narain moduli are given by
\equ{
e_{(\text{a})} ~=~\pmtrx{
 R_{(\text{a})} & w_{(\text{a})}\\[1ex]
 0   & -\frac{\sqrt{3}}{2 R_{(\text{a})}}
} \quad\text{and}\quad B_{(\text{a})} ~=~ -\frac{1}{2} \epsilon~,
}
where parameters $R_{(\text{a})}$ and $w_{(\text{a})}$ are unconstrained. Furthermore, the twist 
$\gTh_{(\text{a})}$ is specified by
\equ{
\gth_{(\text{a})\text{r}} ~=~ \pmtrx{
-\frac{1}{2}        &  \frac{\sqrt{3}}{2}\\[1ex]
-\frac{\sqrt{3}}{2} & -\frac{1}{2}
} \quad\text{and}\quad \gth_{(\text{a})\text{l}} ~=~ \gth_{(\text{a})\text{r}}^2~. 
}
Since $\gth_{(\text{a})\text{r}} \neq \gth_{(\text{a})\text{l}}$, this seems to indicate that this 
an asymmetric Narain orbifold. However, it is equivalent to the symmetric orbifold \Z{3}-I of 
Table~\ref{TabZN2DNarainOrbifolds}:

To see this, we describe this symmetric \Z{3}-I orbifold (labelled with a subscript (s)) in 
some detail: The defining twist in the lattice basis is given by 
\equ{
\widehat{\rho}_{(\text{s})} ~=~ \pmtrx{
  0 & -1 &  0 &  0\\
  1 & -1 &  0 &  0\\
  0 &  0 & -1 & -1\\
  0 &  0 &  1 &  0
} ~\in~ \text{O}_{\widehat{\get}}(2,2;\Intr)~, 
}
from which we obtain
\begin{subequations} 
\equ{
\widehat{M}_{(\text{s})1} ~=~ (G_{(\text{s})}-B_{(\text{s})}) \rho_{(\text{s})\text{r}}~, 
\qquad 
\widehat{M}_{(\text{s})2} ~=~ -\left(\rho_{(\text{s})\text{r}}\right)^{-T}~, 
\\[1ex] 
\rho_{(\text{s})\text{r}} ~=~ e_{(\text{s})}^{-1}\,\theta_{(\text{s})\text{r}}\,e_{(\text{s})} ~=~ \left(\widehat{\gr}_{(\text{s})}\right)_{11} ~=~ \pmtrx{
 0 & -1\\
 1 & -1
}~.
}
\end{subequations} 
In this case, $\rho_{(\text{s})\text{r}}$ acts cryptographically on $e$, i.e.~the first column 
$e_1$ of $e$ is mapped to the second column $e_2$ and $e_2$ is mapped to $-e_1-e_2$. Furthermore, 
the Narain moduli are given by
\equ{
e_{(\text{s})} ~=~ R_{(\text{s})}\,\pmtrx{
 \sqrt{2} & -\frac{1}{\sqrt{2}}\\[1ex]
 0        & \sqrt{\frac{3}{2}}
} \quad\Rightarrow\quad 
G_{(\text{s})} ~=~ R_{(\text{s})}^2\, \pmtrx{
 2 & -1\\[1ex]
-1 &  2
} \quad\text{and}\quad B_{(\text{s})} ~=~ b_{(\text{s})}\, \pmtrx{
 0 & 1\\[1ex]
-1 & 0
}~,
}
where $R_{(\text{s})}$ and $b_{(\text{s})}$ are unconstrained. Thus, the vielbein $e_{(\text{s})}$ 
spans the root lattice of $\SU{3}$ multiplied by an arbitrary radius $R_{(\text{s})}$. Furthermore, 
the twist $\gTh_{(\text{s})}$ is specified by
\equ{
\gth_{(\text{s})\text{r}} ~=~ \gth_{(\text{s})\text{l}} ~=~ \pmtrx{
-\frac{1}{2}        & -\frac{\sqrt{3}}{2}\\[1ex]
 \frac{\sqrt{3}}{2} & -\frac{1}{2}
}~.
}

Clearly, these two descriptions look very different: The parametrization of the moduli does not 
seem to be alike, since, for example, in case (\text{a}) the $B$-field is fixed while in case 
(\text{s}) it is a modulus. Moreover, the twist seems to be asymmetric for case (\text{a}) but 
symmetric for case (\text{s}). However, their Narain point groups $\widehat{\mathbf{P}}_{(\text{s})}$ 
and $\widehat{\mathbf{P}}_{(\text{a})}$ belong to the same $\Intr$-class (and consequently also to 
the same $\Ratl$-class); they are equivalent up to a discrete $T$-duality transformation.

Explicitly, the discrete $T$-duality transformation that relates $\widehat{\mathbf{P}}_{(\text{s})}$ 
and $\widehat{\mathbf{P}}_{(\text{a})}$ reads
\equ{\label{eqn:Z3ExamplesZEquiv}
\widehat{M} ~=~ \pmtrx{
 -1 & -1 &  0 &  0\\
  0 &  0 & -1 &  1\\
  0 &  0 & -1 &  0\\
  0 &  1 &  0 &  0} ~\in~ \text{O}_{\widehat{\get}}(2,2;\Intr) \quad\text{with}\quad \widehat{M}\, \widehat{\rho}_{(\text{a})} ~=~ \widehat{\rho}_{(\text{s})}\, \widehat{M}~,
}
where we used that $\widehat{\rho}_{(\text{s})}$ and $\widehat{\rho}_{(\text{a})}$ are both defined 
with respect to the same Narain metric $\widehat{\get}$. This implies that also the moduli 
$(R_{(\text{s})}, b_{(\text{s})})$ and $(R_{(\text{a})},w_{(\text{a})})$ can be mapped explicitly 
by exploiting  the transformation formula~\eqref{eqn:ODDModuliTrafo}: We Use
\equ{
\widehat{U}_{\widehat{M}}\, \widehat{E}(e_{(\text{a})},B_{(\text{a})}) ~=~ \widehat{E}(e_{(\text{s})},B_{(\text{s})}) \, \widehat{M}
}
with $\widehat{M}$ given in eqn.~\eqref{eqn:Z3ExamplesZEquiv} to relate the moduli in both 
descriptions as 
\begin{subequations}
\begin{eqnarray}
G_{(\text{a})} & = & \frac{1}{2 R_{(\text{s})}^2} \pmtrx{
 b_{(\text{s})}^2 + 3R_{(\text{s})}^4   & b_{(\text{s})}+b_{(\text{s})}^2+3R_{(\text{s})}^4  \\[2ex]
 b_{(\text{s})}+b_{(\text{s})}^2+3R_{(\text{s})}^4 & (1+b_{(\text{s})})^2 + 3R_{(\text{s})}^4}~, \\[2ex]
B_{(\text{a})} & = & -\frac{1}{2} \epsilon~.
\end{eqnarray}
\end{subequations}
This results in
\equ{
w_{(\text{a})} ~=~ \frac{1}{\sqrt{2} R_{(\text{s})}}\, \frac{b_{(\text{s})}+b_{(\text{s})}^2+3R_{(\text{s})}^4}{\sqrt{b_{(\text{s})}^2 + 3R_{(\text{s})}^4}} \quad\text{and}\quad R_{(\text{a})} ~=~ \frac{1}{\sqrt{2} R_{(\text{s})}}\, \sqrt{b_{(\text{s})}^2 + 3R_{(\text{s})}^4}~.
}
In addition, we compute $u_\text{r}$ and $u_\text{l}$ from eqns.~\eqref{eqn:ODD_ul} and~\eqref{eqn:ODDModuliTrafo} to obtain
\equ{
u_\text{l} ~=~ \pmtrx{
\frac{1}{2}        & \frac{\sqrt{3}}{2} \\[1ex]
\frac{\sqrt{3}}{2} & -\frac{1}{2}}\, u_\text{r}~; \quad 
u_\text{r} ~=~ \frac{1}{2\sqrt{b_{(\text{s})}^2 + 3R_{(\text{s})}^4}}\,\pmtrx{
b_{(\text{s})} - 3R_{(\text{s})}^2 & -\sqrt{3} (b_{(\text{s})} + R_{(\text{s})}^2) \\[1ex]
-\sqrt{3} (b_{(\text{s})} + R_{(\text{s})}^2) & -b_{(\text{s})} + 3R_{(\text{s})}^2}~.
}
Note that $\text{det}(u_\text{l}) = +1$ but $\text{det}(u_\text{r}) = -1$. This corresponds to the 
matrix $U_\text{B}$ from eqn.~\eqref{eqn:QZequivChangeOfTheta} that maps the symmetric twist from 
point group $\mathbf{P}_{(\text{s})}$ to the seemingly asymmetric twist from point group 
$\mathbf{P}_{(\text{a})}$.

Let us close this subsection with the comment that for Narain orbifolds of order 3, we were able to 
distinguish between three $\Ratl$-classes, where each $\Ratl$-class contains only a single 
$\Intr$-class. In the nomenclature of Table~\ref{TabZN2DNarainOrbifolds} the two-dimensional Narain 
orbifold \Z{3}-I is a symmetric orbifold, while the other two, \Z{3}-II and \Z{3}-II, are asymmetric. 
In fact, they are each others mirrors in the sense that their $\gth_\text{l}$ and $\gth_\text{r}$ 
are interchanged.

\subsection[Symmetric $\Z{2}$ Narain orbifolds from inequivalent $\Intr$-classes]{Symmetric $\boldsymbol{\Z{2}}$ Narain orbifolds from inequivalent $\boldsymbol{\Intr}$-classes}
\label{SecZ2InequivZclasses} 

For the examples considered so far, we found that each Narain $\Ratl$-class contains just a single 
Narain $\Intr$-class. This might convey the impression that the notion of $\Intr$-classes for 
Narain orbifolds is obsolete. To emphasize that this is not the case, we consider two symmetric 
$\Z{2}$ Narain point groups in $D=2$ next. Both correspond geometrically to the M\"obius strip, 
where the $B$-field is either turned on or off. We will show that even though these two 
Narain point groups belong to the same Narain $\Ratl$-class, they live in different Narain 
$\Intr$-classes, hence they are physically inequivalent.

Consider the symmetric \Z{2}-II-2 Narain orbifold of Table~\ref{TabZN2DNarainOrbifolds} without a 
$B$-field: In detail, we choose
\equ{
\widehat{\rho}_{(1)}  ~=~ \pmtrx{
  0 & 1 & 0 &  0\\
  1 & 0 & 0 &  0\\
  0 & 0 & 0 &  1\\
  0 & 0 & 1 &  0
} ~\in~ \text{O}_{\widehat{\get}}(2,2;\Intr)
}
and obtain
\equ{
\widehat{M}_{(1)1} ~=~ (G_{(1)} -B_{(1)} ) \pmtrx{
 0 & 1\\
 1 & 0
} \qquad\text{and}\qquad \widehat{M}_{(1)2} ~=~ -\pmtrx{
 0 & 1\\
 1 & 0
}~.
}
In this case, the Narain moduli are given by
\equ{
e_{(1)}  ~=~ \pmtrx{
 R_1 & R_1\\[1ex]
-R_2 & R_2
} 
\quad\Rightarrow\quad 
G_{(1)}  ~=~ \pmtrx{
R_1^2+R_2^2 & R_1^2-R_2^2\\[1ex]
R_1^2-R_2^2 & R_1^2+R_2^2
} 
\quad\text{and}\quad B_{(1)}  ~=~ \pmtrx{
 0 & 0\\[1ex]
 0 & 0
}~,
}
for $R_1R_2 \neq 0$. Furthermore, the twist $\gTh$ is specified by
\equ{
\gth_{\text{r}} ~=~ \gth_{\text{l}} ~=~ \pmtrx{
1 & 0\\[1ex]
0 & -1
}~.
}
This orbifold geometrically corresponds to the M\"obius strip, see Figure~\ref{fig:Moebius}.

\begin{figure}[t]
\centering
\includegraphics[width=0.5\textwidth]{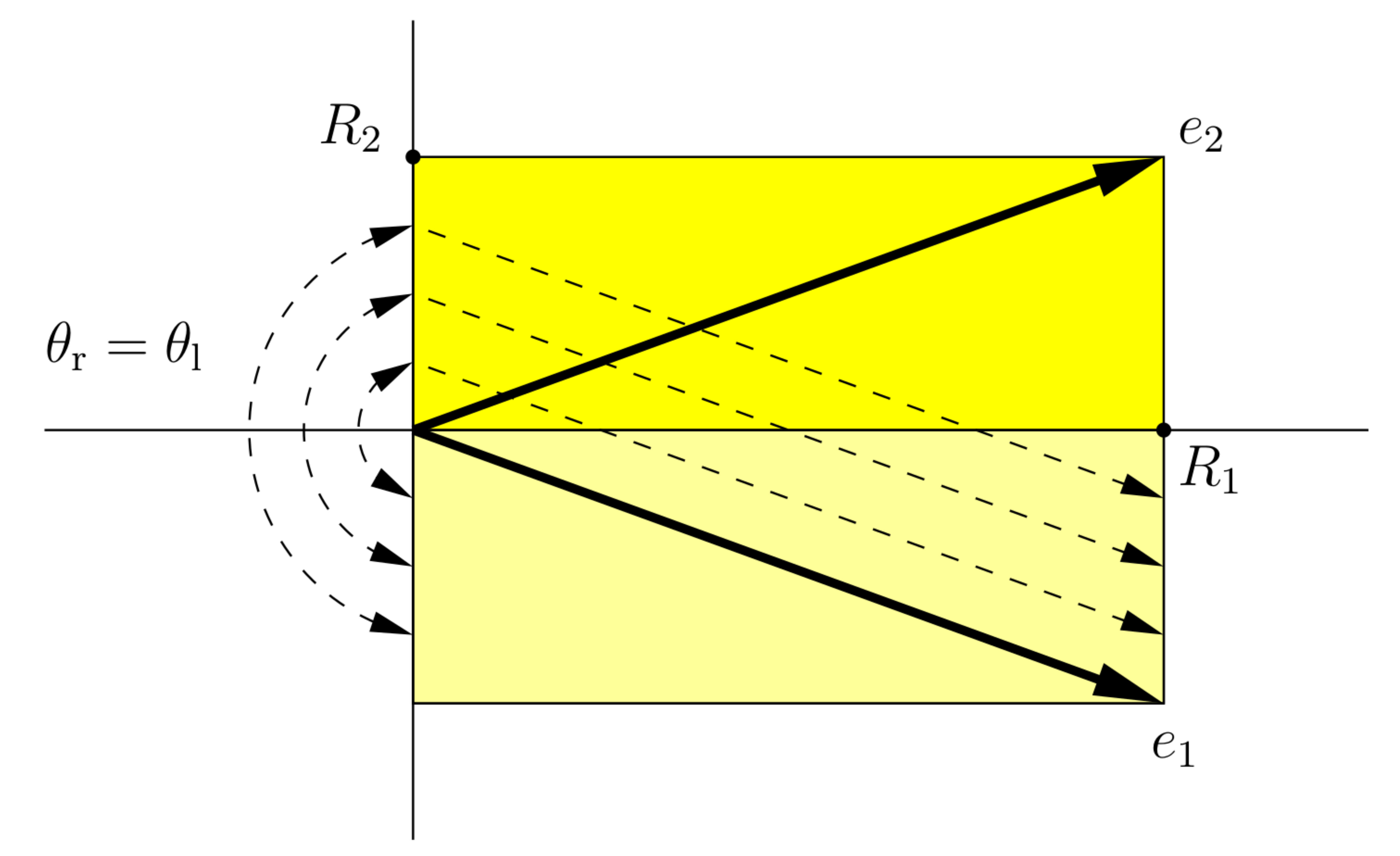}
\vspace*{-4mm}
\caption{Two--dimensional M\"obius strip as $\Z{2}$ orbifold: The underlying two-torus lattice is
spanned by $e_1$ and $e_2$. The upper and lower yellow regions combined give a convenient choice 
for the fundamental domain of the resulting two-torus. The symmetric twist 
$\gth_\text{r}=\gth_\text{l}$ gives a reflection at the horizontal axis. Consequently, we may take 
the lower yellow region to represent a fundamental domain of the resulting $\Z{2}$ orbifold. In 
this picture the 3+3 dashed arrows illustrate how the left and right side of the lower yellow 
region get glued together, hence this orbifold corresponds to the M\"obius strip.}
\label{fig:Moebius}
\end{figure}

Another symmetric $\Z{2}$ orbifold has a non-vanishing $B$-field: For this \Z{2}-II-3 Narain orbifold in Table~\ref{TabZN2DNarainOrbifolds} we choose
\equ{\label{eqn:Z2ExampleForZClass}
\widehat{\rho}_{(2)} ~=~ \pmtrx{
  0 & 1 & 0 &  0\\
  1 & 0 & 0 &  0\\
 -1 & 0 & 0 &  1\\
  0 & 1 & 1 &  0
} ~=~ \widehat{M}_B(\Delta B)^{-1}\, \widehat{\rho}_{(1)}\, \widehat{M}_B(\Delta B)~\in~ \text{O}_{\widehat{\get}}(2,2;\Intr)~,
}
where $\widehat{\rho}_{(1)}$ is the twist from the $\Z{2}$ orbifold discussed just above and 
$\widehat{M}_B(\Delta B)$ is a fractional $B$-field shift given by
\equ{
\widehat{M}_B(\Delta B) ~=~ \pmtrx{
 \Id_2 & 0\\
 \Delta B & \Id_2
}\qquad\text{for}\qquad \Delta B ~=~ \frac{1}{2}\,\epsilon~.
}
Now, we obtain
\equ{
\widehat{M}_{(2)1} ~=~ 
\pmtrx{
 1 & 0\\
 0 &-1
} + (G-B) \pmtrx{
 0 & 1\\
 1 & 0
} 
\qquad\text{and}\qquad 
\widehat{M}_{(2)2} ~=~ -\pmtrx{
 0 & 1\\
 1 & 0
}~.
}
In this case, the Narain moduli are given by
\equ{
e_{(2)} ~=~ \pmtrx{
 R_1 & R_1\\[1ex]
-R_2 & R_2
} 
\quad\Rightarrow\quad 
G_{(2)} ~=~ \pmtrx{
R_1^2+R_2^2 & R_1^2-R_2^2\\[1ex]
R_1^2-R_2^2 & R_1^2+R_2^2
} 
\quad\text{and}\quad 
B_{(2)} ~=~ \pmtrx{
 0 & \frac{1}{2}\\[1ex]
-\frac{1}{2} & 0
}~,
}
for $R_1R_2 \neq 0$. Furthermore, the twist $\gTh$ remains unchanged, i.e.
\equ{
\gth_\text{r} ~=~ \gth_\text{l} ~=~ \pmtrx{
1 & 0\\[1ex]
0 & -1
}~.
}
Note, that the metric $G_{(2)}$ is identical to $G_{(1)}$ from the case above; the only 
difference is that we now have a non-vanishing $B$-field.

The conjugation of the generator $\widehat{\rho}_{(1)}$ with $\widehat{M}_B(\Delta B)$ in 
eqn.~\eqref{eqn:Z2ExampleForZClass} tells us that these two Narain point groups belong to the same 
$\Ratl$-class. However, it turns out that they are from different $\Intr$-classes: There is no 
$\widehat{M}\in\text{O}_{\widehat{\get}}(D,D;\Intr)$ that can relate $\widehat{\gr}_{(1)}$ to 
$\widehat{\gr}_{(2)}$. Since the transformation~\eqref{eqn:Z2ExampleForZClass} is a conjugation 
with a discrete fractional $B$-field transformation, the $\Intr$-classes under investigation 
can be used to parametrize the inequivalent choices for the $B$-field for the given geometrical 
setting. As can be inferred from Table~\ref{TabZN2DNarainOrbifolds} we identified three 
inequivalent $\Intr$-classes for the $\Ratl$-class \Z{2}-II, where \Z{2}-II-1 and \Z{2}-II-2 
both have vanishing $B$-field but are based on inequivalent lattices.

\newpage

\appendix 
\def\theequation{\thesection.\arabic{equation}}

\section{Moduli deformations and the generalized metric} 
\label{sec:ModuliDeformations}
\setcounter{equation}{0}

Choose a specific generalized metric $\cH_0$. Next, consider the finite group of all discrete 
$T$-duality transformations that leaves this generalized metric invariant and choose a subgroup 
$\widehat{\mathbf{H}}$ thereof. Then, the general question, which we are addressing in this 
section, reads: What infinitesimal moduli deformations are allowed such that the deformed 
generalized metric stays invariant under all transformations from $\widehat{\mathbf{H}}$? 
We will answer this question in three steps. First, we define the group $\widehat{\mathbf{H}}$ in 
Appendix~\ref{sec:DefH}. Second, we parametrize all infinitesimal moduli deformations in 
Appendix~\ref{sec:allmoduli}. Third, in Appendix~\ref{sec:Hinvariantmoduli} we restrict them to the 
ones which are compatible with the action of $\widehat{\mathbf{H}}$. In addition, in 
Appendix~\ref{sec:numberofmoduli} we derive a closed expression which counts the number of moduli 
that are compatible with the action of $\widehat{\mathbf{H}}$. We use the results form this appendix in Section~\ref{sec:IdentifyModuli}, where we set 
$\widehat{\mathbf{H}} = \widehat{\mathbf{P}}$, i.e.~equal to the point group in the lattice basis. 
By doing so, we identify the moduli in Narain orbifolds.

\subsection[$T$-duality transformations that leave a generalized metric invariant]{$\boldsymbol{T}$-duality transformations that leave a generalized metric invariant}
\label{sec:DefH}

Consider a subgroup $\widehat{\mathbf{H}}$ of the group of all $\text{O}_{\hat\get}(D,D+16; \Intr)$ 
transformations which leave a specific generalized metric $\cH_0 = E_0^T E_0$ invariant, i.e.
\equ{ \label{Hinvariant}
\widehat{\mathbf{H}} ~\subseteq~ \Big\{ \widehat{M} ~\in~ \text{O}_{\hat\get}(D,D+16; \Intr) ~\Big|~ 
\widehat{M}^T \cH_0\, \widehat{M} = \cH_0 \Big\}~.
}
The following discussion is independent of whether $\widehat{\mathbf{H}}$ is Abelian or 
non-Abelian. Since the elements $\widehat{M} \in \widehat{\mathbf{H}}$ preserve both $\hat\get$ and 
$\cH_0$ we find that the corresponding element $\gTh(\widehat{M})$ as a function of $\widehat{M}$ 
is given by
\equ{ \label{DefUM} 
\gTh(\widehat{M}) ~=~ E_0\, \widehat{M}\, E_0^{-1} \quad\text{with}\quad 
\gTh(\widehat{M})^T \gTh(\widehat{M}) = \Id~, \quad
\gTh(\widehat{M})^T \get\, \gTh(\widehat{M}) = \get~.
}
Hence,
\equ{
\gTh(\widehat{M}) ~=~ \pmtrx{ \gth_\text{r}(\widehat{M})  & 0 \\ 0 & \gTh_\text{L}(\widehat{M}) } ~\subset~ \text{O}(D;\Real) \times \text{O}(D+16;\Real)~,
}
and $\gTh(\widehat{M})$ is a group homomorphism from $\widehat{\mathbf{H}}$ to a finite subgroup of 
$\text{O}(D;\Real) \times \text{O}(D+16;\Real)$.

\subsection[Infinitesimal moduli deformations of the Narain lattice]{Infinitesimal moduli deformations of the Narain lattice}
\label{sec:allmoduli}

We want to determine which parameters $\gd E$ in the generalized vielbein can be deformed 
infinitesimally, i.e.\ $E_0 \rightarrow E_0 + \gd E$ to first order in the perturbations. Since the 
generalized vielbein with $(2D+16)^2$ components is parametrized in terms of $D(D+16)$ parameters 
(i.e. the vielbein $e$, the $B$-field and the Wilson lines $A$), not all components of $\gd E$ are 
independent. To characterize the infinitesimal moduli perturbations without choosing a particular 
parametrization, we expand the constraint $E_0^T\, \get\, E_0 = \hat\get$ from eqn.~\eqref{Minkowski} 
to first order in $\gd E$ and obtain
\equ{\label{eqn:Eperturbations}
\gd E^T \get\, E_0 + E_0^T\get\, \gd E ~=~ 0~. 
}
This can be cast into the form
\equ{ 
\gd \frak{e}^T \get + \get\, \gd \frak{e} ~=~ 0~, 
}
where we have defined $\gd \frak{e} =  \gd E\, E_0^{-1}$. The general solution reads
\equ{ \label{DeltaFrake} 
\gd \frak{e} ~=~ \gd E\, E_0^{-1} = 
\frac 12\, \pmtrx{ \gd \frak{u}_D & \gd \frak{m} \\ 
\gd \frak{m}^T & \gd \frak{u}_{D+16} }~,
}
with $\gd \frak{m} \in M_{D\times (D+16)}(\Real)$. Furthermore, 
$\gd \frak{u}_D^T = - \gd \frak{u}_D$ and $\gd \frak{u}_{D+16}^T = - \gd \frak{u}_{D+16}$ generate 
$O(D;\Real)$ and $O(D+16;\Real)$, respectively. These orthogonal groups correspond to the $U$ 
transformation in eqn.~\eqref{eqn:MostGeneralGeneralizedVielbein}. Next, we consider the 
perturbations of the generalized metric $\gd \cH = \gd E^T E_0 + E_0^T \gd E$ to first order. Using 
eqn.~\eqref{eqn:Eperturbations} one can see that the constraint 
$(\hat\get^{-1} (\cH_0 + \gd \cH))^2 = \Id$ from eqn.~\eqref{ConstraintGenMetric} is fulfilled. In 
fact, we may write $\gd \frak{h} =  \gd \frak{e}^T + \gd \frak{e}$,  where $\gd\frak{e} = \frac 12\, \gd\frak{h} + \frac 12\, \gd\frak{u}$ with 
\equ{ \label{DeltaFraku}
\gd\frak{u} ~=~ - \gd\frak{u}^T ~=~ 
\pmtrx{ \gd \frak{u}_D & 0 \\  0 & \gd \frak{u}_{D+16} }~.
}
Hence, the infinitesimal moduli are uniquely identified by $\gd\frak{m}$, i.e.\ $\gd\frak{m}$ 
encodes the deformations of the metric $\gd G$, the $B$-field $\gd B$ and the Wilson lines $\gd A$. 
This can be stated explicitly as follows. We can determine $\gd \frak{e}$ by using 
eqn.~\eqref{DeltaFrake} with $E_0=R \widehat{E}$ and the expression for $\hat E$ given in 
eqn.~\eqref{NarainModuli}. Thereby we directly confirm that $\gd \frak{u}_D$ and 
$\gd \frak{u}_{D+16}$ are anti-symmetric and we derive that $\gd \frak{m}$ is given at linear order 
in the moduli perturbations $\gd G$, $\gd B$ and $\gd A$ as given in eqn.~\eqref{DeltaFrakh}, using 
$(e_0+\gd e)^{-1} \approx e^{-1} - e_0^{-1} \gd e\, e_0^{-1}$.

\subsection[$\widehat{H}$-invariant infinitesimal moduli deformations]{$\boldsymbol{\widehat{\mathbf{H}}}$-invariant infinitesimal moduli deformations}
\label{sec:Hinvariantmoduli}

In order to determine which of the Narain moduli are compatible with the action of 
$\widehat{\mathbf{H}}$ we consider the first order perturbation of eqn.~\eqref{Hinvariant} and 
obtain
\equ{ \label{PerturbationInv} 
\widehat{M}^{\,T} \gd\cH\, \widehat{M} ~=~ \gd \cH 
\quad \Leftrightarrow \quad 
\gTh(\widehat{M})^T \gd \frak{h}\, \gTh(\widehat{M}) ~=~  \gd \frak{h}~.
}
This reads on the level of the moduli deformations
\equ{ \label{ModuliInvariance}
\gth^T_\text{r}(\widehat{M})\, \gd \frak{m}\, \gTh_\text{L}(\widehat{M}) ~=~ \gd \frak{m}~,
} 
for each $\widehat{M} \in \widehat{\mathbf{H}}$. Eqn.~\eqref{ModuliInvariance} can be written as 
\equ{
\big( \gth_\text{r}(\widehat{M}) \otimes \gTh_\text{L}(\widehat{M})\big)\, \gd \frak{m} ~=~ \gd\frak{m}~.
}
Here, we interpret $\gd\frak{m}$ as a vector with $D(D+16)$ components using the standard tensor 
product notation $\otimes$. To solve this condition we introduce the projection operator 
$\cP_{\widehat{\mathbf{H}}}$ that projects the moduli perturbations on their 
$\widehat{\mathbf{H}}$-invariant subspace, i.e.
\equ{ \label{HinvProjection}
\cP_{\widehat{\mathbf{H}}} ~=~ \frac1{|\widehat{\mathbf{H}}|} \sum_{\widehat{M} \in \widehat{\mathbf{H}}}
 \gth_\text{r}(\widehat{M}) \otimes \gTh_\text{L}(\widehat{M}) 
 \quad\text{with}\quad 
 \big( \gth_\text{r}(\widehat{M}) \otimes \gTh_\text{L}(\widehat{M})\big)\, \cP_{\widehat{\mathbf{H}}} ~=~ \cP_{\widehat{\mathbf{H}}}~.
}
Using that $\gTh(\widehat{M})$ defines a group homomorphism, it is not difficult to show that this 
indeed defines a projection operator, i.e.\ $\cP_{\widehat{\mathbf{H}}}^2 = \cP_{\widehat{\mathbf{H}}}$. 
Thus, the $\widehat{\mathbf{H}}$-invariant moduli space is given by
\equ{
\cM_{\widehat{\mathbf{H}}} ~=~ \Big\{ \gd\frak{m}_{\widehat{\mathbf{H}}} ~=~ \cP_{\widehat{\mathbf{H}}} \gd\frak{m} \Big\}~.
}

\subsection[The number of $\widehat{H}$-invariant Narain moduli]{The number of $\boldsymbol{\widehat{\mathbf{H}}}$-invariant Narain moduli}
\label{sec:numberofmoduli}

The dimension of the $\widehat{\mathbf{H}}$-invariant Narain moduli space is determined by the 
trace of the projection operator $\cP_{\widehat{\mathbf{H}}}$, i.e.
\equ{ \label{eqn:NumberOfHModuli}
\dim(\cM_{\widehat{\mathbf{H}}}) 
~=~ 
\text{tr}(\cP_{\widehat{\mathbf{H}}}) 
~=~ 
 \frac1{|\widehat{\mathbf{H}}|} \sum_{\widehat{M} \in \widehat{\mathbf{H}}} \chi\left(\gth_\text{r}(\widehat{M})\right)\, \chi\left(\gTh_\text{L}(\widehat{M})\right)^* 
  ~=~ 
   \frac1{|\mathbf{H}|} \sum_{\gTh \in \mathbf{H}} \chi_\text{r}(\gTh)\, \chi_\text{L}(\gTh)^*~. 
}
Here, we have used the linearity of the trace, $\text{tr}(A\otimes B) = \text{tr}(A) \text{tr}(B)$ 
and we have used the definition~\eqref{ConstraintGenMetric}. In addition, we have included a 
complex conjugate in eqn.~\eqref{eqn:NumberOfHModuli} for later use. Furthermore, we have 
introduced the group characters
\begin{subequations}
\begin{eqnarray} 
\chi_\text{r}(\gTh) & = & \chi\left(\gth_\text{r}(\widehat{M})\right) ~=~ \text{tr}(\gth_\text{r}(\widehat{M})) ~=~ \text{tr}\Big[ \frac{\Id - \cZ}2\, \widehat{M}\Big]~,  
\\[1ex] 
\chi_\text{L}(\gTh) & = & \chi\left(\gTh_\text{L}(\widehat{M})\right) ~=~ \text{tr}(\gTh_\text{L}(\widehat{M})) ~=~ \text{tr}\Big[ \frac{\Id + \cZ}2\, \widehat{M}\Big]~,  
\end{eqnarray}
\end{subequations}
which are real for the real representations $\gth_\text{r}(\widehat{M})$ and 
$\gTh_\text{L}(\widehat{M})$, respectively.

%
%

\providecommand{\href}[2]{#2}\begingroup\raggedright\endgroup

\end{document}